\documentclass{article}
\usepackage[german, english]{babel}
\usepackage[a4paper, left=2.5cm, right=2.5cm, top=3.5cm, bottom=3cm]{geometry}
\usepackage{amssymb}
\usepackage{amsmath}
\usepackage{bbm}
\usepackage{amscd}
\usepackage{latexsym}
\usepackage{graphicx}
\usepackage{ifthen}
\usepackage{epsfig}
\usepackage{setspace}
\usepackage{subcaption}
\usepackage{hyperref}
\usepackage{dsfont}
\usepackage{mathrsfs}
\usepackage{color}
\hypersetup{
 colorlinks=true,
 linkcolor=blue,
 citecolor=magenta,
}

\catcode`@=11
\renewcommand{\section}{\@startsection{section}{1}{0pt}{\medskipamount}
{\medskipamount}{\large\bf}}
\numberwithin{equation}{section}
\catcode`@=12

\newcommand{\N}{\mathds N}
\newcommand{\R}{\mathds R}

\newcommand{\Acal}{{\cal A}}
\newcommand{\Fcal}{{\cal F}}
\newcommand{\Ecal}{{\cal E}}
\newcommand{\Bcal}{{\cal B}}

\newcommand{\Ical}{{\cal I}}

\newcommand{\cn}{{\mathrm{cn}\bigl(\sfrac{\tau}{\epsilon},k\bigr)}}
\newcommand{\sn}{{\mathrm{sn}\bigl(\sfrac{\tau}{\epsilon},k\bigr)}}
\newcommand{\dn}{{\mathrm{dn}\bigl(\sfrac{\tau}{\epsilon},k\bigr)}}

\def\im{\mathrm{i}}
\def\ep{\mathrm{e}}
\def\pa{\mbox{$\partial$}}
\def\diff{\mathrm{d}}
\def\tr{\mathrm{tr}}
\def\sfrac#1#2{{\textstyle\frac{#1}{#2}}}
\def\]{\right]}
\def\[{\left[}
\def\){\right)}
\def\({\left(}
\def\>{\rangle}
\def\<{\langle}
\def\+{\dagger}

\def\={\ =\ }
\newcommand{\unity}{\mathbbm{1}}
\def\und{\quad\textrm{and}\quad}
\def\with{\quad\textrm{with}\quad}
\def\for{\quad\textrm{for}\quad}
\def\3j#1#2#3#4#5#6{\begin{pmatrix} #1&#2&#3\\#4&#5&#6 \end{pmatrix}}
\def\s3j#1#2{\begin{pmatrix} #1\\#2 \end{pmatrix}}
\def\rep#1#2{\bigl[\begin{smallmatrix}#1\\#2\end{smallmatrix}\bigr]}

\begin{document}

\title{\bf\huge Instability of cosmic Yang-Mills fields}
\date{~}

\author{
{\Large Kaushlendra Kumar}\ ,\quad
{\Large Olaf Lechtenfeld} \ and \ 
{\Large Gabriel Pican\c{c}o Costa}
\\[24pt]
Institut f\"ur Theoretische Physik and\\ 
Riemann Center for Geometry and Physics\\
Leibniz Universit\"at Hannover \\ 
Appelstra{\ss}e 2, 30167 Hannover, Germany
\\[24pt]
} 

\clearpage
\maketitle
\thispagestyle{empty}

\begin{abstract}
\noindent\large
There exists a small family of analytic SO(4)-invariant but time-dependent 
SU(2) Yang--Mills solutions in any conformally flat four-dimensional spacetime. 
These might play a role in early-universe cosmology for stabilizing the symmetric Higgs vacuum. 
We analyze the linear stability of these ``cosmic gauge fields'' against general 
gauge-field perturbations while keeping the metric frozen, 
by diagonalizing the (time-dependent) Yang--Mills fluctuation operator around them and 
applying Floquet theory to its eigenfrequencies and normal modes.
Except for the exactly solvable SO(4) singlet perturbation, which is found to be
marginally stable linearly but bounded nonlinearly,
generic normal modes often grow exponentially due to resonance effects.
Even at very high energies, all cosmic Yang--Mills backgrounds are rendered linearly unstable.
\end{abstract}

\newpage
\setcounter{page}{1} 

\section{Introduction: a tale of three anharmonic oscillators}

\noindent
Classical Yang--Mills fields play a central role in various areas of theoretical physics,
from QCD confinement to spin-orbit interactions in condensed matter theory and early-universe cosmology.
Concerning the latter, scenarios have been proposed and analyzed for isotropic inflation driven by non-Abelian
gauge fields, such as gauge-flation or chromo-natural inflation, by employing a homogeneous and isotropic
Yang--Mills background in a spatially flat Friedmann--Lema\^{i}tre--Robertson--Walker (FLRW) universe
(for a review, see~\cite{gaugeflation}. More minimalistically, Friedan has recently put forward an evolution of
the electroweak epoch based on the Standard Model and general relativity alone, where an oscillating isotropic
SU(2) gauge field stabilizes the symmetric Higgs vacuum in a spatially closed FLRW spacetime~\cite{Friedan}.
It is therefore of natural interest to establish the stability features of such ``cosmic Yang--Mills fields'' against
classical and quantum perturbations. While this has been done using cosmological perturbation theory in the
context of gauge-flation, the issue remains unclear in the pre-inflation scenario of Friedan,
despite some early partial analysis of Hosotani~\cite{Hosotani}.

Here, we address this matter by performing a complete stability analysis of the only known family 
of analytic SU(2) Yang--Mills solutions in a closed FLRW universe. The situation differs from that of 
gauge-flation, since we do not break conformal invariance, and our homogeneous and isotropic 
gauge background would conformally map to an inhomogeneous Yang--Mills configuration in 
spatially flat FLRW spacetime. Our analysis, however, is partial in that we investigate only 
gauge-field fluctuations while keeping the metric fixed, for two reasons. Firstly,
the background FLRW dynamics does not influence the gauge field, since the gauge sector 
is conformally invariant and we conformally map to a static metric. Secondly, the time scale of the 
gauge-field dynamics in the electroweak epoch is supposed to be hugely shorter than that of 
the spacetime geometry, thus we do not expect (non-conformal) metric fluctuations to exert 
a sizeable influence on the stability of the gauge-field configuration.
Our results should therefore be relevant to Friedan's scenario. 
Nevertheless, a full cosmological perturbation theory of the combined Einstein--Yang--Mills system 
requires turning on also metric perturbations, something we reserve for a future task.

Even without coupling to gravity and independent of potential cosmological applications, 
the perturbation theory around any analytic classical field configuration is of general interest 
for assessing its relevance for quantum properties, since the determinant of the second variation 
of the action yields the leading quantum correction to a saddle point in the semiclassical analysis 
of the path integral. This aspect has been investigated for many classical Yang--Mills solutions
but not yet for the ones studied here, to our knowledge.

It is generally impossible to find analytic solutions to the coupled Einstein--Yang--Mills system of equations,
in part because they are coupled both ways. However, in a homogeneous and isotropic universe, 
where the metric is conformally flat, the Yang--Mills equations decouple 
due to their conformal invariance in four spacetime dimensions.
Thus, if one can find isotropic Yang--Mills solutions on Minkowski, de Sitter, or anti de Sitter space, 
then their energy-momentum tensor will be compatible with any FLRW metric (of the same topology)
and allow for an analytic computation of the scale factor from the Friedmann equation.

Fortunately, for the de Sitter case and a gauge group SU(2), 
such Yang--Mills configurations with finite energy and action are available 
\cite{AFF,Luescher,Schechter}.
They are most easily constructed on the cylinder $(0,\pi)\times S^3$, 
which is related to de Sitter space by a purely temporal reparametrization
and Weyl rescaling~\cite{Hosotani,Volkov,ILP1,Friedan},
\begin{equation} \label{deSittermetric}
\diff s_{\textrm{dS}_4}^2 \= -\diff t^2 + \ell^2\cosh^2\!\sfrac{t}{\ell}\,\diff\Omega_3^2
\= \sfrac{\ell^2}{\sin^2\!\tau} \bigl(-\diff\tau^2 + \diff\Omega_3^2\bigr)
\qquad\textrm{for}\quad t\in(-\infty,+\infty)\ \Leftrightarrow\ \tau \in (0,\pi)\ ,
\end{equation}
where $\diff\Omega_3^{\ 2}$ is the round metric on $S^3$, and $\ell$ is the de Sitter radius. 
This provides an explicit relation between co-moving time~$t$ and conformal time~$\tau$,
and it fixes the cosmological constant to~$\Lambda=3/\ell^2$.
One employs the identification SU$(2)\simeq S^3$ to write an $S^3$-symmetric ansatz
for the SU(2) gauge potential~$A_\mu$,
which produces a solvable ODE for a parameter function~$\psi(\tau)$ of conformal time.
This ODE has the form of Newton's equation for a mass point in a double-well potential
\begin{equation} \label{Vpot}
V(\psi) \= \sfrac12 (\psi^2-1)^2\ ,
\end{equation}
yielding a first anharmonic oscillator.
A two-parameter family of (in general time-dependent) solutions can be given 
in terms of Jacobi elliptic functions and describe SO(4) invariant Yang--Mills fields on 
the Lorentzian cylinder over~$S^3$.

These Yang--Mills solutions then exist (at least locally) in any conformally related spacetime,
but the conformal transformation will ruin isotropy unless we restrict ourselves to
spatially closed FLRW metrics,
\begin{equation} \label{FLRW}
\diff s^2 \= -\diff t^2 + a(t)^2\,\diff\Omega_3^{\ 2}
\= a(\tau)^2\bigl(-\diff\tau^2 + \diff\Omega_3^{\ 2}\bigr)
\qquad\textrm{for}\quad t\in(0,t_{\textrm{max}})\ \Leftrightarrow\ \tau \in \Ical \equiv (0,T')\ ,
\end{equation}
where we impose a big-bang initial condition~$a(0){=}0$, so that
\begin{equation} 
\diff\tau \= \frac{\diff t}{a(t)} \qquad\textrm{with}\qquad 
\tau(t{=}0) = 0 \quad\und\quad \tau(t{=}t_{\textrm{max}}) =: T'<\infty\ .
\end{equation}
The lifetime $t_{\textrm{max}}$ of the universe 
can be infinite (big rip, $a(t_{\textrm{max}}){=}\infty$) 
or finite (big crunch, $a(t_{\textrm{max}}){=}0$).
Bouncing cosmologies as in (\ref{deSittermetric}) are also allowed but will not be pursued here.
Since the energy-momentum tensor of our Yang--Mills configurations is SO(4) symmetric,
their gravitational backreaction will keep us inside the FLRW framework and merely modify the
cosmic scale factor~$a(\tau)$. 
The latter is fully determined by the Friedmann equation in the presence of
the Yang--Mills energy-momentum and a cosmological constant~$\Lambda$, whose value may be dialed.
It is well known that the Friedmann equation takes the form of another Newton equation.
Its (cosmological) potential for the case at hand reads
\begin{equation} \label{Wpot}
W(a) \= \sfrac12 a^2 - \sfrac{\Lambda}{6} a^4\ ,
\end{equation}
which is our second anharmonic oscillator (although inverted). 
Each pair $(\psi,a)$ of solutions to the two systems (\ref{Vpot}) and~(\ref{Wpot}) 
yields an exact classical Einstein--Yang--Mills configuration. 
One parameter in $\psi$ is the conserved mechanical energy $E$ in the potential~$V$, 
which in turn determines the mechanical energy~$\tilde{E}$ for $a$ in the potential~$W$. 
This one-way coupling is the only relation between the two anharmonic oscillators.

For a physical embodiment of the cosmological constant, we may add a third player,
for instance a complex scalar Higgs field~$\phi$ in the fundamental SU(2) representation.
The standard-model Higgs potential
\begin{equation}
U(\phi) \= \sfrac12\,\lambda^2\,\bigl(\phi^\+\phi-\sfrac12 v^2\bigr)^2\ ,
\end{equation}
where $v/\sqrt{2}$ is the Higgs vev and $\lambda v$ is the Higgs mass, 
gives us a third anharmonic oscillator.
The dictate of SO(4) invariance, however, allows only the zero solution, $\phi\equiv0$,
which provides us with a definite positive cosmological constant of
\begin{equation}
\Lambda \= \kappa\,U(0) \= \sfrac18\,\kappa\,\lambda^2 v^4\ ,
\end{equation}
where $\kappa$ is the gravitational coupling.
The full Einstein--Yang--Mills--Higgs action (in standard notation),
\begin{equation}
S \= \int\!\diff^4 x\ \sqrt{-g}\ \Bigl\{ \sfrac{1}{2\kappa} R + \sfrac{1}{8g^2} \tr F_{\mu\nu}F^{\mu\nu}
-D_\mu\phi^\+ D^\mu\phi - U(\phi) \Bigr\}\ ,
\end{equation}
reduces in the SO(4)-invariant sector to
\begin{equation}
S[a,\psi,\Lambda] \= 12\pi^2 \int_0^{T'}\!\!\diff\tau\ \Bigl\{
\sfrac{1}{\kappa} \bigl(-\sfrac12\dot{a}^2+W(a)\bigr) + 
\sfrac{1}{2g^2} \bigl(\sfrac12\dot{\psi}^2-V(\psi)\bigr) \Bigr\}\ ,
\end{equation}
where $g$ is the gauge coupling, and the overdot denotes a derivative with respect to conformal time.

For large enough ``gauge energy''~$E$, the universe undergoes an eternal expansion, 
which is accompanied by rapid fluctuations of the gauge field. 
The latter's coupling to the Higgs field stabilizes the symmetric vacuum $\phi\equiv0$ 
at the local maximum of~$U$ as a parametric resonance effect, 
as long as $a$ is not too large. 
Eventually, when $a$ exceeds a critical value $a_{\textrm{EW}}$,
the Higgs field will begin to roll down towards a minimum of~$U$,
breaking the SO(4) symmetry. 
The corresponding time $t_{\textrm{EW}}$ signifies the electroweak phase transition in the early universe.
This scenario was put forward recently by D.~Friedan~\cite{Friedan}.

The goal of the current paper is a stability analysis of these classical oscillating ``cosmic'' Yang--Mills fields.
To begin with, Section~2 describes the geometry of $S^3$
and reviews the classical configurations~$(A_\mu,g_{\mu\nu})$ 
in terms of Newtonian solutions~$(\psi,a)$ for the anharmonic oscillator pair~$(V,W)$.
To investigate arbitrary small perturbations of the gauge field 
departing from the time-dependent background~$A_\mu$ parametrized by the ``gauge energy''~$E$, 
Section~3 linearizes the Yang--Mills equation around it and diagonalizes the fluctuation operator 
to obtain a spectrum of time-dependent natural frequencies.
To decide about the linear stability of the cosmic Yang--Mills configurations we have to
analyze the long-time behavior of the solutions to Hill's equation for all these normal modes.
In Section~4 we employ Floquet theory to learn that their growth rate is determined 
by the stroboscopic map or monodromy, which is easily computed numerically for any given mode. 
We do so for a number of low-frequency normal modes and find, when varying~$E$, 
an alternating sequence of stable (bounded) and unstable (exponentially growing) fluctuations.
The unstable bands roughly correspond to the parametric resonance frequencies.
With growing ``gauge energy'' the runaway perturbation modes become more prominent,
and some of them persist in the infinite-energy limit, where we detect universal 
natural frequencies and monodromies.
A special role is played by the SO(4)-invariant fluctuation, 
which merely shifts the parameter~$E$ of the background.
We treat it exactly and beyond the linear regime in Section~5.
This ``singlet'' mode turns out to be marginally stable, i.e.~it has a vanishing Lyapunov exponent. 
Its linear growth, however, gets limited by nonlinear effects of the full fluctuation equation,
whose analytic solutions exhibit wave beat behavior. 
Finally, some explicit data for the first few natural frequencies are collected in an Appendix.


\section{Cosmic Yang--Mills solutions}

\noindent
In order to describe the classical Yang--Mills solutions
we need to develop some elements of the spatial $S^3$ geometry.
Taking advantage of the fact that
\begin{equation}
S^3 \simeq \textrm{SU}(2) \quad\und\quad so(4) \simeq su(2)_L\oplus su(2)_R
\end{equation}
we introduce a basis $\{L_a\}$ for $a=1,2,3$ of left-invariant vector fields on~$S^3$
generating the right multiplication on~SU(2) and forming the $su(2)_L$ algebra
\begin{equation}\label{vecfields}
\[ L_a , L_b \] \= 2\,\varepsilon_{ab}^{\ \ c}\,L_c \ .
\end{equation}
It is dual to a basis $\{e^a\}$ of left-invariant one-forms on~$S^3$, 
i.e.~$e^a(L_b)=\delta^a_{\ b}$, subject to
\begin{equation}
\diff e^a + \varepsilon^a_{\ bc}\,e^b\wedge e^c \=0 \quad\und\quad e^a e^a \= \diff\Omega_3^{\ 2}\ .
\end{equation}
One may obtain this basis by expanding the left Cartan one-form 
\begin{equation}
\Omega_{_L}(g) \= g^{-1}\, \diff g \= e^a\,L_a \ .
\end{equation}
Here, the group element~$g$ provides the identification map
\begin{equation}
g: \ S^3 \rightarrow \textrm{SU}(2) \qquad\textrm{via}\quad
(\alpha,\beta) \mapsto -\im \begin{pmatrix} \beta & \alpha^* \\ \alpha & -\beta^* \end{pmatrix}
\quad\with |\alpha|^2+|\beta|^2=1\ ,
\end{equation}
which sends the $S^3$ north pole $(0,\im)$ to the group identity $\mathds{1}_2$. 
We shall coordinatize $S^3$ by an SU(2) group element~$g$.
The $su(2)_R$ half of the three-sphere's $so(4)$ isometry is provided by right-invariant vector fields~$R_a$
belonging to the left multiplication on the group manifold and obeying
\begin{equation}
\[ R_a , R_b \] \= 2\,\varepsilon_{ab}^{\ \ c}\,R_c\ .
\end{equation}
The differential of a function~$f$ on~${\cal I}\times S^3$ is then conveniently taken as
\begin{equation}
\diff f \= \diff\tau\,\pa_\tau f\,+\,e^a L_a f\ .
\end{equation}

Functions on $S^3$ can be expanded in a basis of harmonics~$Y_j(g)$ with $2j\in\N_0$, 
which are eigenfunctions of the scalar Laplacian,\footnote{
The SO(4) spin of these functions is actually $2j$, but we label them with half their spin, for reasons to be clear below.}
\begin{equation}
-\mathop{}\!\mathbin\bigtriangleup_3 Y_j \= 2j(2j{+}2)\,Y_j \= 4j(j{+}1)\,Y_j \= -\sfrac12(L^2+R^2)\,Y_j \ ,
\end{equation}
where $L^2=L_a L_a$ and $R^2=R_a R_a$ are (minus four times) the Casimirs of $su(2)_L$ and $su(2)_R$, respectively,
\begin{equation} 
-\sfrac14 L^2\,Y_j \= -\sfrac14 R^2\,Y_j \= -\sfrac14\mathop{}\!\mathbin\bigtriangleup_3 Y_j \= j(j{+}1)\,Y_j\ .
\end{equation}
The left-right (or toroidal) harmonics~$Y_{j;m,n}$ are eigenfunctions of $L^2=R^2$, $L_3$ and~$R_3$,
\begin{equation} \label{Y-action1}
\sfrac{\im}{2}\,L_3\,Y_{j;m,n} \= m\,Y_{j;m,n} \quad\und\quad
\sfrac{\im}{2}\,R_3\,Y_{j;m,n} \= n\,Y_{j;m,n} 
\end{equation}
for $m,n=-j,-j{+}1,\ldots,{+}j$, 
and hence the corresponding ladder operators 
\begin{equation}
L_\pm \= (L_1\pm\im L_2)/\sqrt{2} \quad\und\quad R_\pm \= (R_1\pm\im R_2)/\sqrt{2}
\end{equation}
act as
\begin{equation} \label{Y-action2}
\sfrac{\im}{2}\,L_\pm\,Y_{j;m,n} \= \sqrt{(j{\mp}m)(j{\pm}m{+}1)/2}\,Y_{j;m\pm1,n} \quad\!\und\!\quad
\sfrac{\im}{2}\,R_\pm\,Y_{j;m,n} \= \sqrt{(j{\mp}n)(j{\pm}n{+}1)/2}\,Y_{j;m,n\pm1}\ .
\end{equation}

The  gauge potential is an $su(2)$-valued one-form on our spacetime.
We use the ${\cal I}\times S^3$ parametrization and write
\begin{equation} \label{Acalfull}
\Acal \= \Acal_\tau(\tau,g)\,\diff\tau\ +\ \sum_{a=1}^3 \Acal_a(\tau,g)\,e^a(g)
\quad\with g \in \textrm{SU}(2)\ .
\end{equation}
It has been shown~\cite{Luescher, Friedan} that 
the requirement of SO(4) equivariance enforces the form
\begin{equation}\label{Aansatz}
\Acal_\tau(\tau,g) \= 0 \quad\und\quad  
\Acal_a(\tau,g) \= \sfrac12\bigl(1+\psi(\tau)\bigr)\,T_a
\end{equation}
with some function $\psi: {\cal I}\to\R$,
where $T_a$ denotes the $su(2)$ generators subject to
\begin{equation}
\[ T_a , T_b \] \= 2\,\varepsilon_{ab}^{\ \ c}\,T_c\ ,
\end{equation}
so that the adjoint representation produces $\tr(T_a T_b)=-8\,\delta_{ab}$.
The corresponding field strength reads
\begin{equation}\label{2-form}
\begin{aligned}
\Fcal &\= \diff\Acal + \Acal\wedge\Acal 
\= \partial_\tau\Acal_a\,\diff\tau{\wedge}e^a + \sfrac12\bigl(
R_{[b}\Acal_{c]} - 2\varepsilon_{bc}^{\ \ a}\Acal_a + [\Acal_b,\Acal_c]\bigr)e^b{\wedge}\,e^c\\[4pt]
&\= \sfrac12\dot\psi\,T_a\,\diff\tau{\wedge}e^a
\ +\ \sfrac14\varepsilon^a_{\ bc} (\psi^2{-}1)\,T_a\,e^b{\wedge}e^c
\end{aligned}
\end{equation}
where $\dot{\psi}\equiv\partial_\tau\psi$.
The Yang--Mills action on this ansatz simplifies to
\begin{equation}\label{YMaction}
S \= \frac{-1}{4g^2}\int_{\Ical\times S^3} \!\!\! \tr\ \Fcal\wedge *\Fcal \=
\frac{6\pi^2}{g^2} \int_{\cal I} \!\diff\tau\ \bigl[ \sfrac12\dot\psi^2 - V(\psi) \bigr]
\qquad\textrm{with}\quad V(\psi)\=\sfrac12(\psi^2{-}1)^2\ ,
\end{equation}
where ${\cal I}=[0,T']$ and $g$ here denotes the gauge coupling.
Due to the principle of symmetric criticality~\cite{Palais}, solutions to the mechanical problem 
\begin{equation} \label{Newton}
\ddot{\psi} + V'(\psi) \= 0
\end{equation}
will, via (\ref{Aansatz}), provide Yang--Mills configurations which extremize the action.
Conservation of energy implies that
\begin{equation} \label{energylaw}
\sfrac12\dot{\psi}^2 +V(\psi) \= E \= \textrm{constant}\ ,
\end{equation}
and the generic solution in the double-well potential~$V$ is periodic in~$\tau$ with a period~$T(E)$.

Hence, fixing a value for~$E$ and employing time translation invariance to set $\dot{\psi}(0)=0$
uniquely determines the classical solution~$\psi(\tau)$ up to half-period shifts.
Its explicit form is
\begin{equation} \label{backgrounds}
\psi(\tau) \= \begin{cases}
\ \sfrac{k}{\epsilon}\,\mathrm{cn}\bigl(\sfrac{\tau}{\epsilon},k\bigr)  
\qquad\qquad\ \ \,\textrm{with}\quad T=4\,\epsilon\,K(k) 
& \textrm{for}\quad \sfrac12<E<\infty \\[4pt]
\ 0 \qquad\qquad\qquad\qquad\quad\! \textrm{with}\quad T=\infty 
& \textrm{for}\quad E=\sfrac12 \\[4pt]
\ \pm\sqrt{2}\,\mathrm{sech}\bigl(\sqrt{2}\,\tau\bigr) 
\qquad\ \textrm{with}\quad T=\infty 
& \textrm{for}\quad E=\sfrac12 \\[4pt]
\ \pm\sfrac{k}{\epsilon}\,\mathrm{dn}\bigl(\sfrac{k\,\tau}{\epsilon},\sfrac1k\bigr)  
\qquad\quad\ \textrm{with}\quad T=2\,\sfrac{\epsilon}{k}\,K(\sfrac1k) \quad
& \textrm{for}\quad 0<E< \sfrac12 \\[4pt]
\ \pm 1 \qquad\qquad\qquad\qquad \textrm{with}\quad T=\pi
&  \textrm{for}\quad E=0 
\end{cases}\ ,
\end{equation}
where cn and dn denote Jacobi elliptic functions, $K$ is the complete elliptic integral of the first kind, and
\begin{equation}
2\,\epsilon^2 \= 2k^2{-}1 \= 1/\sqrt{2E} \qquad\textrm{with}\quad
k=\sfrac{1}{\sqrt{2}},1,\infty \quad\Leftrightarrow\quad E=\infty,\sfrac12,0\ .
\end{equation}
For $E{\gg}\frac12$, we have $k^2{\to}\frac12$, and the solution is well approximated by 
$\frac{2}{\epsilon}\cos\bigl(\frac{2\sqrt{\pi^3}}{\Gamma(1/4)^2}\frac{\tau}{\epsilon}\bigr)$.
At the critical value of $E{=}\sfrac12$ ($k{=}1$), the unstable constant solution coexists 
with the celebrated bounce solution, and below it the solution bifurcates into oscillations 
in the left or right well of the double-well potential, which halfens the oscillation period. 
The two constant minima $\psi=\pm1$ correspond to the vacua $\Acal=0$ and $\Acal=g^{-1}\diff g$.
Actually, the time translation freedom is broken by the finite range of~$\Ical$,
so that time-shifted solutions differ in their boundary values $\psi(0)$ and $\psi(T')$
and also in their value for the  action.
\begin{figure}[h!]
\centering
\includegraphics[width = 0.35\paperwidth]{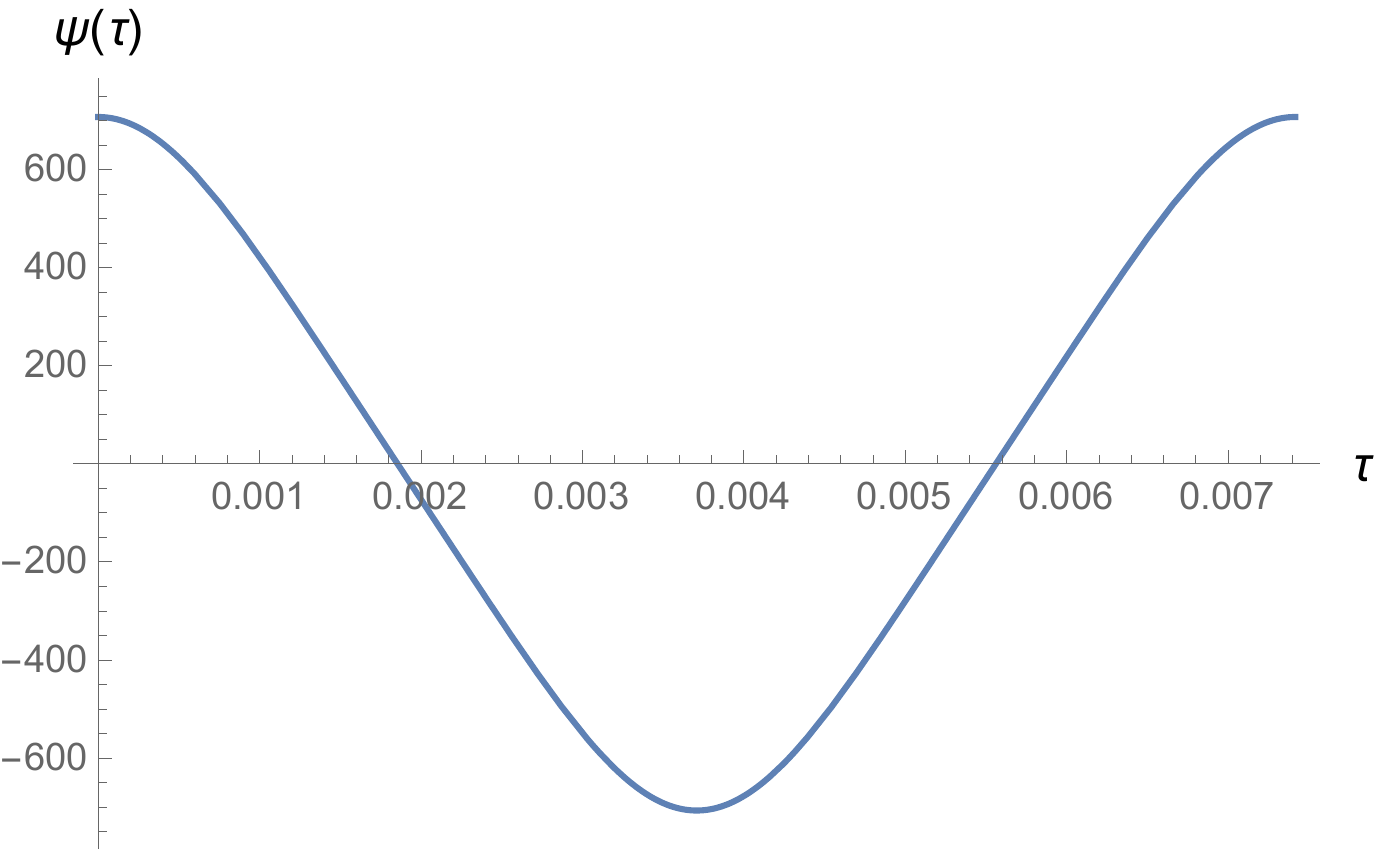} \qquad\quad
\includegraphics[width = 0.35\paperwidth]{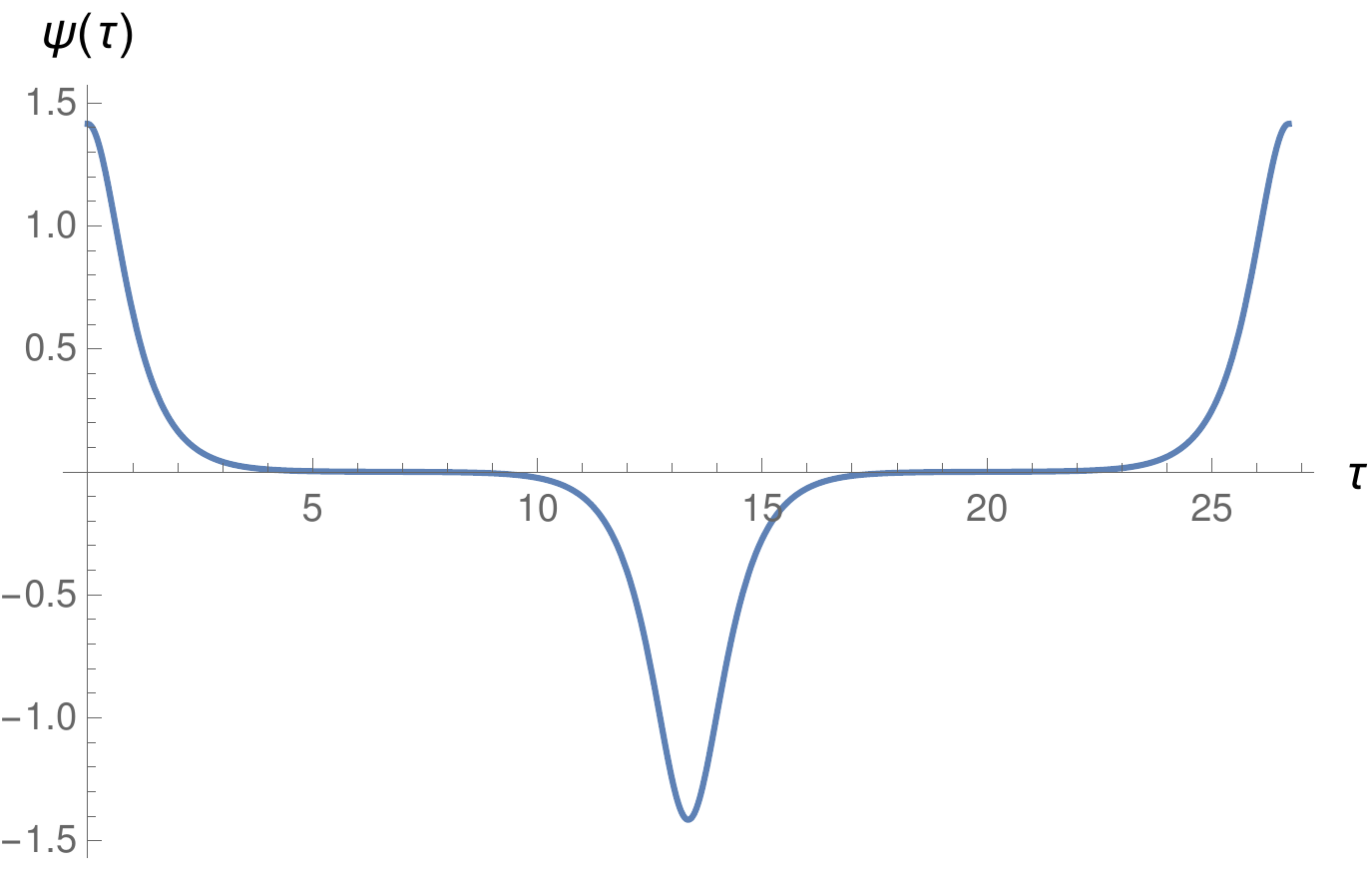} \\[12pt]
\includegraphics[width = 0.35\paperwidth]{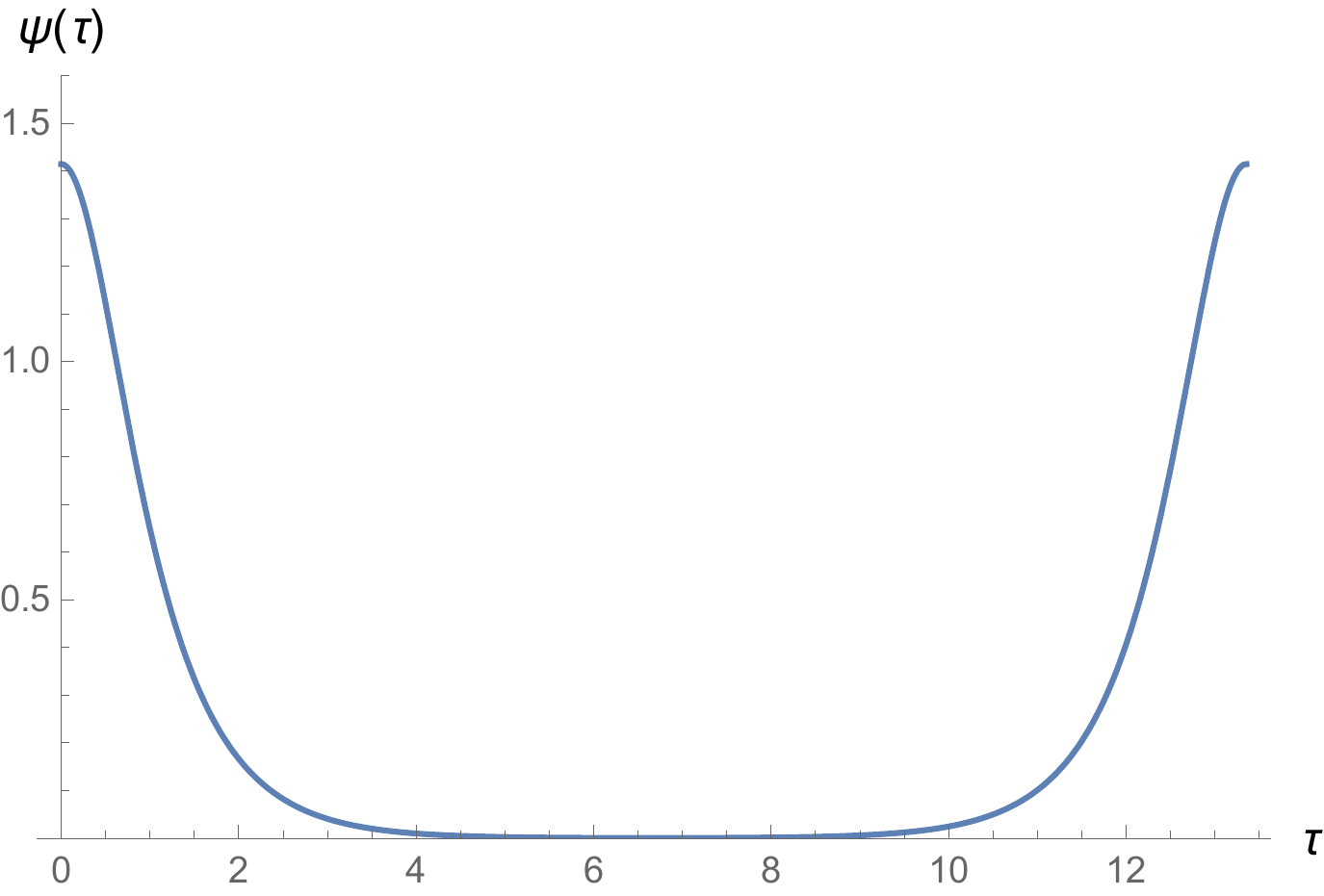} \qquad\quad
\includegraphics[width = 0.35\paperwidth]{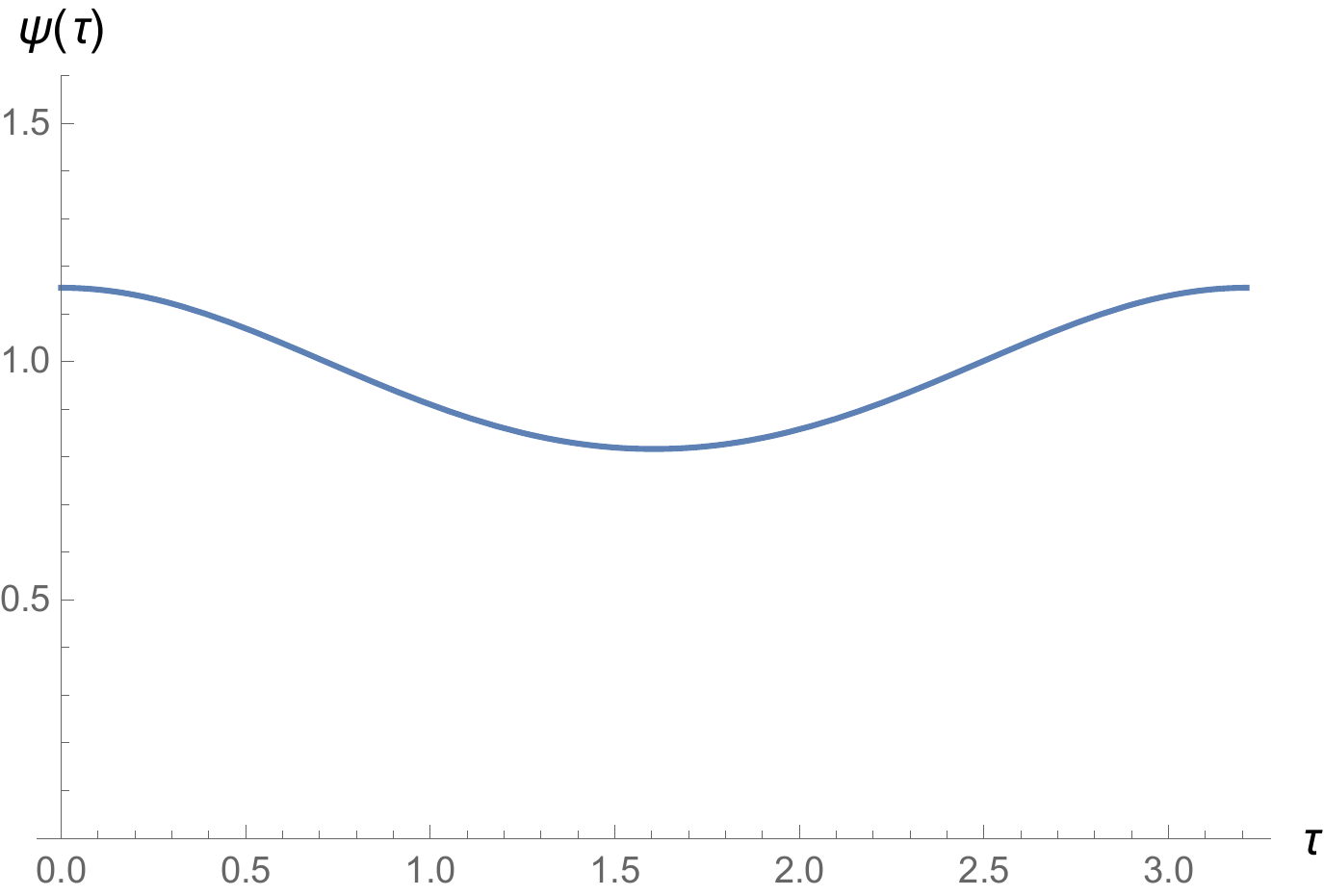}
\caption{Plots of $\psi(\tau)$ over one period, for different values of $k^2$: \newline
\indent\qquad\qquad\
$0.500001$ (top left), $0.9999999$ (top right), $1.0000001$ (bottom left) and $2$ (bottom right).}
\end{figure}

The corresponding color-electric and -magnetic field strengths read
\begin{equation}
\Ecal_a \= \Fcal_{0a} \= \sfrac12\dot\psi\,T_a \quad\und\quad
\Bcal_a \= \sfrac12\varepsilon_a^{\ bc}\Fcal_{bc} \= \sfrac12(\psi^2{-}1)\,T_a\ ,
\end{equation}
which yields a finite total energy (on the cylinder) of $6\pi^2 E/g^2$ and a finite action~\cite{ILP1,ILP2}
\begin{equation}
g^2 S[\psi] \= 6\pi^2\int_\Ical \!\diff\tau\ \bigl[E-(\psi^2{-}1)^2\bigr]
\=6\pi^2\int_\Ical \!\diff\tau\ \bigl[\dot{\psi}^2-E\bigr]\ \ge\ -3\pi^2 T'\ .
\end{equation}
The energy-momentum tensor of our SO(4)-symmetric Yang--Mills solutions is readily found as
\begin{equation}
T \= \frac{3\,E}{g^2 a^2} \bigl( \diff\tau^2 + \sfrac13\,\diff\Omega_3^2 \bigr)\ ,
\end{equation}
which is traceless as expected.

The Einstein equations for a closed FLRW universe with cosmological constant~$\Lambda$
reduce to two independent relations, which can be taken to be its trace and its time-time component.
In conformal time one gets, respectively,
\begin{equation} \label{friedmann}
\left.\begin{cases}
\quad -R+4\,\Lambda \= 0 \\[4pt]
\quad R_{\tau\tau}+\sfrac12R\,a^2-\Lambda\,a^2\= \kappa\,T_{\tau\tau}
\end{cases} \right\}
\qquad\Leftrightarrow\qquad
\left.\begin{cases}
\quad \ddot{a}+W'(a)\=0 \\[4pt]  \quad \sfrac12\dot{a}^2+W(a) \= \frac{\kappa}{2g^2}E \ =:\ E'
\end{cases} \right\}
\end{equation}
with a gravitational coupling $\kappa=8\pi G$, 
a gravitational energy $E'$
and a cosmological potential 
\begin{equation} \label{gravpot}
W(a) \= \sfrac12 a^2 - \sfrac{\Lambda}{6} a^4\ .
\end{equation}
The two anharmonic oscillators, with potential~$V$ for the gauge field and potential~$W$ for gravity,
are coupled only via the balance of their conserved energies,
\begin{equation}
\frac1\kappa \bigl[ \sfrac12\dot{a}^2+W(a)\bigr] \= \frac1{2g^2} \bigl[ \sfrac12\dot{\psi}^2+V(\psi)\bigr]\ ,
\end{equation}
which is nothing but the Wheeler--DeWitt constraint~$H=0$.
\begin{figure}[h!]
\centering
\includegraphics[width = 0.35\paperwidth]{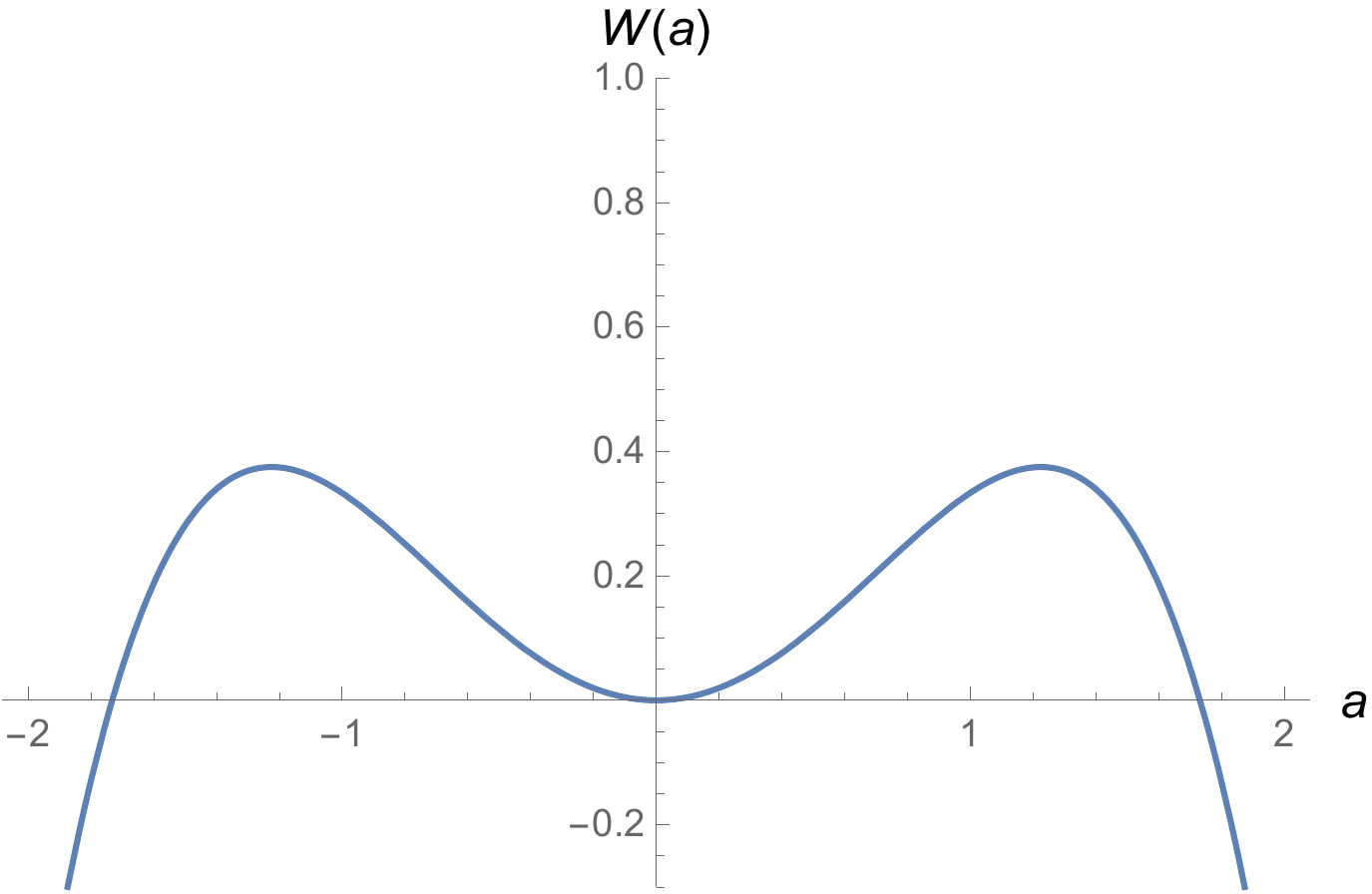} \qquad\quad
\includegraphics[width = 0.35\paperwidth]{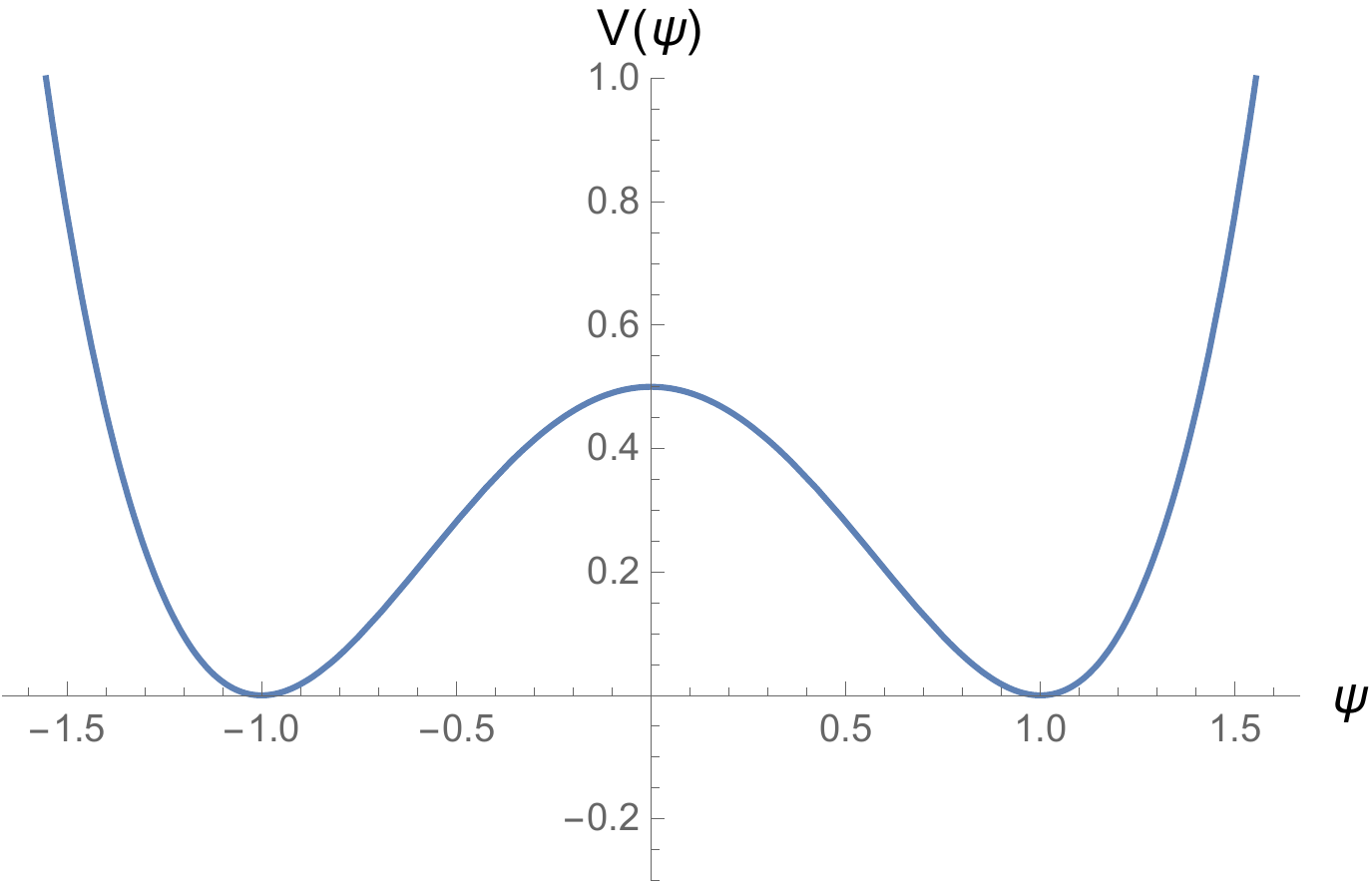} 
\caption{Plots of the cosmological potential $W(a)$ for $\Lambda{=}1$ and the double-well potential $V$.}
\end{figure}

The Friedmann equation~(\ref{friedmann}), being a mechanical system with an inverted anharmonic
potential~(\ref{gravpot}), is again easily solved analytically,
\begin{equation} \label{universalscale}
a(\tau) \= \begin{cases}
\ \sqrt{\sfrac{3}{\Lambda}}\,\sfrac{1}{2\epsilon'}\,
\sqrt{\frac{1-\mathrm{cn}\bigl(\sfrac{\tau}{\epsilon'},k'\bigr)}{1+\mathrm{cn}\bigl(\sfrac{\tau}{\epsilon'},k'\bigr)}}
\qquad\qquad  \textrm{with}\quad T'=2\,\epsilon'K(k') 
& \textrm{for}\quad \sfrac{3}{8\Lambda}<E'<\infty \\[4pt]
\ \sqrt{\sfrac{3}{2\Lambda}}\,\tanh\bigl(\tau/\sqrt{2}\bigr) 
\qquad\qquad\quad\  \textrm{with}\quad T'=\infty 
& \textrm{for}\quad E'=\sfrac{3}{8\Lambda} \\[4pt]
\ \sqrt{\sfrac{3}{\Lambda}}\,\sfrac{1}{2\epsilon'}\,
\sqrt{\frac{1-\mathrm{dn}\bigl(\sfrac{k'\tau}{\epsilon'},\sfrac{1}{k'}\bigr)}
{1+\mathrm{dn}\bigl(\sfrac{k'\tau}{\epsilon'},\sfrac{1}{k'}\bigr)}}
\qquad\quad \textrm{with}\quad T'=2\,\sfrac{\epsilon'}{k'}K(\sfrac{1}{k'}) \quad
& \textrm{for}\quad 0<E'< \sfrac{3}{8\Lambda} \\[10pt]
\ 0 \qquad\qquad\qquad\qquad\qquad\qquad\ \; \textrm{with}\quad T'=\pi
&  \textrm{for}\quad E'=0 
\end{cases}\ ,
\end{equation}
where we abbreviated~\footnote{
Our ${k'}^2$ should not be confused with the dual modulus $1{-}k^2$, which is often denoted this way.}
\begin{equation}
2\,{\epsilon'}^2 \= 2{k'}^2{-}1 \= 1/\sqrt{\smash{\sfrac{8\Lambda}{3}}\, E'}
\qquad\textrm{so that}\quad
k'=\sfrac{1}{\sqrt{2}},1,\infty \quad\Leftrightarrow\quad E'=\infty,\sfrac{3}{8\Lambda},0\ .
\end{equation}
\begin{figure}[h!]
\centering
\includegraphics[width = 0.35\paperwidth]{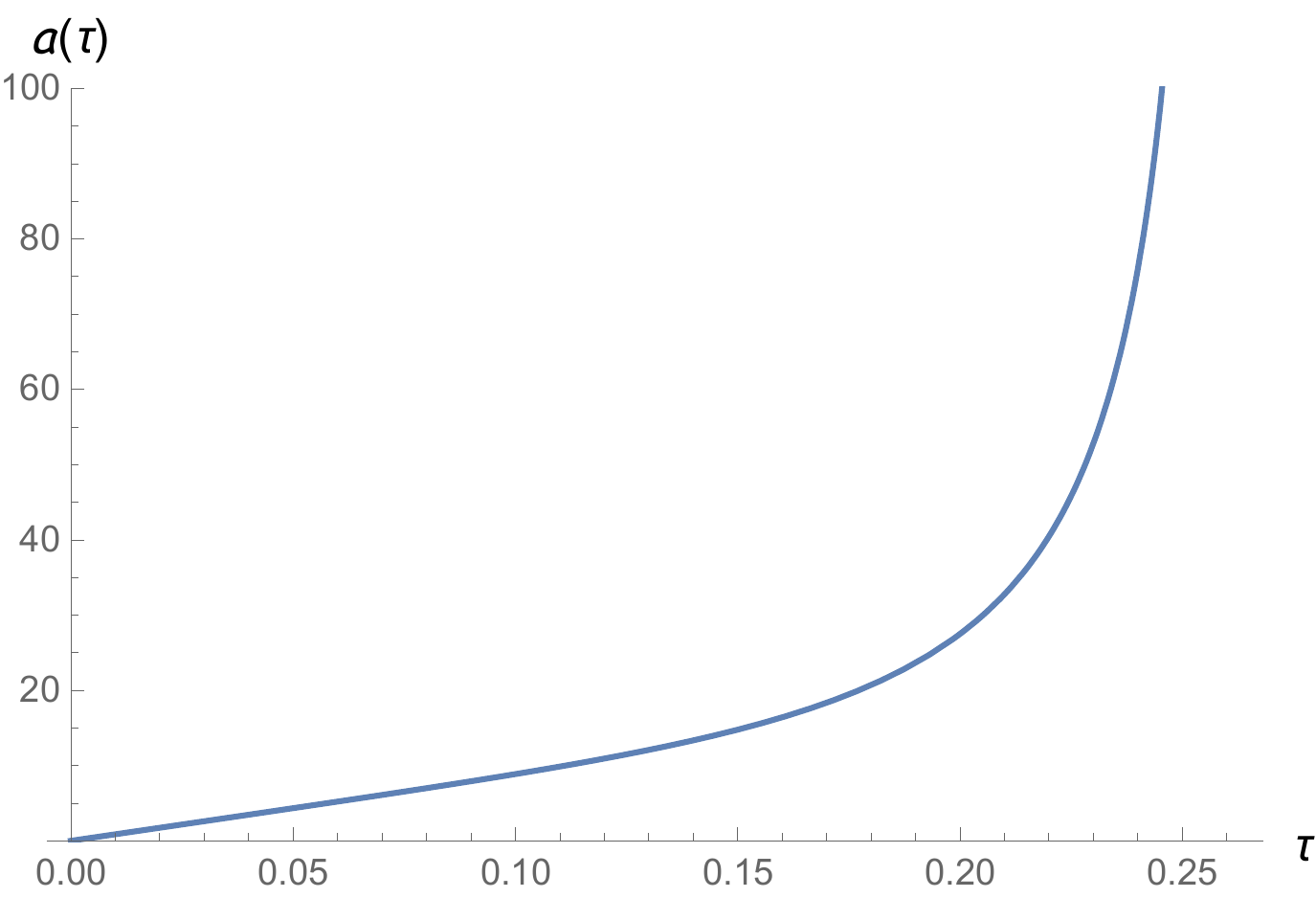} \qquad\quad
\includegraphics[width = 0.35\paperwidth]{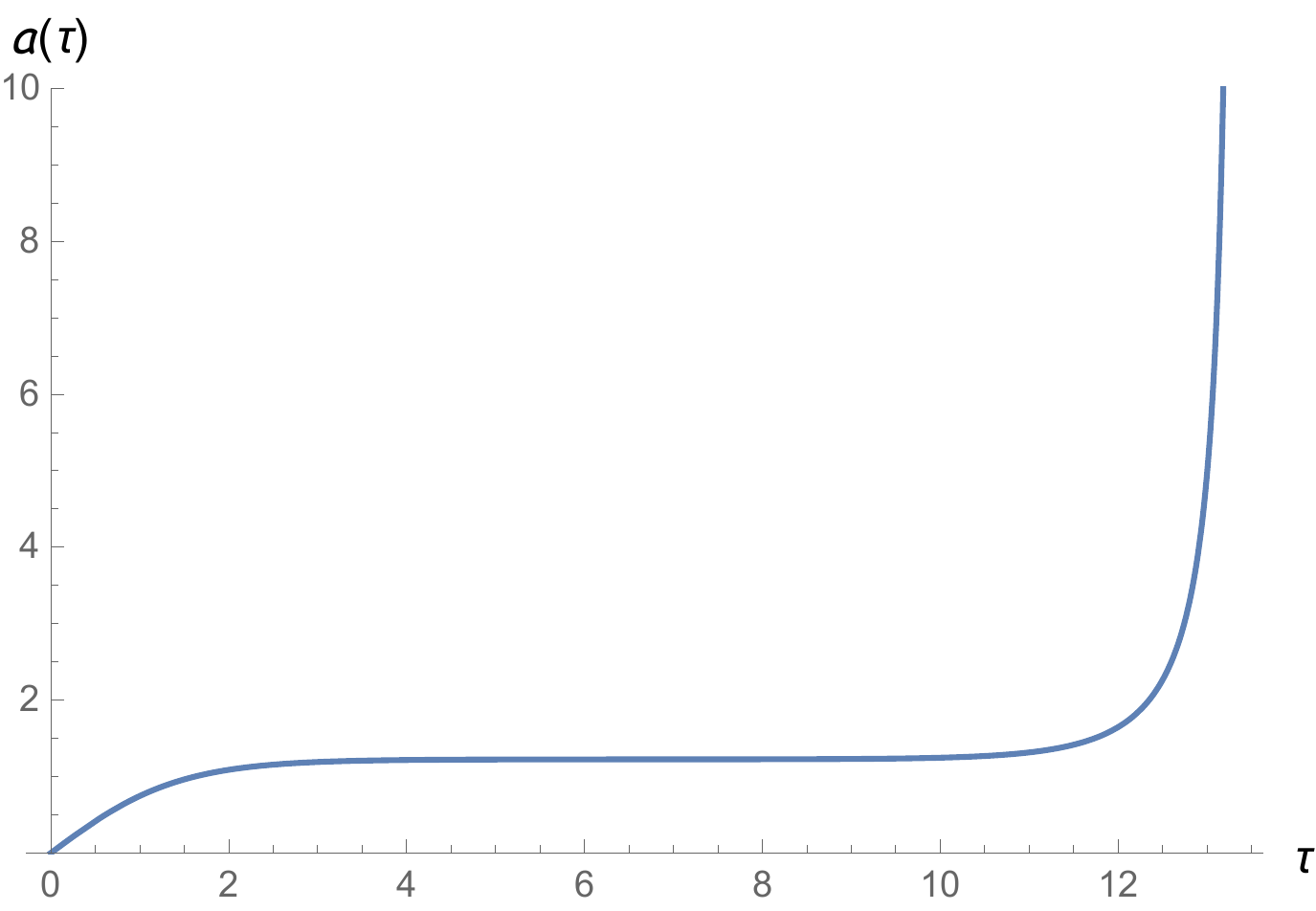} \\[12pt]
\includegraphics[width = 0.35\paperwidth]{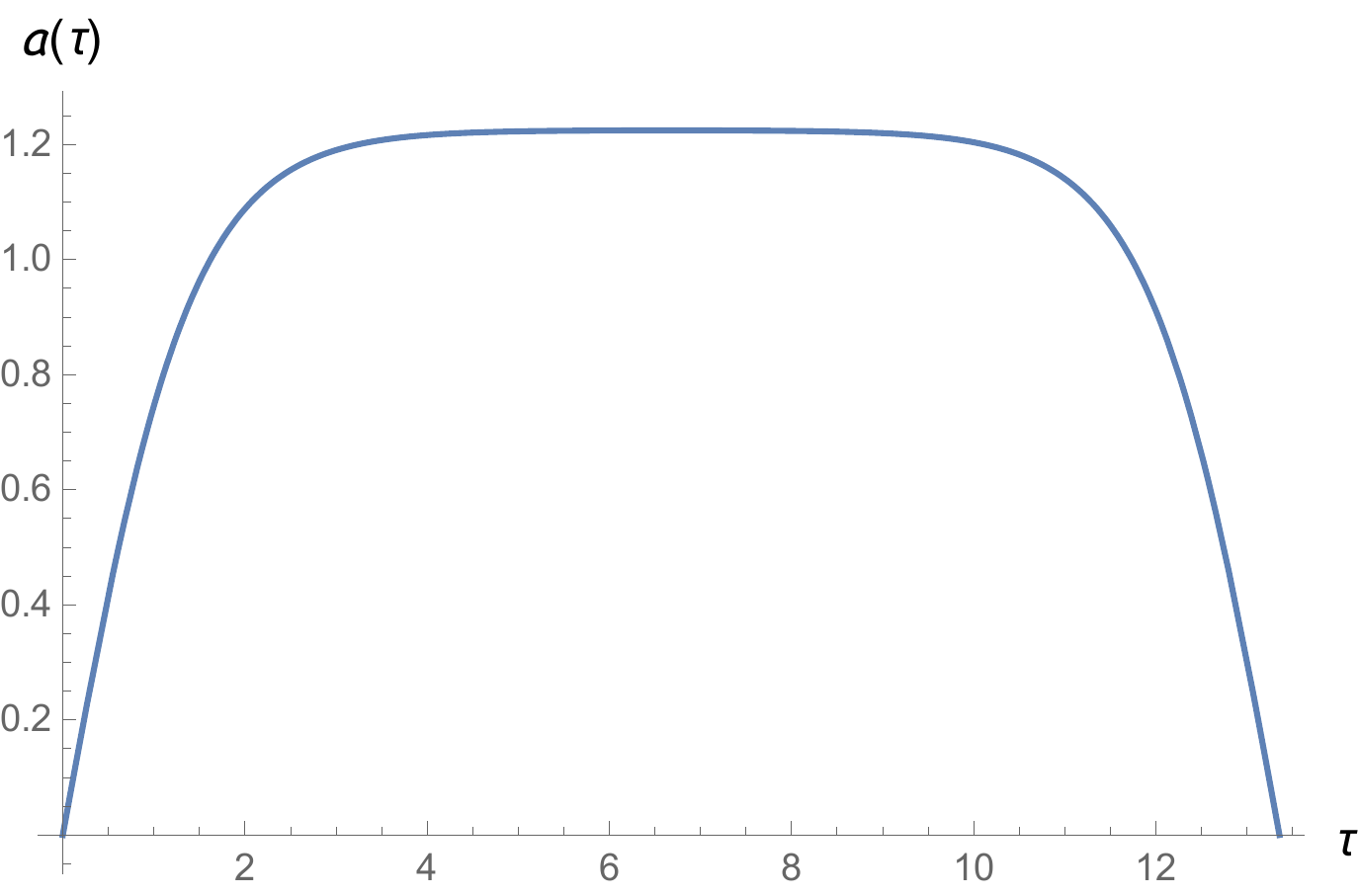} \qquad\quad
\includegraphics[width = 0.35\paperwidth]{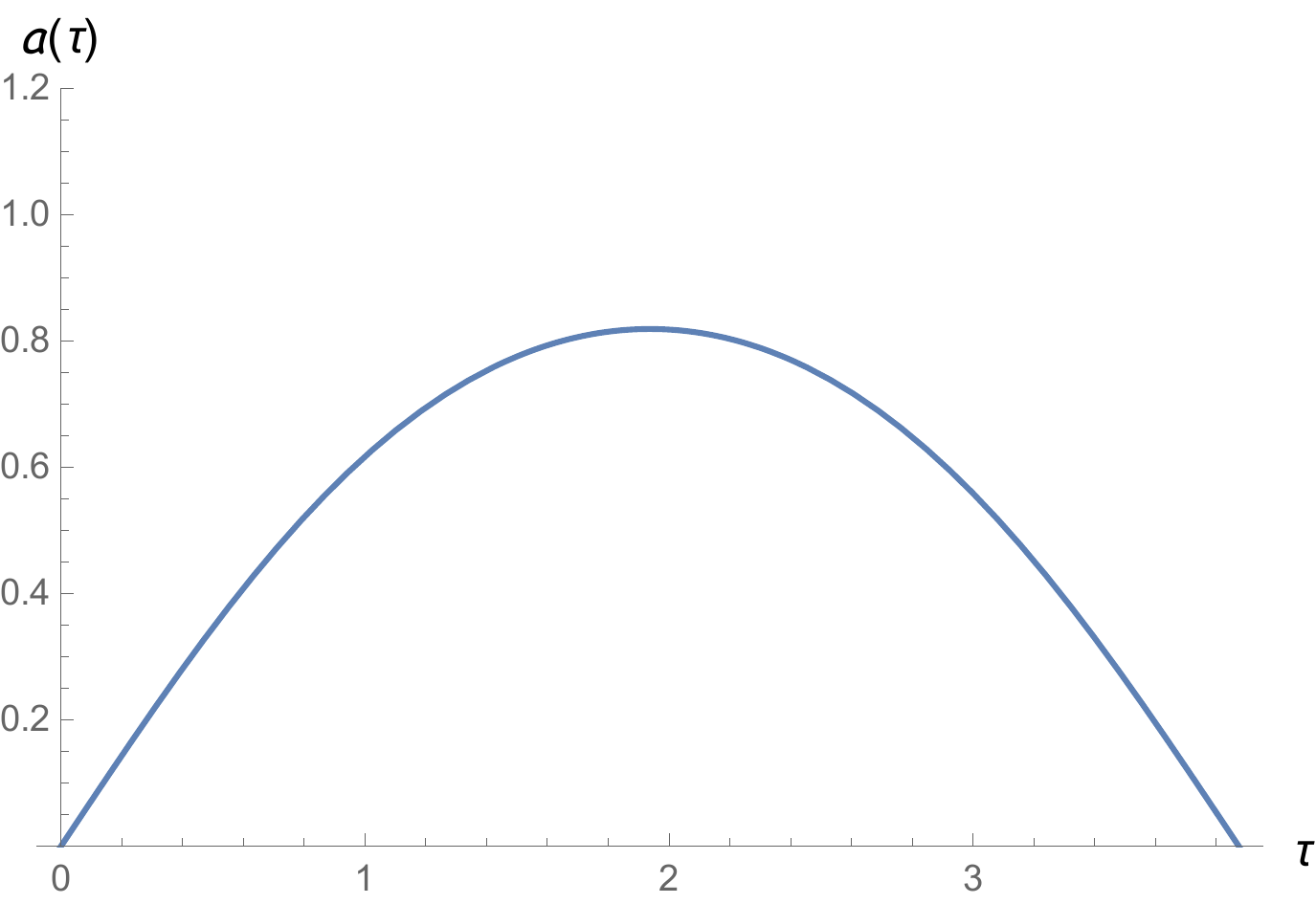}
\caption{Plots of $a(\tau)$ over one lifetime, for $\Lambda{=}1$ and different values of ${k'}^2$: \newline
\indent\qquad\qquad\
$0.505$ (top left), $0.9999999$ (top right), $1.0000001$ (bottom left) and $1.1$ (bottom right).}
\end{figure}
For $E'{\gg}\frac{3}{8\Lambda}$, we have ${k'}^2{\to}\frac12$, and the solution is well approximmated by
$\sqrt{\frac{3}{\Lambda}}\frac{1}{2\epsilon'}\tan\bigl(\frac{\sqrt{\pi^3}}{\Gamma(1/4)^2}\frac{\tau}{\epsilon'}\bigr)$.
We only listed solutions with initial value $a(0)=0$ (big bang). 
There exist also (for $E'<\sfrac{3}{8\Lambda}$) bouncing solutions,
where the universe attains a minimal radius 
$a_{\textrm{min}}=\bigl[\frac{3}{2\Lambda}(1+\sqrt{1-\smash{\sfrac{8\Lambda}{3}}E'})\bigr]^{1/2}$
between infinite extension in the far past ($t{=}{-}\infty\leftrightarrow\tau{=}0$) and the
far future ($t{=}{+}\infty\leftrightarrow\tau{=}T'$). For $E'{>}0$ they are obtained by 
sending $\mathrm{dn}\to-\mathrm{dn}$ in~(\ref{universalscale}) above.
The quantity~$T'$ listed there is the (conformal) lifetime of the universe, 
from the big bang until either the big rip (for $E'>\frac{3}{8\Lambda}$) 
or the big crunch of an oscillating universe (for $E'<\frac{3}{8\Lambda}$).
The solution relevant to our Einstein--Yang--Mills system is entirely determined by the Newtonian energy~$E$
characterizing the cosmic Yang--Mills field: above the critical value of
\begin{equation}
E_{\textrm{crit}} \= \frac{2\,g^2}{\kappa}\,\frac{3}{8\,\Lambda}
\end{equation}
the universe expands forever (until $t_{\textrm{max}}{=}\infty$), 
while below this value it recollapses (at $t_{\textrm{max}}{=}\int_0^{T'}\!\diff\tau\,a(\tau)$).
It demonstrates the necessity of a cosmological constant (whose role may be played
by the Higgs expectation value) as well as the nonperturbative nature of the cosmic Yang--Mills field,
whose contribution to the energy-momentum tensor is of~$O(g^{-2})$.


\section{Natural perturbation frequencies}

\noindent
Our main task in this paper is an investigation of the stability of the cosmic Yang--Mills solutions
reviewed in the previous section. For this, we should distinguish between global and local stability.
The former is difficult to assess in a nonlinear dynamics but clear from the outset in case of a compact
phase space. The latter refers to short-time behavior induced by linear perturbations around
the reference configuration. We shall look at this firstly, in the present section and the following one. 
Here, we set out to diagonalize the fluctuation operator for our time-dependent Yang--Mills backgrounds
and find the natural frequencies. 

Even though our cosmic gauge-field configurations are SO(4)-invariant, we must allow for all kinds of
fluctuations on top of it, SO(4)-symmetric perturbations being a very special subclass of them.
A generic gauge potential ``nearby'' a classical solution~$\Acal$ on $\Ical\times S^3$ can be expanded as
\begin{equation}
\Acal+\Phi \= \Acal(\tau,g)\ +\ \sum_{p=1}^3 \Phi_0^p(\tau,g)\,T_p\,\diff\tau\
+\ \sum_{a=1}^3\sum_{p=1}^3 \Phi_a^p(\tau,g)\,T_p\,e^a(g)
\end{equation}
with, using $(\mu)=(0,a)$,
\begin{equation}
\Phi_\mu^p(\tau,g) \= \sum_{j,m,n} \Phi_{\mu|j;m,n}^p(\tau)\,Y_{j;m,n}(g)\ ,
\end{equation}
on which we notice the following actions (supressing the $\tau$ and $g$ arguments),
\begin{equation}
(L_a \Phi_\mu^p)_{j;m,n} = \Phi_{\mu|j;m',n}^p \bigl(L_a)^{m'}_{\ m}\ ,\quad
(S_a \Phi)_0^p = 0\ ,\quad
(S_a \Phi)_b^p = -2\varepsilon_{abc}\,\Phi_c^p \ ,\quad
(T_a \Phi)_\mu^p = -2\varepsilon_{apq}\,\Phi_\mu^q \ ,
\end{equation}
where the $L_a$ matrix elements are determined from (\ref{Y-action1}) and~(\ref{Y-action2}),
and $S_a$ are the components of the spin operator.
The (metric and gauge) background-covariant derivative reads
\begin{equation}
D_\tau \Phi \= \partial_\tau \Phi \qquad\und\qquad
D_a \Phi \= L_a \Phi + [ \Acal_a,\Phi ] \qquad\textrm{with}\qquad
\Acal_a \= \sfrac12\bigl(1+\psi(\tau)\bigr)\,T_a\ ,
\end{equation}
which is equivalent to
\begin{equation}
D_a\Phi_b^p \= L_a\Phi_b^p - \varepsilon_{abc}\Phi_c^p + [ \Acal_a,\Phi_b]^p
\qquad\textrm{since}\qquad 
D_a\,e^b \= L_a\,e^b -  \varepsilon_{abc}\,e^c \= \varepsilon_{abc}\,e^c \ .
\end{equation}

The background $\Acal$ obeys the Coulomb gauge condition,
\begin{equation}
\Acal_\tau =0 \qquad\und\qquad L_a\Acal_a = 0\ ,
\end{equation}
but we cannot enforce these equations on the fluctuation~$\Phi$. 
However, we may impose the Lorenz gauge condition,
\begin{equation} \label{gauge}
D^\mu \Phi^p_\mu \= 0 \qquad\Rightarrow\qquad
\partial_\tau \Phi_0^p - L_a\Phi^p_a - \sfrac12(1{+}\psi)(T_a\Phi_a)^p \= 0\ ,
\end{equation}
which is seen to couple the temporal and spatial components of~$\Phi$ in general.
We then linearize the Yang--Mills equations around $\Acal$ and obtain
\begin{equation}
D^\nu D_\nu \Phi_\mu - R_{\mu\nu}\Phi^\nu + 2[\Fcal_{\mu\nu},\Phi^\nu] \= 0
\end{equation}
with the Ricci tensor
\begin{equation}
R_{\mu 0} \= 0 \qquad\und\qquad R_{ab} \= 2\delta_{ab}\ .
\end{equation}
After a careful evaluation, the $\mu{=}0$ equation yields
\begin{equation} \label{temporal}
\bigl[\partial_\tau^2 - L_b L_b + 2(1{+}\psi)^2\bigr] \Phi_0^p 
- (1{+}\psi)L_b(T_b\Phi_0)^p - \dot{\psi}(T_b\Phi_b)^p \= 0\ ,
\end{equation}
while the $\mu{=}a$ equations read
\begin{equation} \label{spatial}
\bigl[\partial_\tau^2 - L_b L_b + 2(1{+}\psi)^2{+}4\bigr]\Phi_a^p 
-(1{+}\psi)L_b(T_b\Phi)_a^p - L_b(S_b\Phi)_a^p - \sfrac12(1{+}\psi)(2{-}\psi)(S_bT_b\Phi)_a^p
- \dot{\psi}(T_a\Phi_0)^p \= 0\ .
\end{equation}
It is convenient to package the orbital, spin, isospin, and fluctuation triplets into formal vectors,
\begin{equation}
\vec{L}=(L_a)\ ,\qquad \vec{S}=(S_a)\ ,\qquad \vec{T}=(T_a)\ ,\qquad \vec{\Phi}=(\Phi_a)\ ,
\end{equation}
respectively, but they act in different spaces, hence on different indices,
such that $\vec{S}^2=\vec{T}^2=-8$ on~$\Phi$. 
In this notation, (\ref{gauge}), (\ref{temporal}) and~(\ref{spatial}) take the compact form
(suppressing the color index~$p$)
\begin{eqnarray}
\partial_\tau\Phi_0 - \vec{L}{\cdot}\vec{\Phi} - \sfrac12(1{+}\psi)\vec{T}{\cdot}\vec{\Phi} \= 0\ ,
\label{gauge2}
\qquad\qquad\qquad\qquad\qquad\qquad\qquad\qquad\qquad\qquad\qquad\qquad\qquad\qquad\ \ \\[4pt]
\bigl[\partial_\tau^2{-}\vec{L}^2+2(1{+}\psi)^2\bigr] {\Phi}_0 
- (1{+}\psi)\vec{L}{\cdot}\vec{T}\,\Phi_0 - \dot{\psi}\,\vec{T}{\cdot}\vec{\Phi} \= 0 \ ,
\label{mu=0}\qquad\qquad\qquad\qquad\qquad\qquad\qquad\qquad\qquad\quad\ \ \\[6pt]
\bigl[\partial_\tau^2{-}\vec{L}^2{-}\sfrac12\vec{S}^2{+}2(1{+}\psi)^2\bigr] {\Phi}_a
- (1{+}\psi)\vec{L}{\cdot}\vec{T}\,\Phi_a - \vec{L}{\cdot}(\vec{S}\,\Phi)_a 
- \sfrac12(1{+}\psi)(2{-}\psi)\vec{T}{\cdot}(\vec{S}\,\Phi)_a - \dot{\psi}\,T_a\Phi_0 \= 0\ . \quad
\label{mu=a}
\end{eqnarray}

A few remarks are in order. 
First, except for the last term, (\ref{mu=0}) is obtained from (\ref{mu=a}) by setting $\vec{S}=0$,
since $\Phi_0$ carries no spin index. 
Second, both equations can be recast as
\begin{equation} \label{2spins}
{}\!\!\!
\bigl[ \partial_\tau^2 - \sfrac{1{-}\psi}{2}\vec{L}^2 - \sfrac{1{+}\psi}{2}(\vec{L}{+}\vec{T})^2 
- 2(1{+}\psi)(1{-}\psi) \bigr] \Phi_0 \= \dot{\psi}\,\vec{T}\cdot\vec{\Phi}\ ,
\qquad\qquad\qquad\qquad\qquad\qquad\qquad\qquad\ \ 
\end{equation}
\begin{equation} \label{3spins}
\bigl[ \partial_\tau^2 - \sfrac{(1{-}\psi)(2{+}\psi)}{4}\vec{L}^2 - \sfrac{\psi(1{+}\psi)}{4}(\vec{L}{+}\vec{T})^2
+ \sfrac{\psi(1{-}\psi)}{4}(\vec{L}{+}\vec{S})^2 - \sfrac{(1{+}\psi)(2{-}\psi)}{4}(\vec{L}{+}\vec{T}{+}\vec{S})^2
- 2(1{+}\psi)(1{-}\psi) \bigr] \vec{\Phi} = \dot{\psi}\vec{T}\Phi_0,
\end{equation}
which reveals a problem of addition of three spins and a corresponding symmetry under
\begin{equation} \label{symmetry}
\psi\ \leftrightarrow\ -\psi\ ,\qquad
\vec{L}\ \leftrightarrow\ \vec{L}{+}\vec{T}{+}\vec{S} \quad\und\quad
\vec{L}{+}\vec{S}\ \leftrightarrow\ \vec{L}{+}\vec{T}\ .
\end{equation}
Third, for constant backgrounds $(\dot{\psi}{=}0)$ the temporal fluctuation $\Phi_0$ decouples 
and may be gauged away. 
Still, the fluctuation operator in~(\ref{3spins}) is easily diagonalized only when 
the coefficient of one of the first three spin-squares vanishes, i.e.~for $\vec{L}{=}0$ ($j{=}0$),
for the two vacua $\psi{=}{\pm}1$, or for the ``meron'' (or ``sphaleron'') $\psi{=}0$. 
The latter case has been analyzed by Hosotani and by Volkov~\cite{Hosotani,Volkov}.

Let us decompose the fluctuation problem (\ref{gauge2})--(\ref{mu=a}) into finite-dimensional blocks
according to a fixed value of the spin~$j\in\sfrac12\N$,
\begin{equation}
\vec{L}^2\, \Phi^p_{\mu|j} \= -4\,j(j{+}1)\,\Phi^p_{\mu|j}
\end{equation}
and suppress the $j$ subscript. We employ the following coupling scheme,\footnote{
Another (less convenient) scheme couples $\vec{L}{+}\vec{S}$, then $(\vec{L}{+}\vec{S}){+}\vec{T}=:\vec{V}$.}
\begin{equation}
\vec{L}{+}\vec{T}=:\vec{U} \qquad\textrm{then}\qquad \vec{U}{+}\vec{S}=(\vec{L}{+}\vec{T}){+}\vec{S}=:\vec{V}\ .
\end{equation}
Clearly, $\vec{U}$ and $\vec{V}$ act on $\vec{\Phi}$ in $su(2)$ representations 
$j\otimes 1$ and $j\otimes 1\otimes 1$, respectively. On $\Phi_0$, we must put~$\vec{S}{=}0$
and have just $\vec{V}{=}\vec{U}$ act in a $j\otimes 1$ representation.
Combining the coupled equations (\ref{mu=0}) and (\ref{mu=a}) to a single linear system for
$(\Phi^p_\mu)=(\Phi^p_0,\Phi^p_a)$, we get a $12(2j{+}1)\times 12(2j{+}1)$ fluctuation matrix~$\Omega^2_{(j)}$,
\begin{equation} \label{fluctop}
\bigl[ \delta_{\mu\nu}^{pq}\,\partial_\tau^2\ +\ (\Omega^2_{(j)})_{\mu\nu}^{pq} \bigr]\,\Phi_\nu^q \= 0\ .
\end{equation}
Actually, there is an additional overall $(2j{+}1)$-fold degeneracy present due to the trivial action of
the $su(2)_{\textrm{R}}$ generators~$R_a$, which plays no role here and will be suppressed.
Roughly speaking, the $3(2j{+}1)$ modes of $\Phi_0$ are related to gauge modes,\footnote{
Strictly, they are gauge modes only when $\dot\psi{=}0$. 
Otherwise, the gauge modes are mixtures with the $\Phi_a$ modes.}
and we still must impose the gauge condition~(\ref{gauge2}), which also has $3(2j{+}1)$ components.
Therefore, a subspace of dimension $6(2j{+}1)$ inside the space of all fluctuations will 
represent the physical gauge-equivalence classes in the end.

Our goal is to diagonalize the fluctuation operator~(\ref{fluctop}) for a given fixed value of~$j$.
It has a block structure,
\begin{equation} \label{block}
\Omega^2_{(j)} \= \begin{pmatrix} \bar{N} & -\dot{\psi}\,T^\top \\[6pt] -\dot{\psi}\,T & N \end{pmatrix}\ ,
\end{equation}
where $\bar{N}$ and $N$ are given by the left-hand sides of (\ref{2spins}) and~(\ref{3spins}), respectively.
We introduce a basis where $\vec{U}^2$, $\vec{V}^2$ and $V_3$ are diagonal, i.e.
\begin{equation}
\vec{U}^2\,|uvm\> \= -4\,u(u{+}1)\,|uvm\> \quad\und\quad 
\vec{V}^2\,|uvm\> \= -4\,v(v{+}1)\,|uvm\> \qquad\textrm{with}\quad m=-v,\ldots,v\ ,
\end{equation}
and denote the irreducible $su(2)_v$ representations with those quantum numbers as $\rep{v}{u}$.
On the $\Phi_0$ subspace, $u$ is redundant since $u{=}v$ as $\vec{S}{=}0$.
Working out the tensor products, we encounter the values
\begin{equation} \label{uvreps}
\begin{aligned}
\rep{v}{u} &\= \rep{j{-}2}{j{-}1}\ ;\!\!\!&\rep{j{-}1}{j{-}1}&\ ,\ \rep{j{-}1}{j}\ ;\ \rep{j}{j{-}1}\ ,
\!\!\!&\rep{j}{j}&\ ,\ \rep{j}{j{+}1}\ ;\ \rep{j{+}1}{j}\ ,\!\!\!&\rep{j{+}1}{j{+}1}&\ ;\ \rep{j{+}2}{j{+}1} 
&\quad\textrm{on}\ \vec{\Phi}\ ,& \\
\rep{v}{u} &\= &\rep{j{-}1}{j{-}1}&\ ; &\rep{j}{j}&\ ; &\rep{j{+}1}{j{+}1}&
&\quad\textrm{on} \ \Phi_0&\ ,
\end{aligned}
\end{equation}
with some representations obviously missing for $j{<}2$.

Let us treat the $\dot{\psi}\,T$ term in~(\ref{block}) as a perturbation and momentarily put it to zero,
so that $\Omega^2_{(j)}$ is block-diagonal for the time being.
Then, it is easy to see from (\ref{2spins}) and (\ref{3spins}) that $[\vec{V},\bar{N}]=[\vec{U},\bar{N}]=0$
and $[\vec{V},N]=0$, even though $[\vec{U},N]\neq0$ because $(\vec{L}{+}\vec{S})^2$ 
is not diagonal in our basis. Therefore, we have a degeneracy in~$m$.
Furthermore, both $\bar{N}$ and $N$ decompose into at most three respectively five blocks 
with fixed values of~$v$ ranging from $j{-}2$ to~$j{+}2$ and separated by semicolons in~(\ref{uvreps}). 
Moreover, the $\bar{N}$~blocks are irreducible and trivially also carry a value of~$u{=}v$. 
In contrast, $N$ is not {\it simply\/} reducible; its $\vec{V}$ representations have multiplicity one, two or three. 
Only the $N$~blocks with extremal $v$~values in~(\ref{uvreps}) are irreducible. 
The other ones are reducible and contain more than one $\vec{U}$~representation, 
hence the $u$-spin distinguishes between their (two or three) irreducible $v$~subblocks.
The only non-diagonal term in~$N$ is the $(\vec{L}{+}\vec{S})^2$ contribution, which couples different copies
of the same $v$-spin to each other, but of course not to any $u{=}v$ block of~$\bar{N}$, 
and does not lift the $V_3{=}m$~degeneracy. 
As a consequence, the unperturbed fluctuation equations 
for $\Phi_0{=}\Phi_{(\bar{v})}$ and $\vec{\Phi}{=}\Phi_{(v,\alpha)}$ take the form 
(suppressing the $m$ index)
\begin{equation}
\begin{aligned}
&\unity_{(\bar{v})}\bigl[ \partial_\tau^2\ +\ \bar\omega^2_{(\bar{v})}\bigr]\,\Phi_{(\bar{v})} \= 0 \qquad\und\qquad
\unity_{(v)}\bigl[ \partial_\tau^2\ +\ \omega^2_{(v,\alpha)}\bigr]\,\Phi_{(v,\alpha)} \= 0
\qquad\textrm{for}\quad \dot{\psi}=0\\[4pt]
&\textrm{with}\qquad \bar{v} \in \{ j{-}1,\ j,\ j{+}1\} \qquad\und\qquad
v \in \{ j{-}2,\ j{-}1,\ j,\ j{+}1,\ j{+}2 \} \ ,
\end{aligned}
\end{equation}
where $\unity_{(v)}$ denotes a unit matrix of size~$2v{+}1$, and $\alpha$ counts the multiplicity 
of the $v$-spin representation in~$N$ (between one and three). 
According to~(\ref{2spins}) the unperturbed frequency-squares for $\bar{N}$ are the eigenvalues
\begin{equation}
\bar{\omega}^2_{(\bar{v})} \= 
2(1{-}\psi)\,j(j{+}1) + 2(1{+}\psi)\,\bar{v}(\bar{v}{+}1) - 2(1{+}\psi)(1{-}\psi)
\end{equation}
with multiplicity  $2\bar{v}{+}1$, hence we get
\begin{equation}
\begin{aligned}
\bar{\omega}^2_{(j-1)} &\= 2\,\psi^2-4j\,\psi+2(2j^2{-}1)\ ,\\
\bar{\omega}^2_{(j)} &\= 2\,\psi^2+2(2j^2{+}2j{-}1)\,\\
\bar{\omega}^2_{(j+1)} &\= 2\,\psi^2+4(j{+}1)\,\psi +2(2j^2{+}4j{+}1)\ .
\end{aligned}
\end{equation}
Considering $N$ in (\ref{3spins}), we can read off the eigenvalues at $v=j{\pm}2$
because in these two extremal cases $(\vec{L}{+}\vec{S})^2=\vec{U}^2$ 
is already diagonal in the $\bigl\{ |uvm\>\bigr\}$ basis. For the other $v$-values
we must diagonalize a $2{\times}2$ or $3{\times}3$ matrix to find
\begin{equation}
\begin{aligned}
\omega^2_{(j-2)} &\= \textrm{root of}\ Q_{j-2}(\lambda)
\= -2(2j{-}1)\,\psi + 2(2j^2{-}2j{+}1)\ ,\\
\omega^2_{(j-1,\alpha)} &\= \textrm{two roots of}\ Q_{j-1}(\lambda)\ ,\\
\omega^2_{(j,\alpha)} &\= \textrm{three roots of}\ Q_j(\lambda)\ ,\\
\omega^2_{(j+1,\alpha)} &\= \textrm{two roots of}\ Q_{j+1}(\lambda)\ ,\\
\omega^2_{(j+2)} &\= \textrm{root of}\ Q_{j+2}(\lambda)
\= 2(2j{+}3)\,\psi + 2(2j^2{+}6j{+}5)\ ,
\end{aligned}
\end{equation}
each with multiplicity $2v{+}1$, where $Q_v$ denotes a linear, quadratic or cubic polynomial.\footnote{
For $j{<}2$ some obvious modifications occur due to the missing of $v{<}0$ representations.}

Let us now turn on the perturbation $\dot{\psi}\,T$, which couples $N$ with~$\bar{N}$,
and consider the characteristic polynomial~${\cal P}_j(\lambda)$ of our fluctuation problem,
\begin{equation}
\begin{aligned}
{\cal P}_j(\lambda) &\ :=\
\det\,\Bigl(\begin{smallmatrix} \bar{N}{-}\lambda & -\dot{\psi}T^\top \\[4pt] 
-\dot{\psi}T & N{-}\lambda \end{smallmatrix} \Bigr)
\= \det (N{-}\lambda) \cdot \det\bigl[ (\bar{N}{-}\lambda) - \dot{\psi}^2\,T^\top (N{-}\lambda)^{-1} T \bigr] \\[4pt]
&\= \bigl[{\textstyle\prod}_v \det (N_{(v)}{-}\lambda)\bigr] \cdot 
\det\bigl[ (\bar{N}{-}\lambda) - \dot{\psi}^2\, T^\top \{{\textstyle\bigoplus}_{v} (N_{(v)}{-}\lambda)^{-1}\}\,T\bigr] \ ,
\end{aligned}
\end{equation}
where we made use of
\begin{equation}
\< u\,v\,m|\,N\,|u'v'm'> \= \bigl(N_{(v)}\bigr)_{uu'}\,\delta_{vv'}\delta_{mm'}\ .
\end{equation}
Since $T$ furnishes an $su(2)$ representation (and not an intertwiner) 
it must be represented by square matrices and thus cannot connect different $v$ representations.
Hence the perturbation does not couple different $v$ sectors but only links $N$ and $\bar{N}$
in a common $\bar{v}{=}v$~sector. Therefore, it does not affect the extremal sectors $v=j{\pm}2$.
Moreover, switching to a diagonal basis $\{|\alpha vm\>\}$ for $N$ we can simplify to
\begin{equation}
T^\top \{{\textstyle\bigoplus}_{v} (N_{(v)}{-}\lambda)^{-1}\}\,T\bigr]
\= {\textstyle\bigoplus}_{\bar{v}}\{T^\top (N{-}\lambda)^{-1} T\}_{(\bar{v})}
\={\textstyle\bigoplus}_{\bar{v}}\Bigl\{ {\textstyle\sum}_\alpha 
(\omega^2_{(\bar{v},\alpha)}{-}\lambda)^{-1} \bigl(T^\top|\alpha\>\!\<\alpha|\,T\bigr)_{(\bar{v})}\Bigr\}\ .
\end{equation}
Observing that 
$\bigl(T^\top|\alpha\>\!\<\alpha|\,T\bigr)_{(\bar{v})}=
-t_{\bar{v},\alpha}\bigl(\vec{T}^2\bigr)_{(\bar{v})}=8\,t_{\bar{v},\alpha}\unity_{(\bar{v})}$
with some coefficient functions $t_{\bar{v},\alpha}(\psi)$, with $\sum_\alpha t_{\bar{v},\alpha}=1$,
we learn that the $V_3$ degeneracy remains intact and arrive at ($\bar{v}\in\{j{-}1,j,j{+}1\}$)
\begin{equation} \label{charP}
\begin{aligned}
{\cal P}_j(\lambda) &\= \bigl[{\textstyle\prod}_v Q_v(\lambda)^{2v+1}\bigr] \cdot
{\textstyle\prod}_{\bar{v}} \bigl\{(\bar{\omega}^2_{(\bar{v})}{-}\lambda) 
- 8\dot{\psi}^2 {\textstyle\sum}_\alpha t_{\bar{v},\alpha} 
(\omega^2_{(\bar{v},\alpha)}{-}\lambda)^{-1} \bigr\}^{2\bar{v}+1}\\[4pt]
&\= (\omega^2_{(j-2)}{-}\lambda)^{2j-3}\cdot(\omega^2_{(j+2)}{-}\lambda)^{2j+5}\cdot
{\textstyle\prod}_{\bar{v}} \bigl\{ (\bar{\omega}^2_{(\bar{v})}{-}\lambda)\,Q_{\bar{v}}(\lambda) 
- 8\dot{\psi}^2 P_{\bar{v}}(\lambda) \bigr\}^{2\bar{v}+1} \\[4pt]
&\= (\omega^2_{(j-2)}{-}\lambda)^{2j-3}\cdot(\omega^2_{(j+2)}{-}\lambda)^{2j+5}\cdot
{\textstyle\prod}_{\bar{v}} R_{\bar{v}}(\lambda)^{2\bar{v}+1}\ ,
\end{aligned}
\end{equation}
where $P_{\bar{v}}=Q_{\bar{v}}\sum_\alpha t_{\bar{v},\alpha} (\omega^2_{(\bar{v},\alpha)}{-}\lambda)^{-1}$
is a polynomial of degree one less than $Q_{\bar{v}}$ since all poles cancel, and $R_{\bar{v}}$ is a
polynomial of one degree more.
We list the polynomials $Q_v$, $P_{\bar{v}}$ and $R_{\bar{v}}$ for $j{\le}2$ in the Appendix.

To summarize, by a successive basis change ($m'=-j,\ldots,j$ and $m=-v,\ldots,v$)
\begin{equation}
\bigl\{ |\mu\,p\,m'\> \bigr\} \quad\Rightarrow\quad 
\bigl\{ |\bar{v} m\>, |u v m\> \bigr\} \quad\Rightarrow\quad
\bigl\{ |\bar{v} m\>, |\alpha v m\> \bigr\} \quad\Rightarrow\quad
\bigl\{ |\beta v m\> \bigr\}
\end{equation}
we have diagonalized~(\ref{fluctop}) to
\begin{equation} \label{diagonalized}
\bigl[ \partial_\tau^2 - \Omega^2_{(j,v,\beta)} \bigr] \, \Phi_{(v,\beta)} \= 0
\qquad\textrm{with}\quad v\in\{j{-}2,j{-}1,j,j{+}1,j{+}2\} \ ,
\end{equation}
where $\Omega^2_{(j,v,\beta)}$ are the distinct roots of the characteristic polynomial~${\cal P}_j$ in~(\ref{charP}),
and (for $j{\ge}2$) the multiplicity label $\beta$ takes $1,3,4,3,1$ values, respectively:
\begin{equation}
\Omega^2_{(j,j\pm2)} = \omega^2_{(j\pm2)}\ ,\quad
\Omega^2_{(j,j\pm1,\beta)} = \textrm{three roots of}\ R_{j\pm1}(\lambda)\ ,\quad
\Omega^2_{(j,j,\beta)} = \textrm{four roots of}\ R_{j}(\lambda)\ .
\end{equation}
The reflection symmetry~(\ref{symmetry}) implies that 
$\Omega^2_{(v,j,\cdot)}(\psi)=\Omega^2_{(j,v,\cdot)}(-\psi)$.
For $j{<}2$, obvious modifications occur due to the absence of some $v$~representations.

We still have to discuss the gauge condition~(\ref{gauge2}), which can be cast into the form
\begin{equation}
0 \= \pa_\tau\Phi_0 - \bigl[ \sfrac12(1{-}\psi)\vec{L}+\sfrac12(1{+}\psi)\vec{U}\bigr]\cdot\vec\Phi
\= \pa_\tau\Phi_{(\bar{v},\bar{m})} - K_{\bar{v},\bar{m}}^{\ v,m,\alpha}(\psi)\,\Phi_{(v,m,\alpha)}
\end{equation}
with a $3(2j{+}1){\times}7(2j{+}1)$ linear (in~$\psi$) matrix function~$K$.\footnote{
We have to bring back the $m$ indices because the gauge condition is not diagonal in them.}
Here the $v$~sum runs over $(j{-}1,j,j{+}1)$ only,
since the gauge condition~(\ref{gauge2}) has components only in the middle three $v$~sectors,
like the gauge-mode equation~(\ref{mu=0}).
It does not restrict the extremal $v$~sectors~$v=j{\pm}2$, since these fluctuations
do not couple to the gauge sector~$\Phi_0$ and are entirely physical.
For the middle three $v$~sectors (labelled by~$\bar{v}$), the $\dot{\psi}\,T$ perturbation leads to
a mixing of the $N$ modes with the $\bar{N}$ gauge modes, so their levels will avoid crossing. 
Performing the corresponding final basis change, the gauge condition takes the form
\begin{equation} \label{gauge4}
\bigl[ L_{\bar{v},\bar{m}}^{\ \bar{v}'\!,\bar{m}'\!,\beta}(\psi)\,\pa_\tau 
- M_{\bar{v},\bar{m}}^{\ \bar{v}'\!,\bar{m}'\!,\beta}(\psi) \bigr]\,\Phi_{(\bar{v}'\!,\bar{m}'\!,\beta)}\=0
\end{equation}
with certain $3(2j{+}1){\times}10(2j{+}1)$ matrix functions $L$ and~$M$.
This linear equation represents conditions on the normal mode functions $\Phi_{(\bar{v},\bar{m},\beta)}$
and defines a $7(2j{+}1)$-dimensional subspace of physical fluctuations, 
which of course still contains a $3(2j{+}1)$-dimensional subspace of gauge modes. 
For $j{<}1$, these numbers are systematically smaller.
Together with the two extremal $v$~sectors, we end up with $(7-3+2)(2j{+}1)=6(2j{+}1)$ 
physical degrees of freedom for any given value of~$j({\ge}2)$, as advertized earlier.

We conclude this section with more details for the simplest examples,
which are constant backgrounds and $j{=}0$ backgrounds.
For the vacuum background, say $\psi=-1$, which is isospin degenerate, one gets
\begin{equation}
\bigl( \partial_\tau^2 -\sfrac12\vec{L}^2 -\sfrac12(\vec{L}{+}\vec{S})^2 \bigr)\,\vec{\Phi} \=0, \qquad
\vec{L}{\cdot}\vec{\Phi} = 0\ ,\qquad \Phi_0 =0\ .
\end{equation}
It yields the positive eigenfrequency-squares
\begin{equation}
\omega^2_{(j,u')} \= 2j(j{+}1) + 2u'(u'{+}1) \=
\begin{cases}
\ 4j^2 \ \textrm{at}\ j{{\ge}1} & \for u'=j{-}1 \\ 
\ 4j(j{+}1) & \for u'=j \\
\ 4(j{+}1)^2 & \for u'=j{+}1 
\end{cases}
\end{equation}
for $j=0,\sfrac12,1,\ldots$,
but the $\vec{L}{\cdot}\vec{\Phi}=0$ constraint removes the $u'{=}j$ modes.
Clearly, all (constant) eigenfrequency-squares are positive, hence the vacuum is stable.

For the ``meron/sphaleron'' background, $\psi\equiv0$, one has
\begin{equation}
\bigl( \partial_\tau^2 -\sfrac12\vec{L}^2 -\sfrac12(\vec{L}{+}\vec{T}{+}\vec{S})^2 -2 \bigr)\,\vec{\Phi} \=0, \qquad
\bigl(\vec{L}+\sfrac12\vec{T}\bigr)\cdot\vec{\Phi} = 0\ ,\qquad \Phi_0 =0\ .
\end{equation}
In this case, we read off 
\begin{equation}
\omega^2_{(j,v)} +2 \= 2j(j{+}1) + 2v(v{+}1) \=
\begin{cases}
\ 4(j^2{-}j{+}1)  & \for v=j{-}2 \qquad (0\ \textrm{to}\ 1\ \textrm{times}) \\ 
\ 4j^2 & \for v=j{-}1 \qquad (0\ \textrm{to}\ 2\ \textrm{times}) \\
\ 4j(j{+}1) & \for v=j \qquad\quad\  (1\ \textrm{to}\ 3\ \textrm{times}) \\
\ 4(j{+}1)^2 & \for v=j{+}1 \qquad (1\ \textrm{to}\ 2\ \textrm{times}) \\
\ 4(j^2{+}3j{+}3)  & \for v=j{+}2 \qquad (1\ \textrm{times})
\end{cases}\ ,
\end{equation}
but the constraint removes one copy from each of the three middle cases (and less when $j{<}1$).
We end up with a spectrum $\{\omega^2\}=\{-2,1,6,7,10,\ldots\}$ with certain degeneracies~\cite{Hosotani,Volkov}.
The single non-degenerate negative mode $\omega^2_{(0,0)}{=}{-}2$ is a singlet, $\Phi_a^p=\delta_a^p\phi(\tau)$, 
and it corresponds to rolling down the local maximum of the double-well potential. 
The meron is stable against all other perturbations.

For a time-varying background, the natural frequencies $\Omega_{(j,v,\beta)}$
inherit a $\tau$ dependence from the background~$\psi(\tau)$.
Direct diagonalization is still possible for $j{=}0$, where we should solve
\begin{equation}
\begin{aligned}
& \partial_\tau\Phi_0 - \sfrac12(1{+}\psi)\vec{T}{\cdot}\vec{\Phi} \= 0 \ ,\\[4pt]
& \bigl[\partial_\tau^2+2(1{+}\psi)^2\bigr] {\Phi}_0 - \dot{\psi}\,\vec{T}{\cdot}\vec{\Phi} \= 0\ ,\\[4pt]
& \bigl[ \partial_\tau^2 +2(3\psi^2{-}1) -\sfrac14(1{+}\psi)(2{-}\psi)(\vec{S}{+}\vec{T})^2 \bigr]\ 
\vec\Phi - \dot{\psi}\,\vec{T}\,\Phi_0 \= 0\ ,
\end{aligned}
\end{equation}
with
\begin{equation}
(\vec{S}{+}\vec{T})^2\=\vec{V}^2\=-4\,v(v{+}1)\=0,-8,-24 \qquad\textrm{for}\quad v=0,1,2\ .
\end{equation}
It implies the unperturbed frequencies (suppressing the $j$ index)
\begin{equation} \label{unperturbed}
\bar\omega_{(1)}^2 = 2(\psi{+}1)^2\ \ (3\times)\ ,\quad
\omega_{(0)}^2 = 2(3\psi^2{-}1)\ \ (1\times)\ ,\quad
\omega_{(1)}^2 = 2(2\psi^2{+}\psi{+}1)\ \ (3\times) ,\quad
\omega_{(2)}^2 = 2(3\psi{+}5)\ \ (5\times)
\end{equation}
for
\begin{equation}
\begin{aligned}
(\Phi_0)^p &\equiv \bigl(\Phi_{(\bar{v}=1)}\bigr)^p\ =:\ \delta^{pb}\bar\phi_b \ ,\\
(\vec\Phi)_a^p \ &\equiv \bigl(\Phi_{(0)}+\Phi_{(1)}+\Phi_{(2)}\bigr)^p_a\ =:\
\phi\,\delta_a^p + \epsilon^p_{\ ab}\,\phi_b + (\phi_{(ab)}{-}\delta_{ab}\phi)\delta^{bp}\ ,
\end{aligned}
\end{equation}
as long as $\dot{\psi}$ is ignored.
There are no $v$-spin multiplicities (larger than one) here.
Turning on $\dot{\psi}$ and observing that $(\vec{T}{\cdot}\vec\Phi)^p\sim\delta^{pb}\phi_b$, 
the characteristic polynomial of the coupled $12{\times}12$ system in the $|uvm\>$ basis reads
\begin{equation}
{\cal P}_0(\lambda) \= \det \begin{pmatrix}
(\bar{\omega}_{(1)}^2{-}\lambda)\unity_3 & 
0 & -\dot{\psi}\,T_{(1)}^\top & 0 \\[4pt]
0 & (\omega_{(0)}^2{-}\lambda)\unity_1 & 0 & 0 \\[4pt]
-\dot{\psi}\,T_{(1)} & 0 & (\omega_{(1)}^2{-}\lambda)\unity_3 & 0 \\[4pt]
0 & 0 & 0 & (\omega_{(2)}^2{-}\lambda)\unity_5 
\end{pmatrix}\ .
\end{equation}
Specializing the general discussion above to $j{=}0$, we find just $t_1{=}1$ so that $P_1{=}1$ and arrive at
\begin{equation}
\begin{aligned}
{\cal P}_0(\lambda) &\= (\omega_{(0)}^2{-}\lambda)^1
(\omega_{(1)}^2{-}\lambda)^3 (\omega_{(2)}^2{-}\lambda)^5
\bigl[(\bar{\omega}_{(1)}^2{-}\lambda) - 8\dot{\psi}^2(\omega_{(1)}^2{-}\lambda)^{-1} \bigr]^3 \\
&\= (\omega_{(0)}^2{-}\lambda) (\omega_{(2)}^2{-}\lambda)^5
\bigl\{ (\bar{\omega}_{(\bar{1})}^2{-}\lambda)(\omega_{(1)}^2{-}\lambda) - 8\dot{\psi}^2 \bigr\}^3 \ .
\end{aligned}
\end{equation}
We see that the frequencies $\Omega_{(0)}^2{=}\omega_{(0)}^2$ and $\Omega_{(2)}^2{=}\omega_{(2)}^2$ 
are unchanged and given by~(\ref{unperturbed}), while the gauge mode $\bar{\omega}_{(\bar{1})}^2$ gets
entangled with the (unphysical) $v{=}1$ mode to produce the pair
\begin{equation}
\Omega^2_{(1,\pm)} \= \sfrac12(\bar\omega_{(\bar{1})}^2{+}\omega_{(1)}^2)
\,\pm\sqrt{\sfrac14(\bar\omega_{(\bar{1})}^2{+}\omega_{(1)}^2)^2-\bar\omega_{(\bar{1})}^2\omega_{(1)}^2+8\dot{\psi}^2} 
\= 3\psi^2{+}3\psi{+}2\, \pm\sqrt{\psi^2(\psi{-}1)^2+8\dot{\psi}^2}
\end{equation}
with a triple degeneracy. There are avoided crossings at $\psi{=}0$ and $\psi{=}1$.
Removing the unphysical and gauge modes in pairs, we remain with the singlet mode
$\Omega_{(0,0)}^2$ and the fivefold-degenerate $\Omega_{(0,2)}^2$. 
For all higher spins $j{>}0$, analytic expressions for the natural frequencies $\Omega_{(j,v,\beta)}$
now require merely solving a few polynomial equations of order four at worst.
We have done so up to $j{=}2$ and list them in the Appendix 
but refrain from giving further explicit examples here.
Below we display the cases of $j{=}0$ and $j{=}2$, with similar coloring for like $v$~values,
whose curves avoid crossing each other.
One can see that some of the normal modes dip into the negative regime, i.e.~their frequency-squares 
become negative, for a certain fraction of the time~$\tau$.
Because of this and, quite generally, due to the $\tau$ variability of the natural frequencies,
it is not easy to predict the long-term evolution of the fluctuation modes.
Clearly, the stability of the zero solution $\Phi{\equiv}0$, 
equivalent to the linear stability of the background Yang--Mills configuration, 
is not simply decided by the sign of the $\tau$-average of the corresponding frequency-square.
\begin{figure}[h!]
\centering
\includegraphics[width = 0.35\paperwidth]{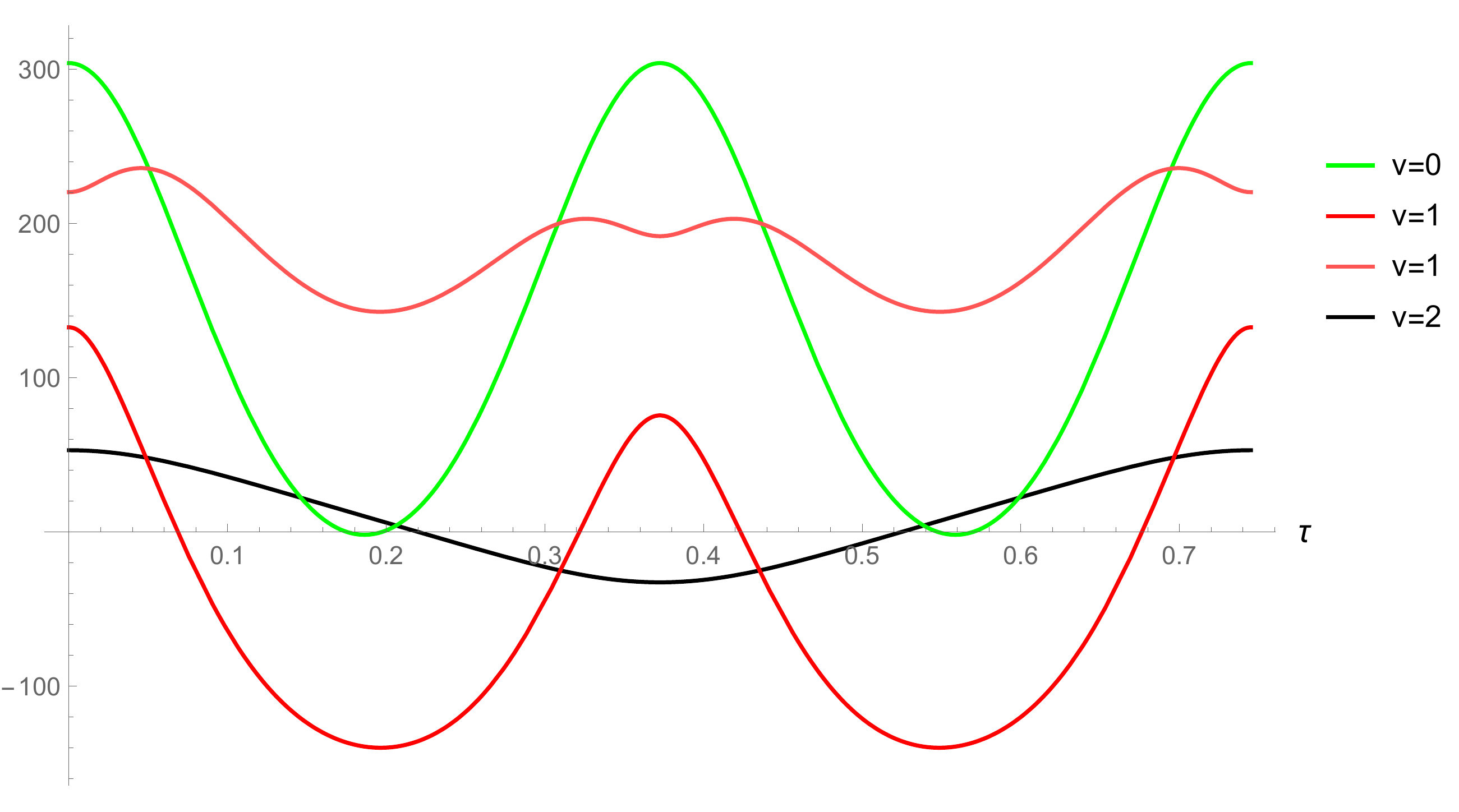} \qquad\quad
\includegraphics[width = 0.35\paperwidth]{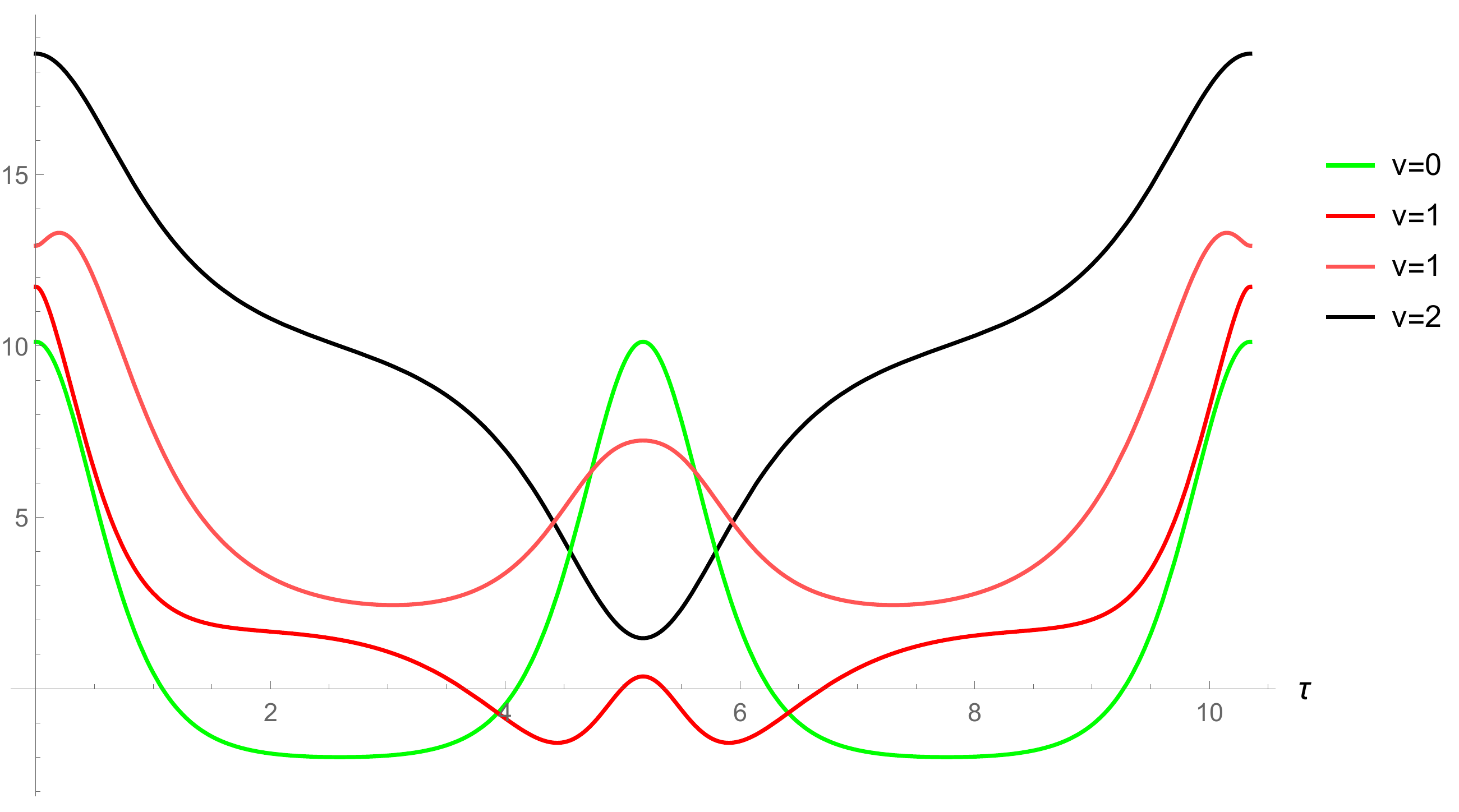} \\[12pt]
\includegraphics[width = 0.35\paperwidth]{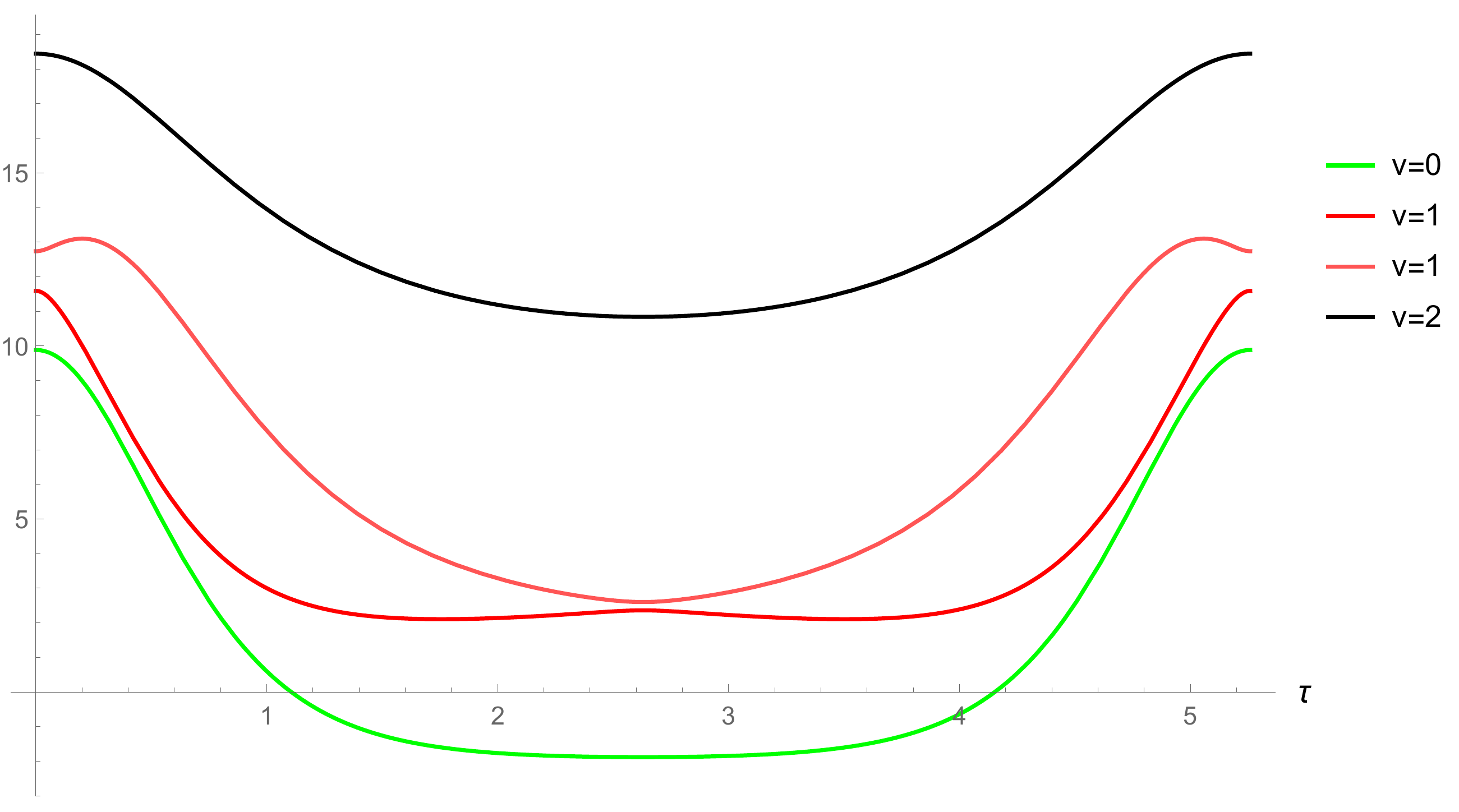} \qquad\quad
\includegraphics[width = 0.35\paperwidth]{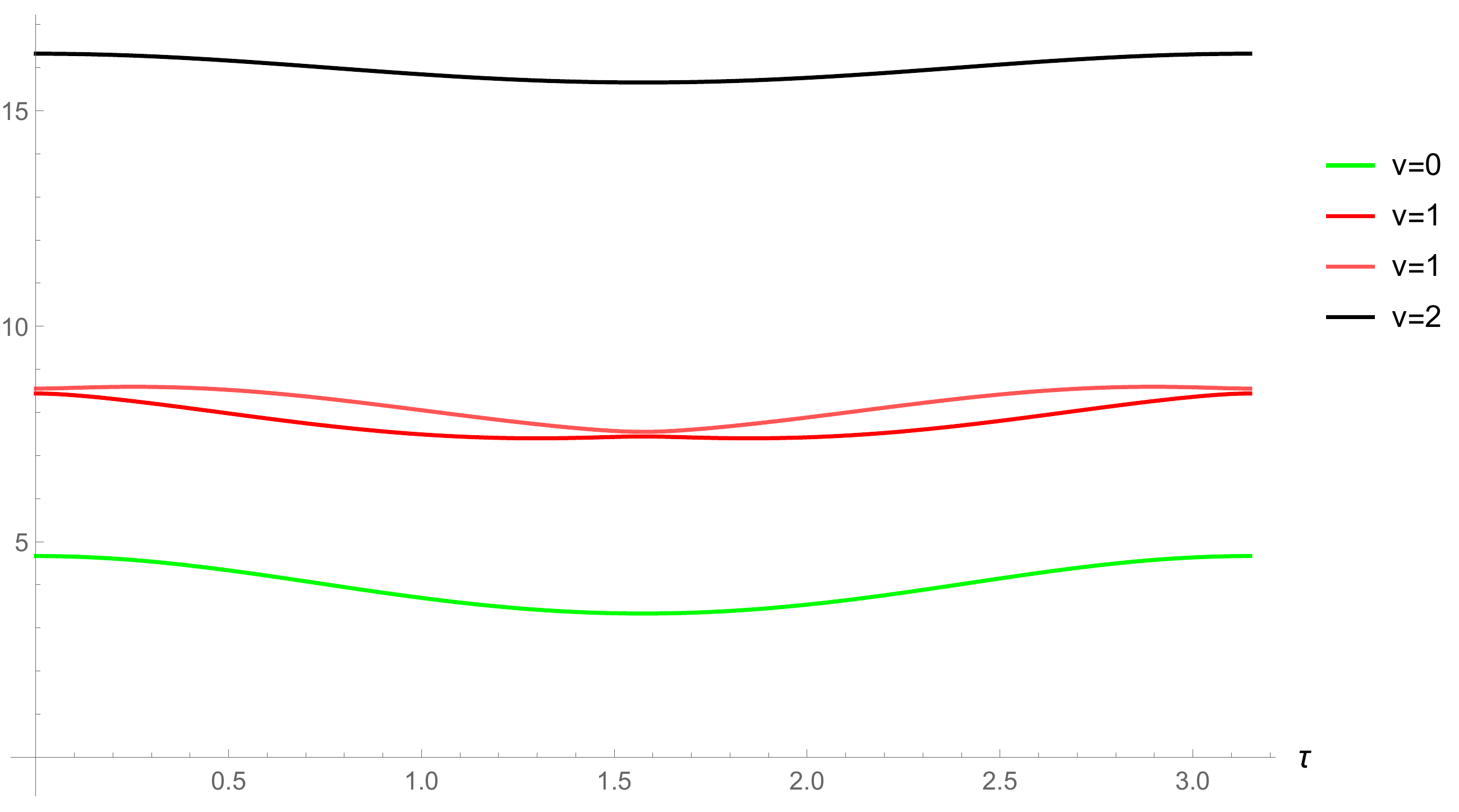}
\caption{Plots of $\Omega^2_{(0,v,\beta)}(\tau)$ over one period, for different values of $k^2$: \newline
\indent\qquad\qquad\
$0.51$ (top left), $0.99$ (top right), $1.01$ (bottom left) and $5$ (bottom right).}
\end{figure}
\begin{figure}[h!]
\centering
\includegraphics[width = 0.35\paperwidth]{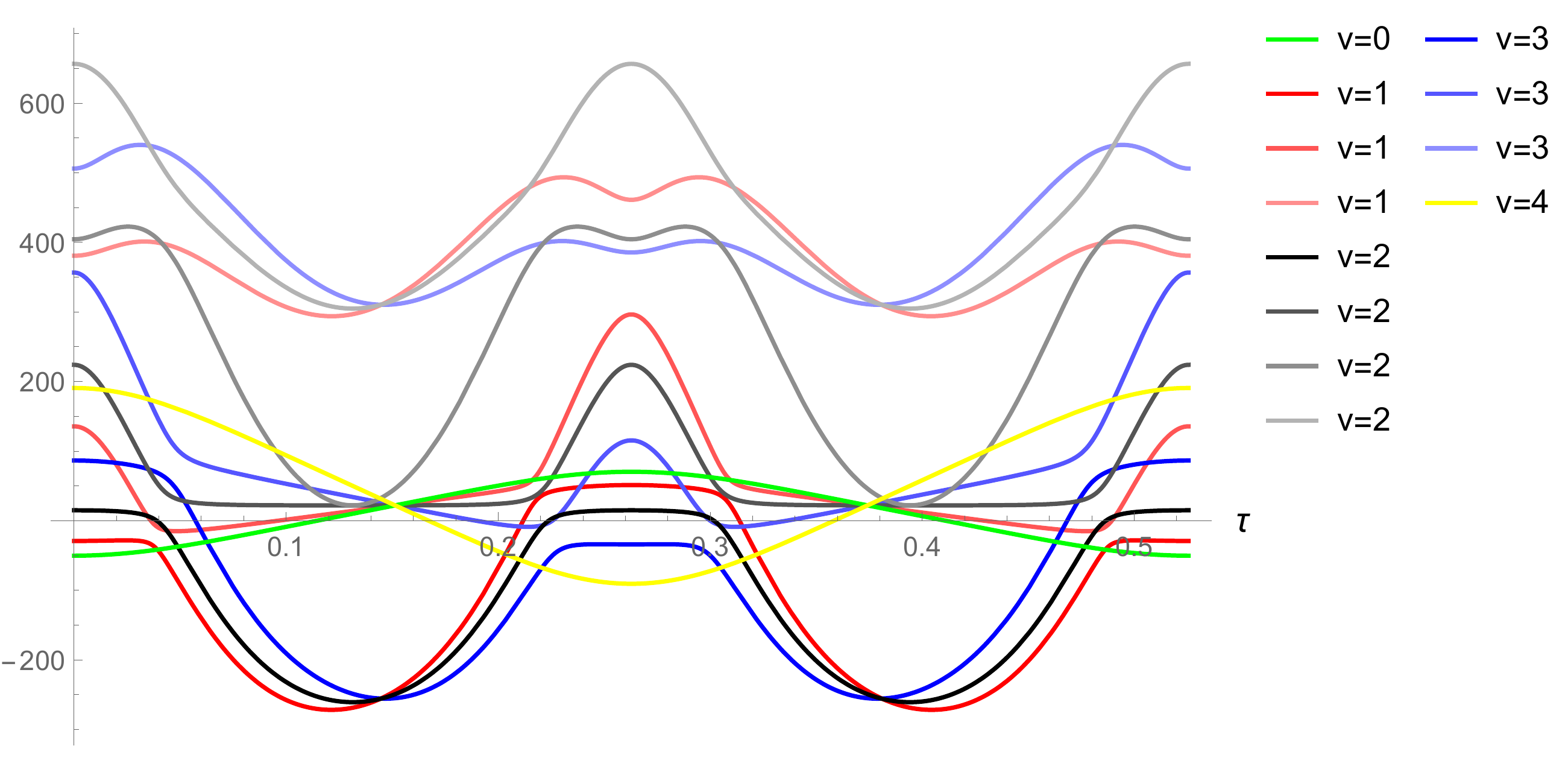} \qquad\quad
\includegraphics[width = 0.35\paperwidth]{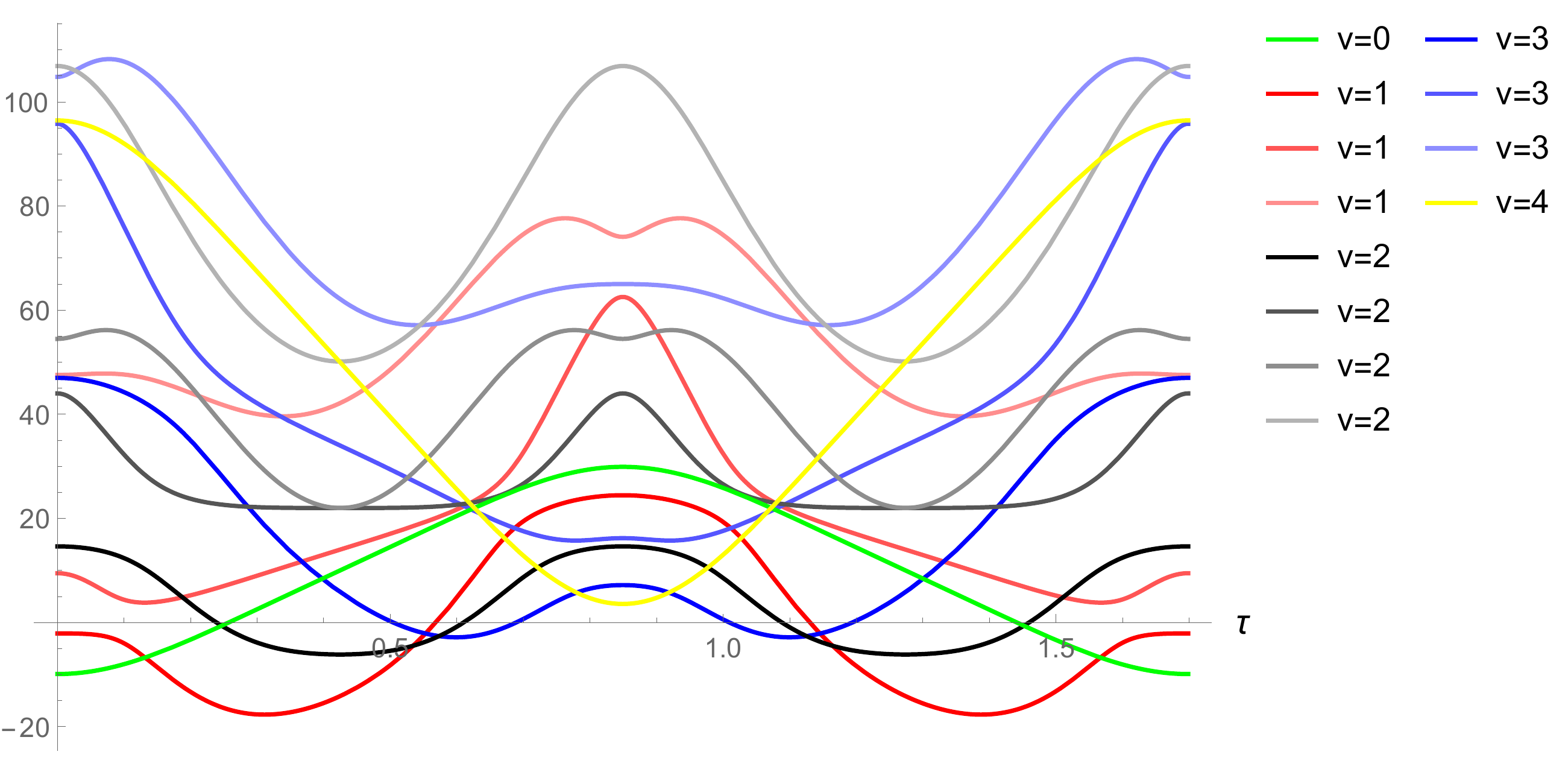} \\[12pt]
\includegraphics[width = 0.35\paperwidth]{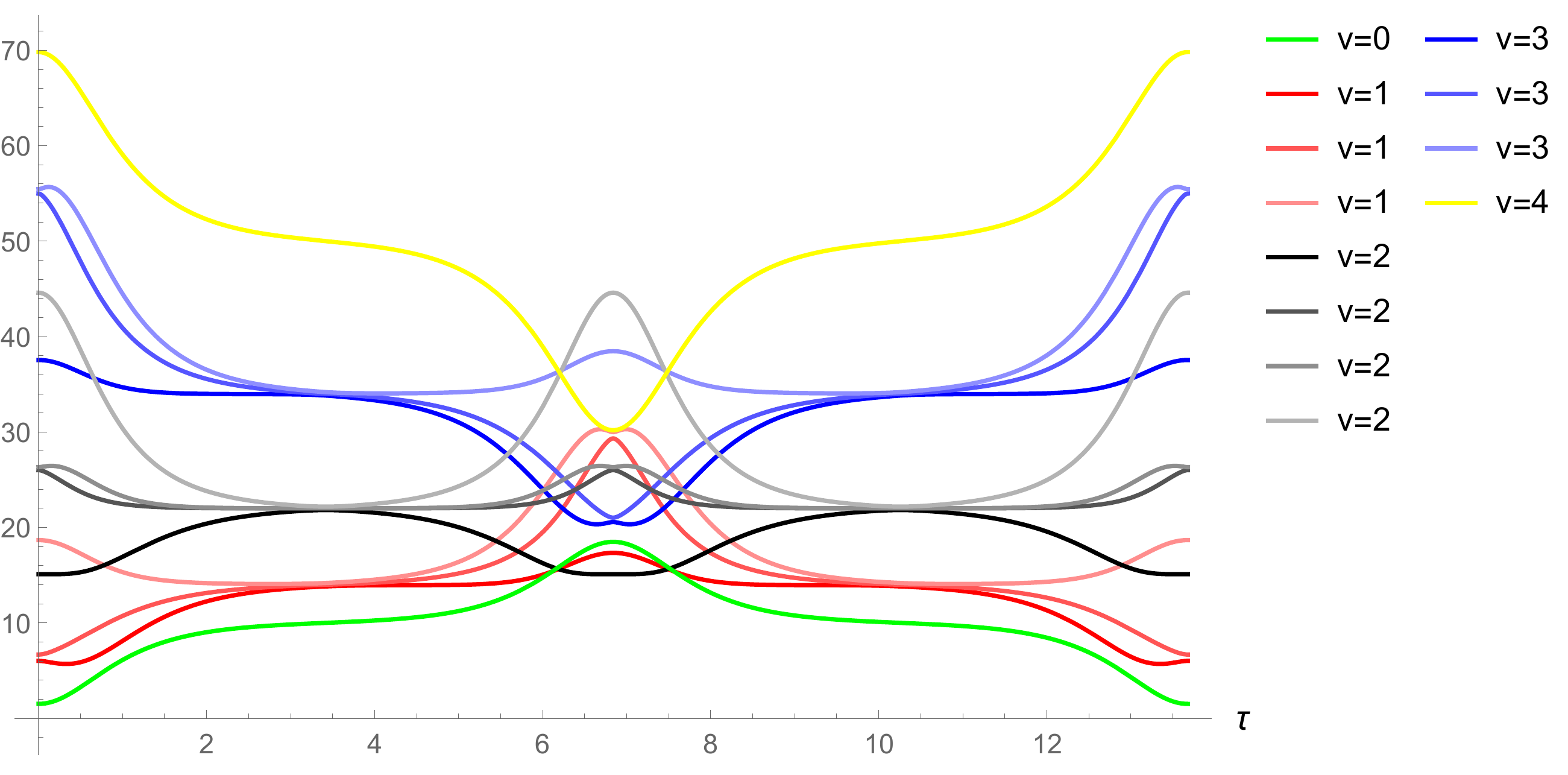} \qquad\quad
\includegraphics[width = 0.35\paperwidth]{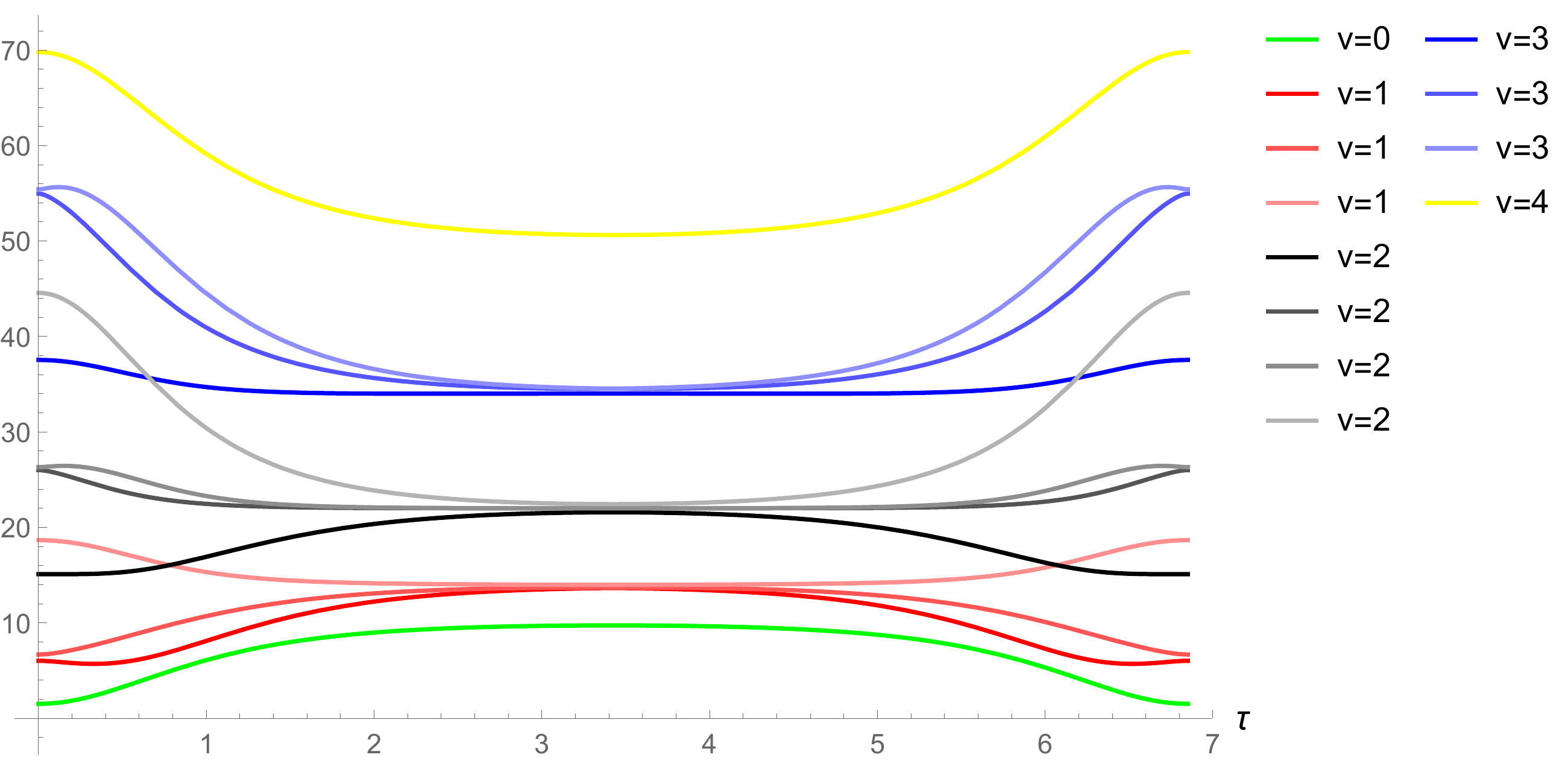}
\caption{Plots of $\Omega^2_{(2,v,\beta)}(\tau)$ over one period, for different values of $k^2$: \newline
\indent\qquad\qquad\
$0.505$ (top left), $0.550$ (top right), $0.999$ (bottom left) and $1.001$ (bottom right).}
\end{figure}

\section{Stability analysis: stroboscopic map and Floquet theory}

\noindent
The diagonalized linear fluctuation equation~(\ref{diagonalized}) represents
a bunch of Hill's equations, where the frequency-squared is a root of a polynomial of order up to four
with coefficients given by a polynomial of twice that order in Jacobi elliptic functions.
A unique solution requires fixing two initial conditions, and so
for each fluctuation~$\Phi_{(j,v,\beta)}$ there is a two-dimensional solution space.
It is well known that Hill's equation, e.g.~in the limit of Mathieu's equation, displays parametric resonance phenomena,
which can stabilize otherwise unstable systems or destabilize otherwise stable ones.

For oscillating dynamical systems with periodically varying frequency, 
there exist some general tools to analyze linear stability.
Switching to a Hamiltonian picture and to phase space, 
it is convenient to transform the second-order differential equation into
a system of two coupled first-order equations (suppressing all quantum numbers),
\begin{equation} \label{first-order}
\bigl[\partial^2_\tau-\Omega^2(\tau)\bigr] \Phi(\tau) \= 0 \qquad\Leftrightarrow\qquad
\partial_\tau \begin{pmatrix} \Phi \\[4pt] \dot\Phi \end{pmatrix} \=
\begin{pmatrix} 0 & 1 \\[4pt] -\Omega^2 & 0 \  \end{pmatrix} 
\begin{pmatrix} \Phi \\[4pt] \dot\Phi \end{pmatrix} \ =:\
\im\,\widehat\Omega(\tau)\,\begin{pmatrix} \Phi \\[4pt] \dot\Phi \end{pmatrix}\ ,
\end{equation}
where the frequency~$\Omega(\tau)$ is $T$-periodic (sometimes $\frac{T}{2}$-periodic) in~$\tau$.
The solution to this first-order system is formally given by
\begin{equation}
\begin{pmatrix} \Phi \\[4pt] \dot\Phi \end{pmatrix}(\tau) \=
{\cal T} \exp\,\Bigl\{ \int_0^\tau \!\diff\tau'\ \im\,\widehat\Omega(\tau') \Bigr\} \,
\begin{pmatrix} \Phi \\[4pt] \dot\Phi \end{pmatrix}(0) \ ,
\end{equation}
where ${\cal T}$ denotes time ordering.
Because of the time dependence of $\Omega$, the time evolution operator above
is not homogeneous thus does not constitute a one-parameter group, except when
the propagation interval is an integer multiple of the period~$T$. For $\tau{=}T$,
one speaks of the stroboscopic map~\cite{Arnold}
\begin{equation}
M\ :=\ {\cal T} \exp\,\Bigl\{ \int_0^T \!\diff\tau\ \im\,\widehat\Omega(\tau) \Bigr\} 
\qquad\Rightarrow\qquad
\begin{pmatrix} \Phi \\[4pt] \dot\Phi \end{pmatrix}(nT) \= M^n
\begin{pmatrix} \Phi \\[4pt] \dot\Phi \end{pmatrix}(0)\ .
\end{equation}
The linear map~$M$ is a functional of the chosen background solution~$\psi$
and hence depends on its parameter~$E$ or~$k$.
This background is Lyapunov stable if the trivial solution $\Phi{\equiv}0$ is,
which is decided by the two eigenvalues $\mu_1$ and $\mu_2$ of~$M$. 
Since the system is Hamiltonian, $\det M{=}1$, and we have three cases:
\begin{equation}
\begin{aligned}
|\tr\,M| > 2 &\quad\Leftrightarrow\quad \mu_i\in\R  
&\quad\;\Leftrightarrow\quad& \textrm{hyperbolic/boost} 
&\quad\Leftrightarrow\quad& \textrm{strongly unstable} \ ,\\
|\tr\,M| = 2  &\quad\Leftrightarrow\quad \mu_i=\pm1  
&\quad\Leftrightarrow\quad& \textrm{parabolic/translation} 
&\quad\Leftrightarrow\quad& \textrm{marginally stable}\ , \\
|\tr\,M| < 2  &\quad\Leftrightarrow\quad \mu_i\in\textrm{U}(1)  
&\quad\Leftrightarrow\quad& \textrm{elliptic/rotation} 
&\quad\Leftrightarrow\quad&\textrm{strongly stable}\ .
\end{aligned}
\end{equation}
Clearly, $|\tr\,M|$ determines the linear stability of our classical solution.

Let us thus try to evaluate the trace of the stroboscopic map~$M$,
making use of the special form of the matrix~$\widehat\Omega$,
\begin{equation}
\begin{aligned}
\tr\,M &\= \sum_{n=0}^\infty \im^n 
\int_0^T\!\!\diff\tau_1\int_0^{\tau_1}\!\!\!\diff\tau_2\ \ldots\int_0^{\tau_{n-1}}\!\!\!\!\diff\tau_n\
\tr\,\bigl[ \widehat\Omega(\tau_1)\,\widehat\Omega(\tau_2)\cdots\widehat\Omega(\tau_n)\bigr] \\[4pt]
&\= 2\ +\  \sum_{n=1}^\infty(-1)^n 
\int_0^T\!\!\diff\tau_1\int_0^{\tau_1}\!\!\!\diff\tau_2\ \ldots\int_0^{\tau_{n-1}}\!\!\!\!\diff\tau_n\
H_n(\tau_1,\tau_2,\ldots,\tau_n)\,\Omega^2(\tau_1)\,\Omega^2(\tau_2)\,\cdots\Omega^2(\tau_n) \\[4pt]
\textrm{with}& \quad H_n(\tau_1,\tau_2,\ldots,\tau_n) \= 
(\tau_1{-}\tau_2)(\tau_2{-}\tau_3)\cdots(\tau_{n-1}{-}\tau_n)(\tau_n{-}\tau_1{+}1)
\und H_1(\tau_1)=1\ .
\end{aligned}
\end{equation}
It is convenient to scale the time variable such as to normalize the period to unity,
\begin{equation}
\tau = T\, x  \qquad\und\qquad \Omega^2(Tx) =: \omega^2(x)\ ,\quad H(\{Tx\})=:h(\{x\}) \ ,
\end{equation}
hence
\begin{equation} \label{Ansum}
\begin{aligned}
\tr\,M &\= 2\ +\  \sum_{n=1}^\infty \bigl(-T^2\bigr)^n 
\int_0^1\!\!\diff x_1\int_0^{x_1}\!\!\!\diff x_2\ \ldots\int_0^{x_{n-1}}\!\!\!\!\diff x_n\
h_n(x_1,x_2,\ldots,x_n)\,\omega^2(x_1)\,\omega^2(x_2)\,\cdots\omega^2(x_n) \\[4pt]
&\= \sum_{n=0}^\infty \frac{2}{(2n)!}\,M_n\,\bigl(-T^2\bigr)^n 
\ =:\  2 - M_1 T^2 + \sfrac{1}{12}M_2 T^4 - \sfrac{1}{360}M_3 T^6 + \sfrac{1}{20160}M_4 T^8 - \ldots\ .
\end{aligned}
\end{equation}

It is impossible to evaluate the integrals $M_n$ without explicit knowledge of~$\omega^2(x)$. 
As a crude guess, we replace the weight function by its (constant) average value
\begin{equation}
\<h_n\> \ :=\ \frac{1}{n!} \int_0^1\!\!\diff x_1\int_0^{x_1}\!\!\!\diff x_2\ \ldots\int_0^{x_{n-1}}\!\!\!\!\diff x_n\ 
h_n(x_1,x_2,\ldots,x_n) \= \frac{2\,n!}{(2n)!}
\end{equation}
and obtain
\begin{equation}
M_n \= \frac{(2n)!}{2}\,\<h_n\> 
\int_0^1\!\!\diff x_1\int_0^{x_1}\!\!\!\diff x_2\ \ldots\int_0^{x_{n-1}}\!\!\!\!\diff x_n\ \prod_{i=1}^n \omega^2(x_i)
\= \Bigl( \int_0^1\!\!\diff x\ \omega^2(x) \Bigr)^n \ =:\ \< \,\omega^2\>^n\ ,
\end{equation}
which yields
\begin{equation}
\tr\,M \= 2\,\sum_{n=0}^\infty \frac{(-1)^n}{(2n)!}\,\<\,\omega^2\>^n\,T^{2n} 
\= 2\,\cos\bigl(\sqrt{\<\,\omega^2\>}\,T\bigr)\ .
\end{equation}
This expression indicates stability as long as $\<\omega^2\>>0$.
However, the result for the $j{=}0$ singlet mode $\omega^2=\Omega^2_{(0,0)}$ in (\ref{j0average}) 
below already shows that the averaged frequency-squared may turn negative in certain domains
thus changing the cos into a cosh there.

To do better, let us look at the individual terms~$M_n$ in~(\ref{Ansum}) for the simplest case
of the SO(4) singlet fluctuation, i.e.~$\Omega^2_{(0,0)}=6\psi^2{-}2$ in~(\ref{unperturbed}). 
Its average frequency-square is easily computed to be
\begin{equation} \label{j0average}
\<\,\Omega_{(0,0)}^2 \> \= \frac{1}{\epsilon^2}\Bigl(6 \frac{E(k)}{K(k)}+4k^2-5\Bigr)\ , 
\end{equation}
where $E(k)$ and $K(k)$ denote the second and first complete elliptic integrals, respectively.
Plotting this expression as a function of the modulus~$k$, we see that it
becomes negative only in a very narrow range around $k{=}1$, namely for $|k{-}1|\lesssim0.00005$.
\begin{figure}[h!]
\centering
\includegraphics[width = 0.35\paperwidth]{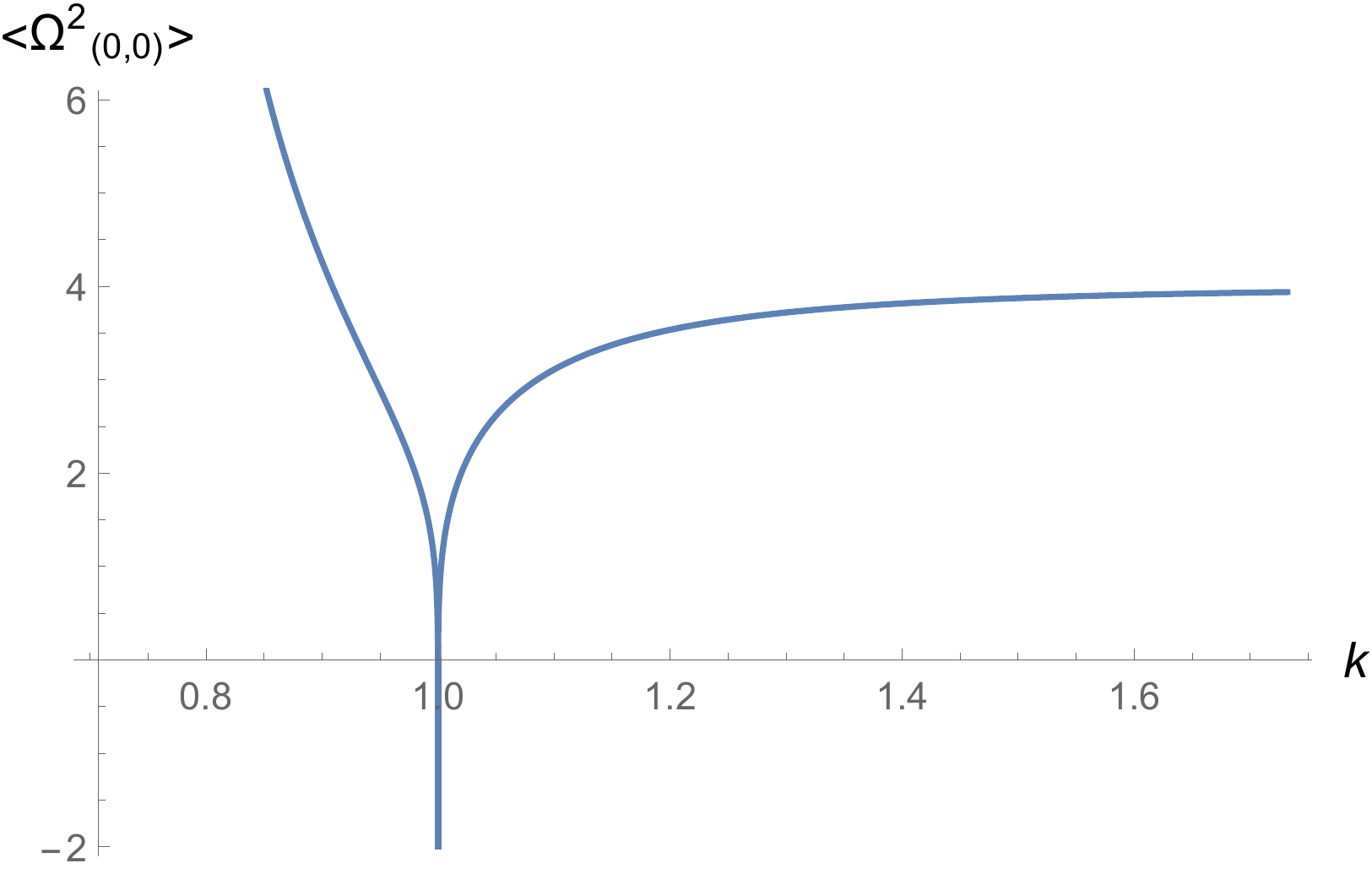} \qquad\quad
\includegraphics[width = 0.35\paperwidth]{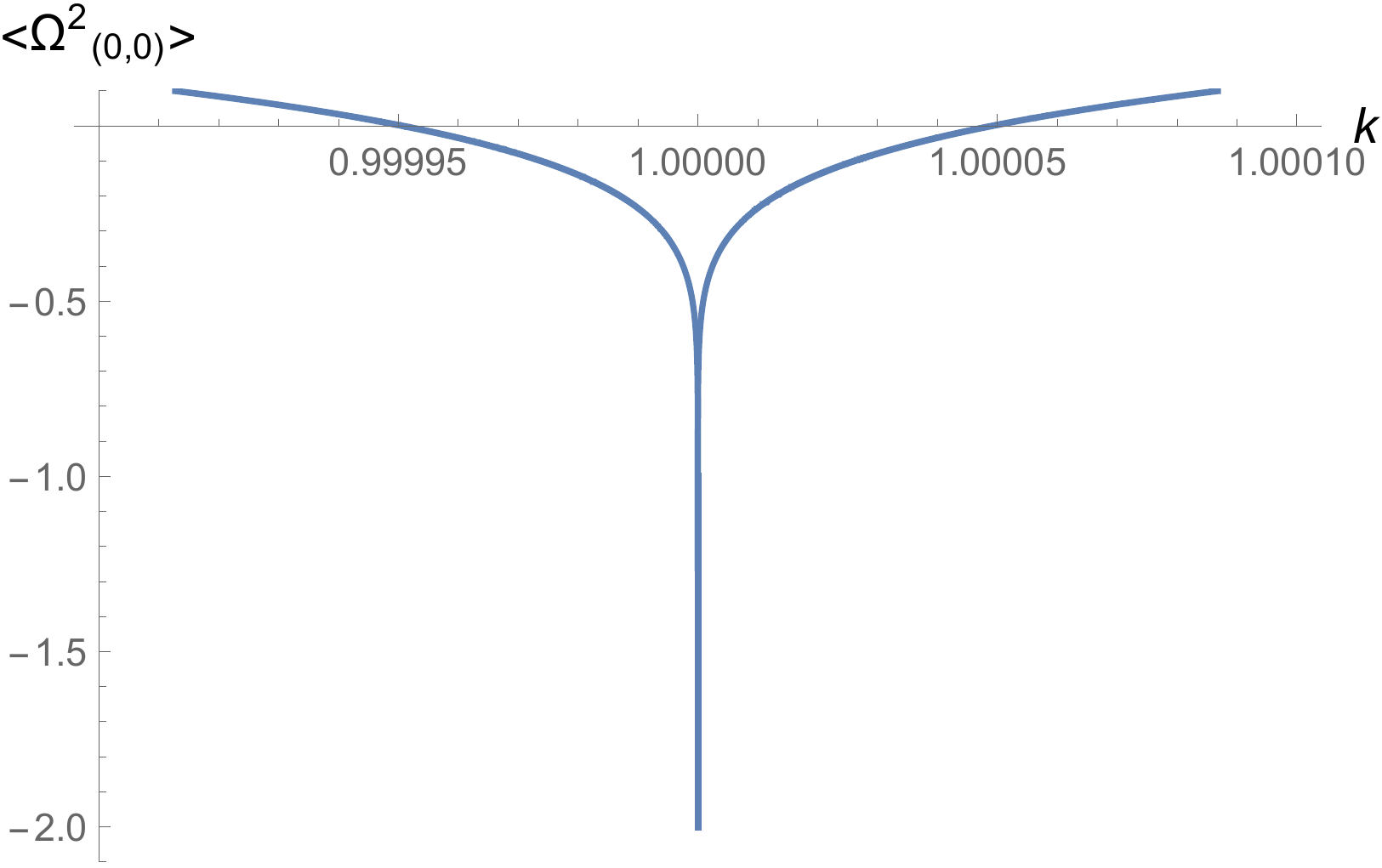}
\caption{Plot of $\<\Omega^2_{(0,0)}\>$ as a function of~$k$, with detail on the right.}
\end{figure}
We have only been able to analytically evaluate (with $k{<}1$ for simplicity)
\begin{equation}
M_1 \= \<\,\Omega_{(0,0)}^2 \>  \qquad\und\qquad
M_2 \= \<\,\Omega_{(0,0)}^2 \> ^2\ -\ \frac{1}{\epsilon^4}\Bigl(
9\frac{2{-}k^2}{K(k)^2} - 27\frac{E(k)}{K(k)^3} + \frac{9\,\pi^2}{4\,K(k)^4} \Bigr)\ ,
\end{equation}
which does not suffice to rule out instability.
Indeed, numerical studies show that $M_n$ as a function of~$k$ looses its positivity in a range
around $k{=}1$ which increases with~$n$, where the series~(\ref{Ansum}) ceases to be alternating.
Moreover, even in the limit of a very large background amplitude, $k^2\to\frac12$, we find that
\begin{equation}
\<\,\Omega_{(0,0)}^2 \>\ \to\ \frac{24\,\pi^2}{\epsilon^2\,\Gamma(\frac14)^4}\ \approx\ \frac{1.37}{\epsilon^2}
\qquad\Rightarrow\qquad \sqrt{\<\,\Omega^2\>}\,T\ \to\ \sqrt{24\,\pi}\ \approx\ 8.68\ ,
\end{equation}
implying that we must push the series in~(\ref{Ansum}) at least to $O(M_{10}T^{20})$,
even though it turns out that $M_n<\<\,\Omega^2_{(0,0)}\>^n$ at $k^2=\frac12$ for $n>1$.

For a more complete analysis of linear stability in an oscillating system with time-dependent frequency
we can take recourse to Floquet theory.
It tells us that a general fundamental matrix solution
\begin{equation}
\widehat\Phi(\tau) \= \begin{pmatrix} \Phi_1 & \Phi_2 \\[4pt] \dot\Phi_1 & \dot\Phi_2 \end{pmatrix}(\tau)
\qquad\Rightarrow\qquad \partial_\tau \widehat\Phi(\tau) \= \im\,\widehat\Omega(\tau)\,\widehat\Phi(\tau)
\end{equation}
of our system~(\ref{first-order}) with some initial condition $\widehat\Phi(0)=\widehat\Phi_0$ 
can be expressed in so-called Floquet normal form as
\begin{equation}
\widehat\Phi(\tau) \= Q(\tau)\;\ep^{\tau R} 
\qquad\textrm{with}\qquad Q(\tau{+}2T) \= Q(\tau)\ ,
\end{equation}
where $Q(\tau)$ and $R$ are real $2{\times}2$ matrices,
so that the time dependence of the frequency can be transformed away by a change of coordinates,
\begin{equation}
\Psi(\tau) \ :=\ Q(\tau)^{-1} \widehat\Phi(\tau) \qquad\Rightarrow\qquad
\partial_\tau \Psi(\tau) \= R\,\Psi(\tau)\ .
\end{equation}
Due to the identity
\begin{equation}
\widehat\Phi(\tau{+}T) \= \widehat\Phi(\tau)\,\widehat\Phi(0)^{-1}\,\widehat\Phi(T) 
\= \widehat\Phi(T)\,\widehat\Phi(0)^{-1}\,\widehat\Phi(\tau) \= M\,\widehat\Phi(\tau)
\end{equation}
we see that our stroboscopic map~$M$ is nothing but the monodromy, and
\begin{equation}
M^2 \= \widehat\Phi(2T)\,\widehat\Phi(0)^{-1} 
\= Q(0)\,\widehat\Phi(0)^{-1}\,\widehat\Phi(2T)\,Q(0)^{-1}
\= Q(0)\,\ep^{2 R T}\,Q(0)^{-1}\ ,
\end{equation}
so that its eigenvalues (or characteristic multipliers)
\begin{equation}
\mu_i = \ep^{\rho_i T} \qquad\textrm{for}\quad i=1,2
\end{equation}
define a pair of (complex) Floquet exponents $\rho_i$ 
whose real parts are the Lyapunov exponents.
Since $\mu_1\mu_2=1$ implies that $\rho_1{+}\rho_2=0$,
our system is linearly stable if and only if both eigenvalues~$\rho_i$ of~$R$
are purely imaginary (or zero).

Generally it is impossible to find analytically the monodromy pertaining to
a normal mode~$\Phi_{(j,v,\beta)}$.\footnote{
An exception is the SO(4) singlet perturbation~$\Phi_{(0,0)}$, to be treated in the following section.}
However, we can evaluate it numerically for a number of examples. 
Before doing so, let us estimate at which energies~$E$
or, rather, moduli~$k$, possible resonance frequencies might occur.
To this end, we determine the period-average of the natural frequency $\Omega_{(j,v,\beta)}$ 
and compare it to its modulation frequency~$\sfrac{2\pi}{T}$. If we model 
\begin{equation}
\Omega^2(\tau) \ \approx\ \<\,\Omega^2\> \,\bigl(1+h(\tau)\bigr)  \qquad\textrm{with}\qquad
\<\,\Omega^2\> = \sfrac{1}{T}\smallint_0^T \!\diff\tau\ \Omega^2(\tau) \quad\und\quad
h(\tau) \ \propto\ \cos(2\pi\tau/T) \ ,
\end{equation}
where $T=4\,\epsilon\,K(k)$, then the resonance condition is met for
\begin{equation}
\sqrt{\<\Omega^2\>} \= \ell\,\frac{\pi}{T} \qquad\Rightarrow\qquad
k=k_\ell(j,v,\beta)\qquad\textrm{for}\quad \ell=1,2,3,\ldots\ .
\end{equation}
Since this model reproduces only the rough features of $\Omega^2(\tau)$, we expect potential instability
due to parametric resonance effects in a band around or near the values~$k_\ell$.

Our expectation is confirmed by precise numerical evaluation of various monodromies as a function of~$k$.
Below we display, together with the would-be resonant values~$k_\ell$,
the function $\tr M(k)$ for the sample cases of $(j,v)=(2,0)$ and $(2,2)$.
\begin{figure}[h!]
\centering
\includegraphics[width = 0.35\paperwidth]{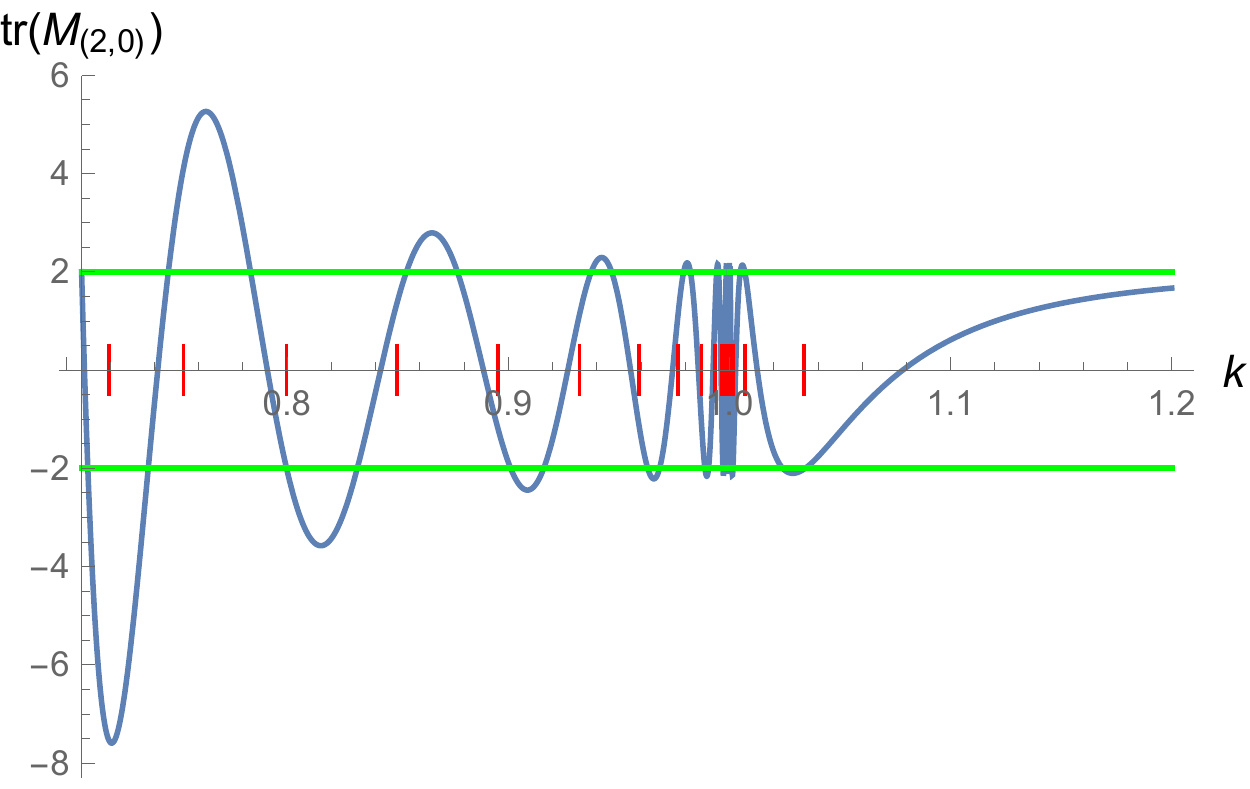} \qquad\quad
\includegraphics[width = 0.35\paperwidth]{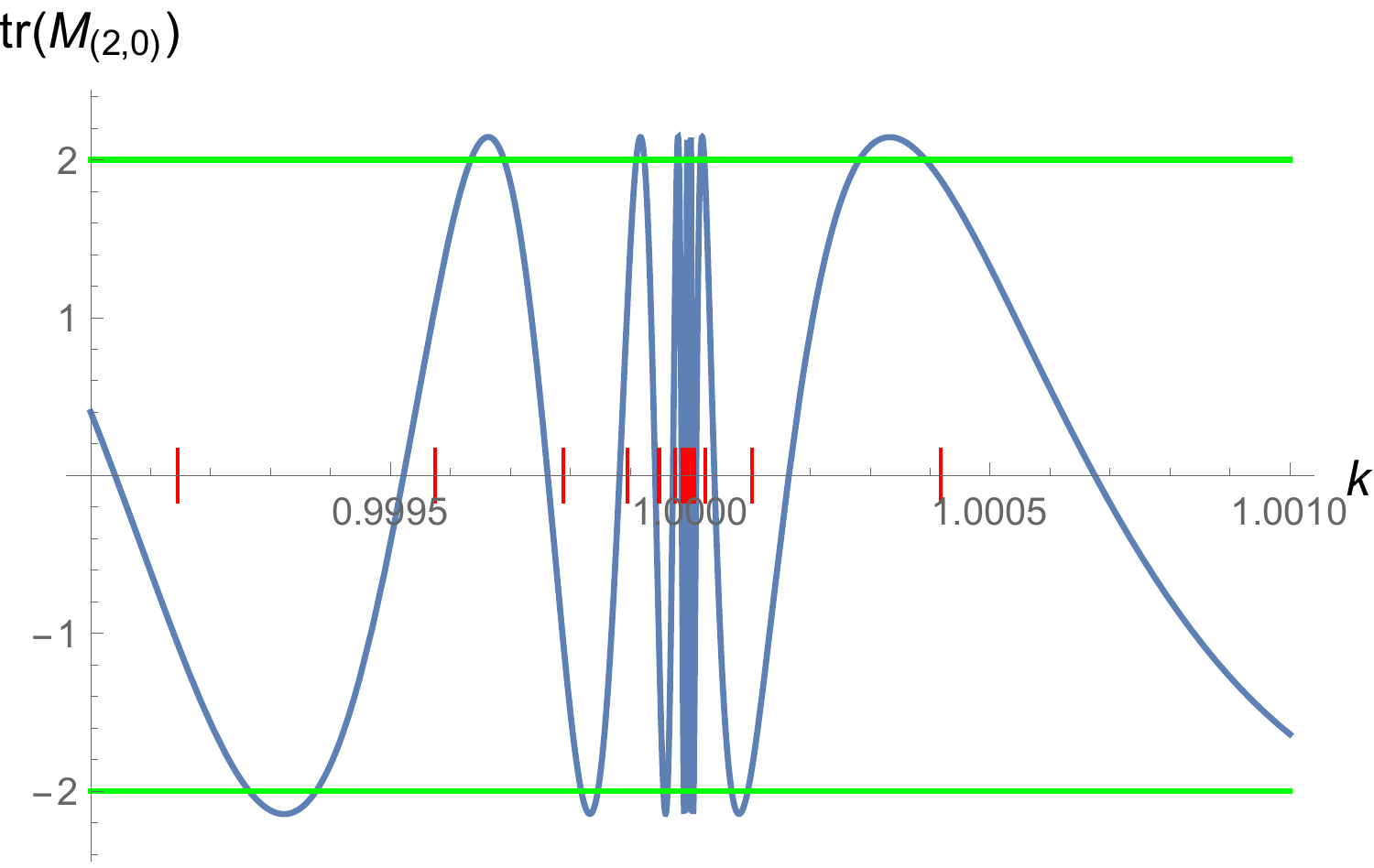} 
\caption{Plot of $\tr\,M(k)$ for $(j,v)=(2,0)$, with detail on the right. Would-be resonances marked in red.}
\end{figure}
\begin{figure}[h!]
\centering
\includegraphics[width = 0.35\paperwidth]{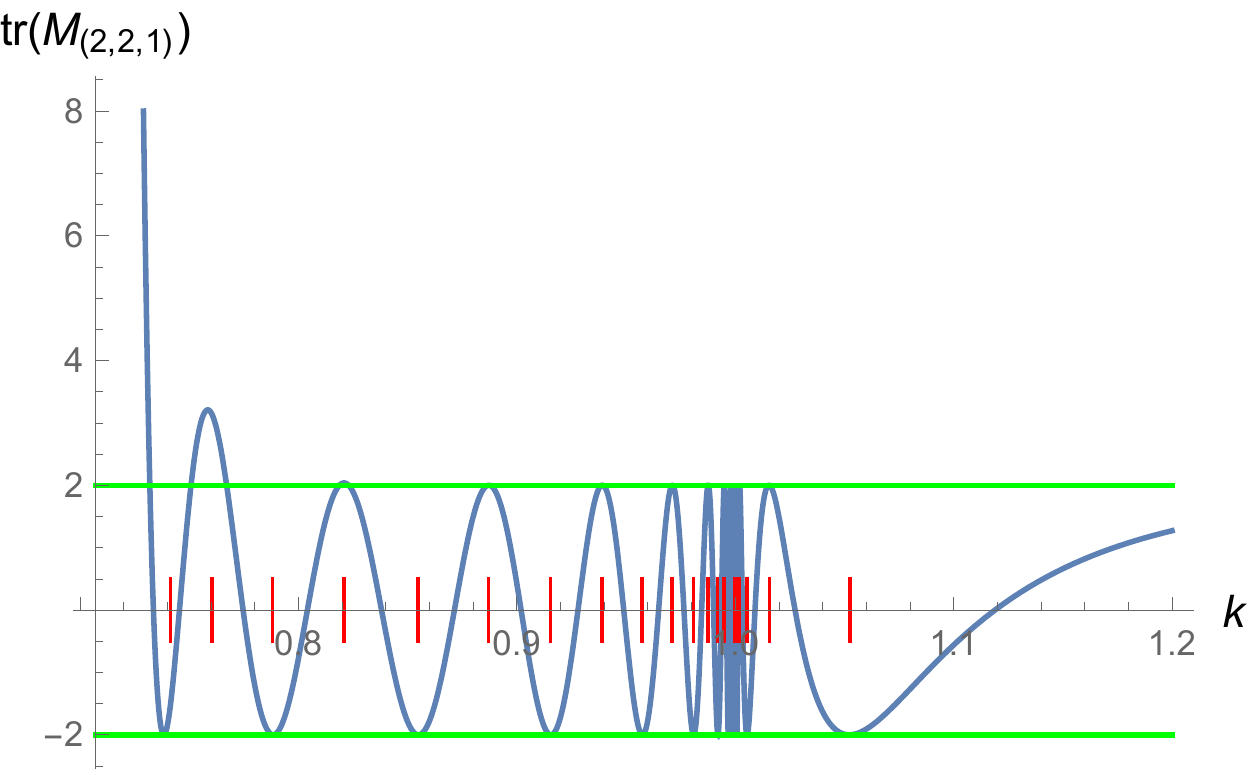} \qquad\quad
\includegraphics[width = 0.35\paperwidth]{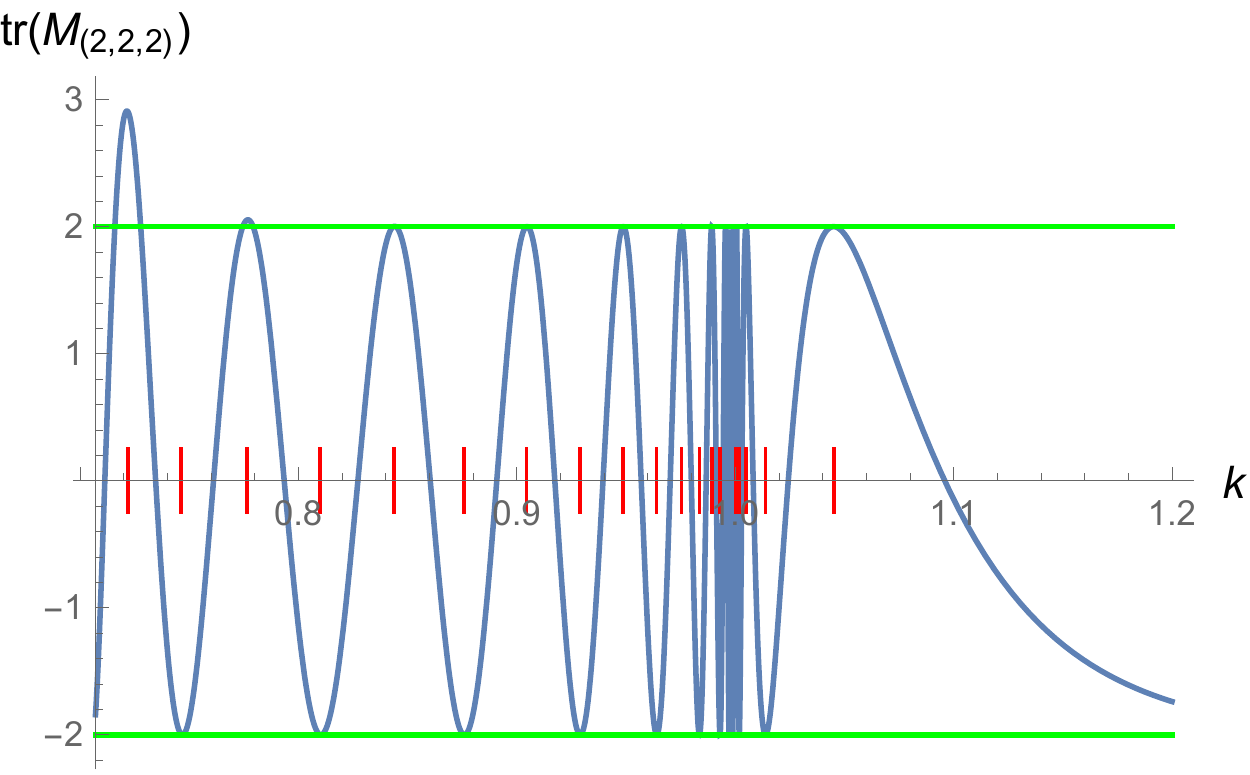} \\[8pt]
\includegraphics[width = 0.35\paperwidth]{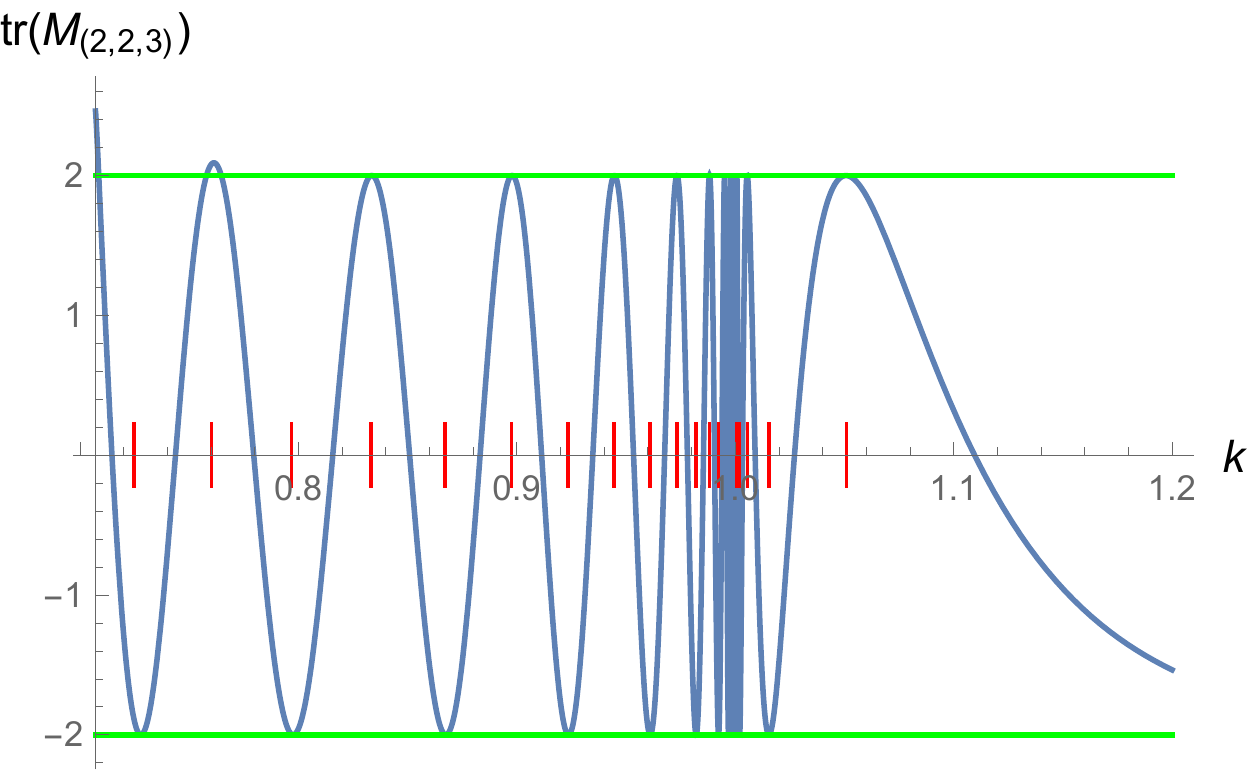} \qquad\quad
\includegraphics[width = 0.35\paperwidth]{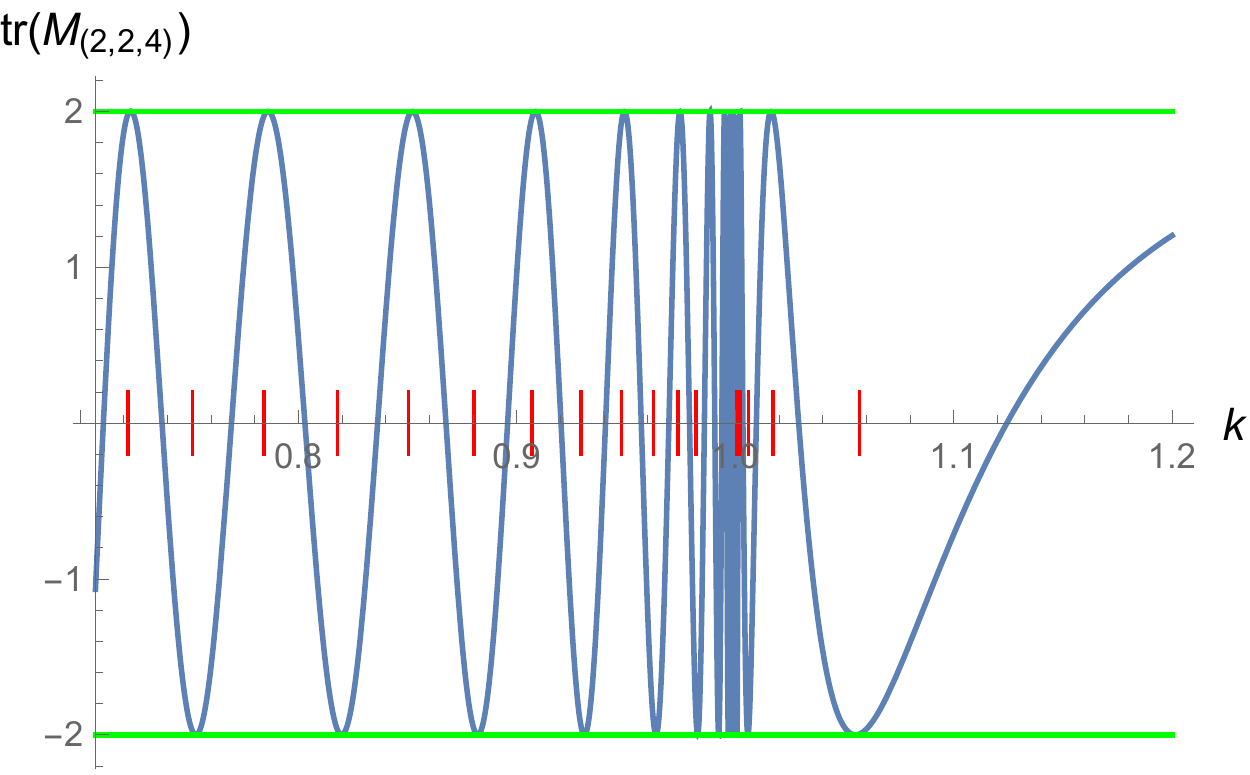} 
\caption{Plots of $\tr\,M(k)$ for $(j,v)=(2,2)$ and $\beta=1,2,3,4$. Would-be resonances marked in red.}
\end{figure}
One sees that, on both sides of the critical value of $E{=}\sfrac12$ (or $k{=}1$), 
corresponding to the double-well local maximum, the $k_\ell$ values accumulate at the critical point.
But while for $k{>}1$ (energy below the critical point) $\tr M(k)$ oscillates between values close to $2$ 
in magnitude and thus exponential growth is rare and mild, for $k{<}1$ (energy above the critical point)
the oscillatory behavior of $\tr M(k)$ comes with an amplitude exceeding~2 and growing with energy.
Hence, in this latter regime stable and unstable bands alternate.
This is supported by long-term numerical integration, as we demonstrate in Figure~9 by plotting
$\Phi(\tau)$ for $(j,v,\beta)=(2,2,1)$ with initial values $\Phi(0){=}1$ and $\dot\Phi(0){=}0$ 
on both sides of the first transition from instability to stability for $\tr\,M_{(2,2,1)}$ shown in Figure~8
(at the highest value of~$E$ or the lowest value of~$k$).
\begin{figure}[h!]
\centering
\includegraphics[width = 0.35\paperwidth]{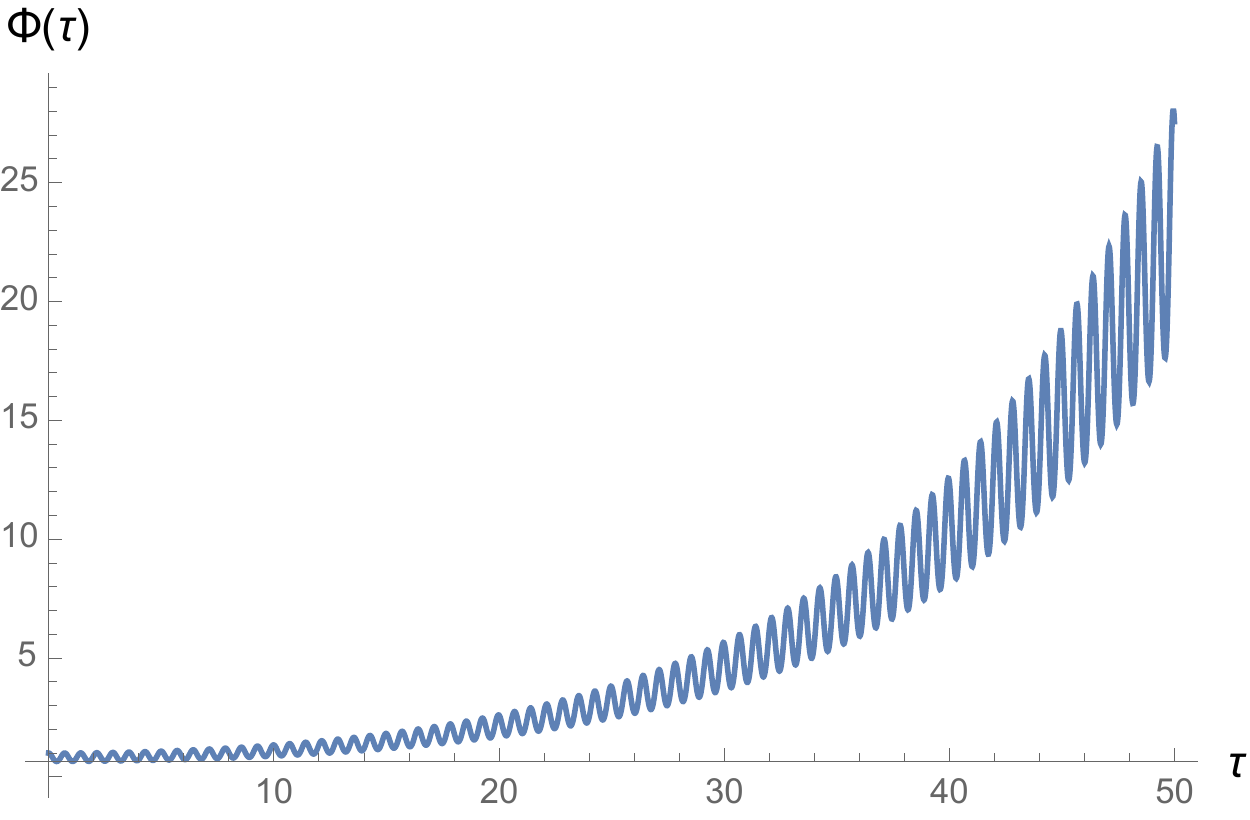} \qquad\quad
\includegraphics[width = 0.35\paperwidth]{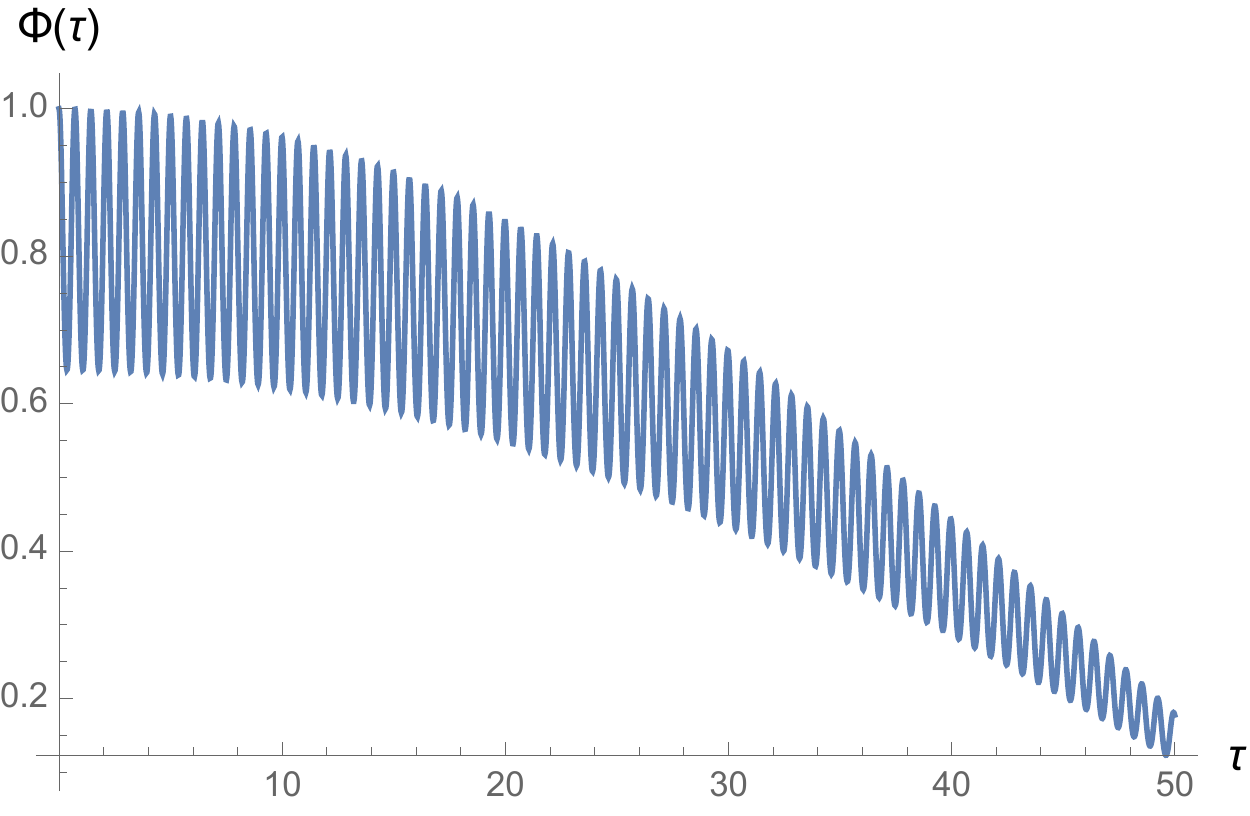} 
\caption{Plot of $\Phi(\tau)$ for $(j,v,\beta)=(2,2,1)$ and $k{=}0.73198$ (left) and $k{=}0.73199$ (right).}
\end{figure}

Most relevant for the cosmological application is the regime of very large energies, $E\to\infty$ (or $k\to 1/\sqrt{2}$).
In this limit, we observe the following universal behavior. 
Because the period~$T$ collapses with $\epsilon{=}\sqrt{k^2{-}1/2}$, we rescale
\begin{equation}
\sfrac{\tau}{\epsilon} = z \in [0,4K(\sfrac12)]\quad ,\qquad
\epsilon\,\psi = \tilde\psi\quad,\qquad 
\epsilon^2\dot\psi = \partial_z\tilde\psi\quad,\qquad 
\epsilon^2\Omega^2 = \tilde\Omega^2\quad,\qquad 
\epsilon^2\lambda = \tilde\lambda
\end{equation}
so that the tilded quantities remain finite in the limit,
and find, with $\bar\omega^2_{(\bar{v})}\to 2\psi^2$,\footnote{
For the cases $(j,v)=(0,1)$ and $(1,0)$, the factor $\tilde\lambda$ is missing; 
for $(j,v)=(0,0)$, one only has $R=Q\sim(\tilde\lambda{-}6\tilde\psi^2)$.}
\begin{equation}
\begin{aligned}
&Q_{(v\pm2)}\ \sim\ \tilde\lambda\ , & \\
&Q_{(v\pm1)}\ \sim\ \tilde\lambda\,(\tilde\lambda-4\tilde\psi^2)\ , &
&R_{(v\pm1)}\ \sim\ \tilde\lambda\,\bigl[(\tilde\lambda-2\tilde\psi^2)(\tilde\lambda-4\tilde\psi^2)-8(\pa_z\tilde\psi)^2\bigr]\ ,\\
&Q_{(v)}\ \quad\sim\ \tilde\lambda\,(\tilde\lambda-4\tilde\psi^2)(\tilde\lambda-6\tilde\psi^2)\ ,\!\quad&
&R_{(v)}\ \quad\sim\ \tilde\lambda\,\bigl[(\tilde\lambda-2\tilde\psi^2)(\tilde\lambda-4\tilde\psi^2)-8(\pa_z\tilde\psi)^2\bigr](\tilde\lambda-6\tilde\psi^2)\ ,
\end{aligned}
\end{equation}
because all $j$-dependent terms in the polynomials are subleading and drop out in the limit.
Factorizing the $R$~polynomials, we find the four universal natural frequency-squares
\begin{equation}
\tilde\Omega^2_1 = 0\ ,\qquad
\tilde\Omega^2_2 = 3\,\tilde\psi^2-\sqrt{\tilde\psi^4+8(\pa_z\tilde\psi)^2}\ ,\qquad
\tilde\Omega^2_3 = 3\,\tilde\psi^2+\sqrt{\tilde\psi^4+8(\pa_z\tilde\psi)^2}\ ,\qquad
\tilde\Omega^2_4 = 6\,\tilde\psi^2\ .
\end{equation}
One must pay attention, however, to the fact that the avoided crossings disappear in the $\epsilon\to0$ limit.
Therefore, the correct limiting frequencies to input into
\begin{equation}
\bigl[ \pa_z - \tilde\Omega^2_{(j,v,\beta)} \bigr]\,\tilde\Phi_{(j,v,\beta)} \= 0
\end{equation}
are
\begin{equation}
\begin{aligned}
&\tilde\Omega^2_{(j,j\pm2)}\ \ = 0\ ,\\
&\tilde\Omega^2_{(j,j\pm1,\beta)} \in \bigl\{ 
\textrm{min}(\tilde\Omega^2_1,\tilde\Omega^2_2),\ \textrm{max}(\tilde\Omega^2_1,\tilde\Omega^2_2),\
\tilde\Omega^2_3 \bigr\}\ ,\\
&\tilde\Omega^2_{(j,j,\beta)} \ \ \ \in \bigl\{ 
\textrm{min}(\tilde\Omega^2_1,\tilde\Omega^2_2),\ \textrm{max}(\tilde\Omega^2_1,\tilde\Omega^2_2),\ 
\textrm{min}(\tilde\Omega^2_3,\tilde\Omega^2_4),\ \textrm{max}(\tilde\Omega^2_3,\tilde\Omega^2_4) \bigr\}\ ,
\end{aligned}
\end{equation}
of which we show below the last list as a function of~$z$. 
\begin{figure}[h!]
\centering
\includegraphics[width = 0.5\paperwidth]{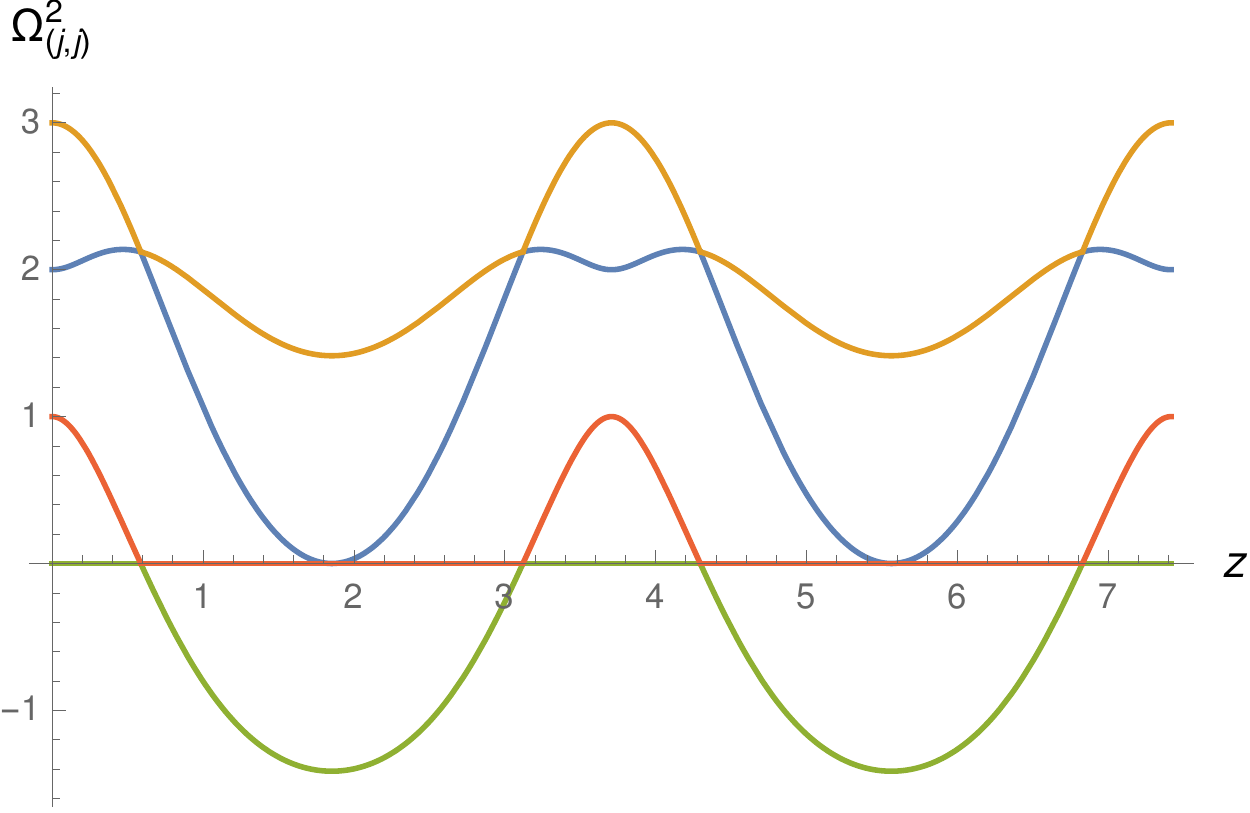} 
\caption{Plot of the universal limiting natural frequency-squares 
$\tilde\Omega^2_{(j,v,\beta)}$ for $v{=}j$ and $\beta=1,2,3,4$.}
\end{figure}
The monodromies are easily computed numerically,\footnote{
For the cases $(j,v)=(0,1)$ and $(1,0)$ one gets $\{56.769,\ -1.659\}$;
for $(j,v)=(0,0)$ we have $\tr\,M=2$.}
\begin{equation}
\begin{aligned}
&\tr\,M_{(j,j\pm2)}(E{\to}\infty)\ \ \ = 2\ ,\\
&\tr\,M_{(j,j\pm1,\beta)}(E{\to}\infty)\ \in \bigl\{ 306.704,\ -1.842,\ -1.659 \bigr\}\ ,\\
&\tr\,M_{(j,j,\beta)}(E{\to}\infty) \ \ \ \ \in \bigl\{ 306.704,\ -1.842,\ 2.462,\ -1.067 \bigr\}\ ,
\end{aligned}
\end{equation}
in agreement with the figures above. In particular, the extremal $v$-values become marginally stable,
while part of the non-extremal cases are unstable for high energies. 

Of course, for each non-extremal value of~$v$ we still have to project out unphysical modes
by imposing the gauge condition~(\ref{gauge4}). However, in the $12(2j{+}1)$-dimensional fluctuation space
the gauge condition has rank~$3(2j{+}1)$ while we see that (for $j{\ge}2$) in total $4(2j{+}1)$ normal modes 
are unstable at high energy. Therefore, the projection to physical modes cannot remove all instabilities. 
We must conclude that, for sufficiently high energy~$E$, some fluctuations grow exponentially,
implying that the solution $\Phi{\equiv}0$ is linearly unstable, and thus is the Yang--Mills background.

\section{Singlet perturbation: exact treatment}

\noindent
Even though the Floquet representation helped to reduce the long-time behavior of the perturbations
to the analysis of a single period~$T$, it normally does not give us an exact solution to Hill's equation.
However, for the SO(4) singlet fluctuation around $\psi(\tau)$, we can employ the fact that 
$\dot{\psi}$ trivially solves the fluctuation equation,
\begin{equation} \label{timeshift}
(\dot\psi)^{\cdot\cdot} = (\ddot\psi)^{\cdot} = -\bigl(V'(\psi)\bigr)^{\cdot} 
= -V''(\psi)\,\dot{\psi} \= -(6\psi^2{-}2)\,\dot\psi \= -\Omega^2_{(0,0)}(\tau)\,\dot\psi\ ,
\end{equation}
with a frequency function which is $\frac{T}{2}$-periodic.
This implies that all fluctuation modes are $T$-periodic.
With the knowledge of an explicit solution to the fluctuation equation 
we can reduce the latter to a first-order equation and solve that one to find a second solution.
The normalizations are arbitrary, so we choose
\begin{equation}
\Phi_1(\tau)\=-\sfrac{\epsilon^3}{k}\,\dot{\psi}(\tau) \qquad\und\qquad
\Phi_2(\tau) \= \Phi_1(\tau)\,\int^\tau\frac{\diff\sigma}{\Phi_1(\sigma)^2}
\= -\sfrac{k}{\epsilon^3}\,\dot{\psi}(\tau)\,\int^\tau \frac{\diff\sigma}{\dot{\psi}^2(\sigma)}\ ,
\end{equation}
which are linearly independent since
\begin{equation}
W(\Phi_1,\Phi_2)\ \equiv\ \Phi_1\dot\Phi_2-\Phi_2\dot\Phi_1 \= 1\ .
\end{equation}

For simplicity, we restrict ourselves to the energy range 
$\sfrac12{<}E{<}\infty$, i.e.~$1{>}k^2{>}\sfrac12$. 
Explicitly, we have
\begin{equation}
\begin{aligned}
\!\!\!\Phi_1(\tau) &\= \epsilon\,\sn\,\dn \ , \\[4pt]
\!\!\!\Phi_2(\tau) &\= \sfrac{1}{1-k^2}\,\cn
\bigl[ (2k^2{-}1)\,\textrm{dn}^2\bigl(\sfrac{\tau}{\epsilon},k\bigr) -k^2 \bigr] 
+ \sn\,\dn \bigl[ \sfrac{\tau}{\epsilon} +
\sfrac{2k^2{-}1}{1{-}k^2}\,E\bigl(\textrm{am}(\sfrac{\tau}{\epsilon},k),k\bigr)\bigr]\;, 
\end{aligned}
\end{equation}
where $\textrm{am}(z,k)$ denotes the Jacobi amplitude and $E(z,k)$ is the elliptic integral of the second kind.
\begin{figure}[h!]
\centering
\includegraphics[width = 0.35\paperwidth]{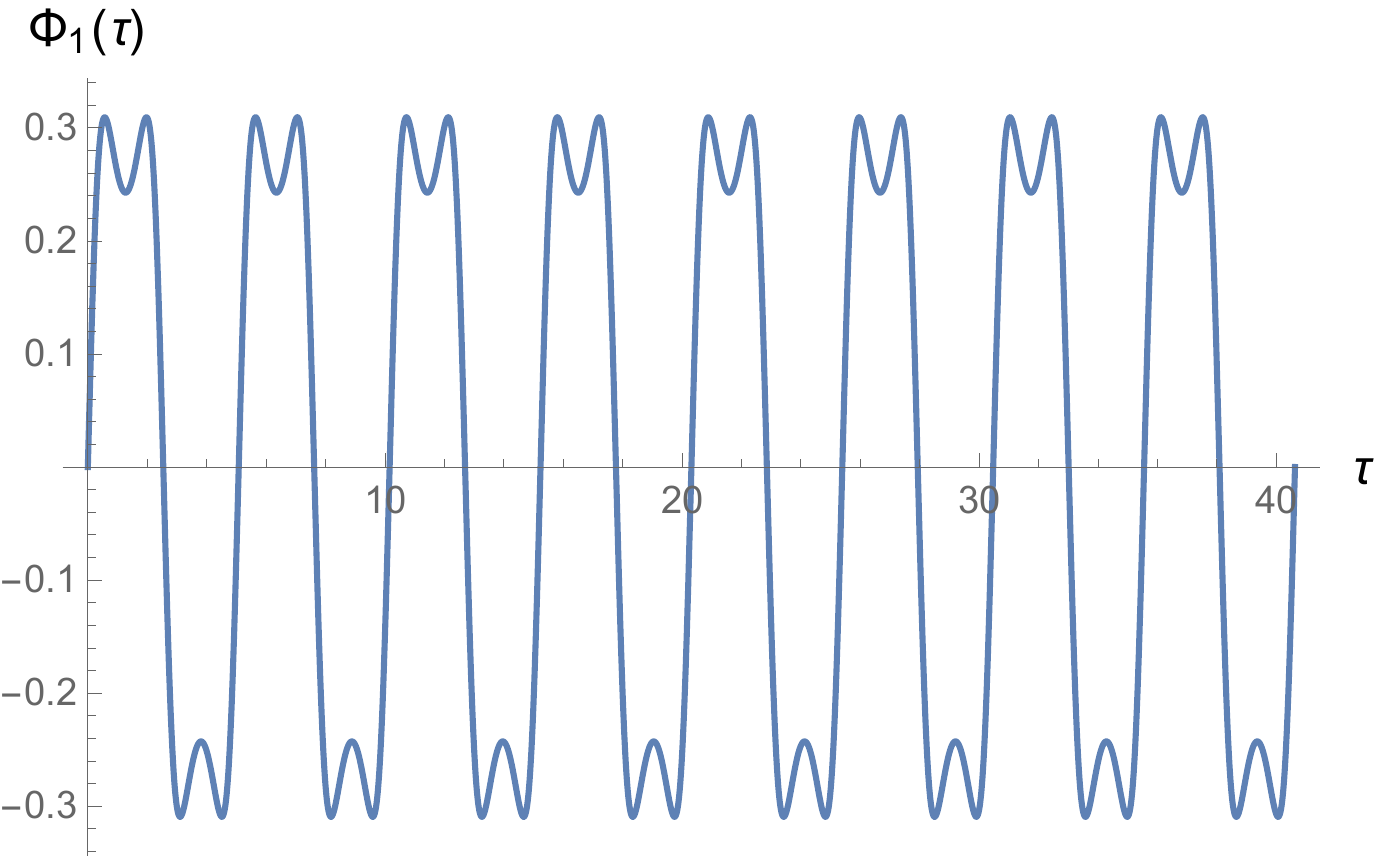} \qquad\quad
\includegraphics[width = 0.35\paperwidth]{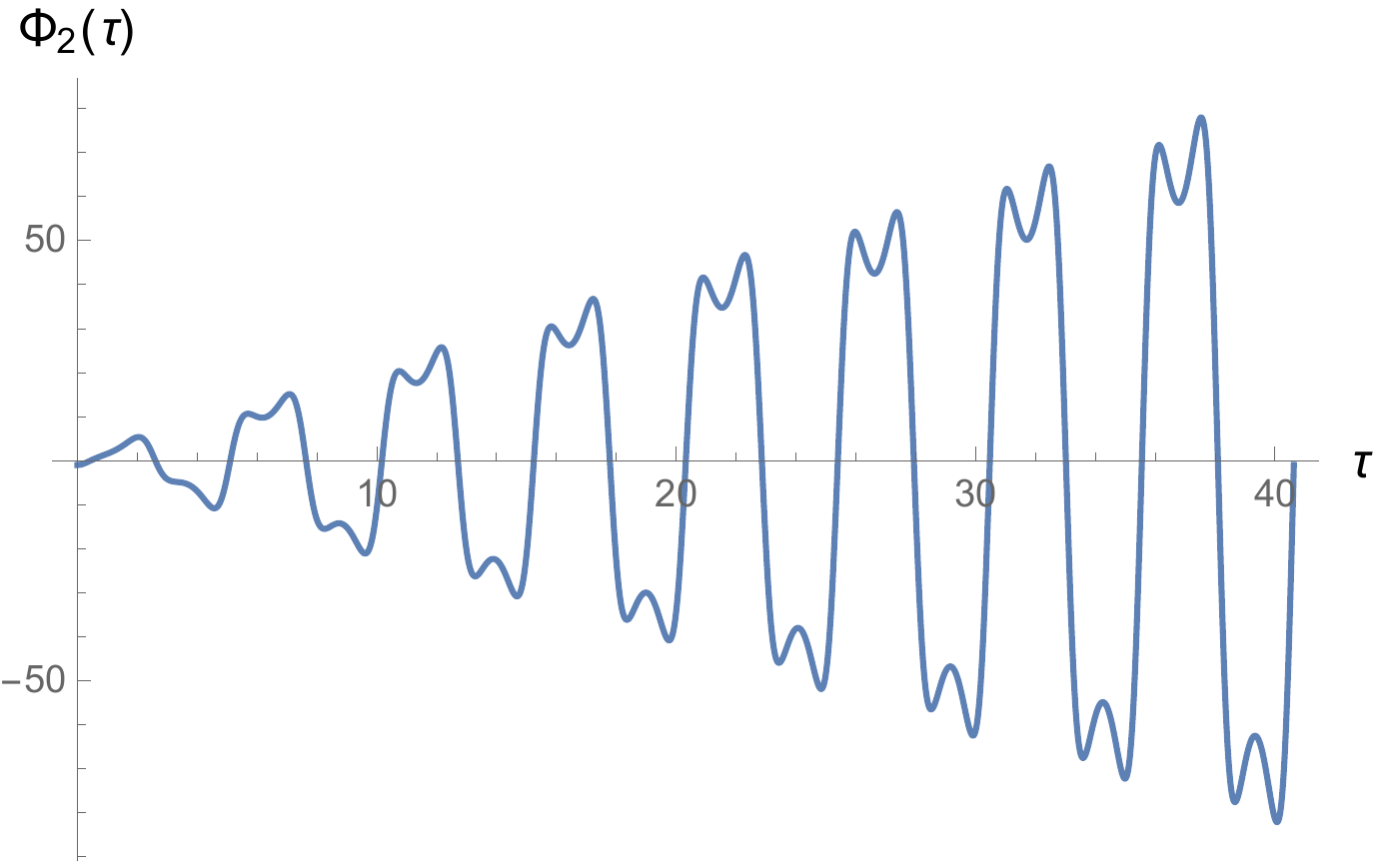} 
\caption{Plot of the SO(4) singlet fluctuation modes $\Phi_1$ and $\Phi_2$ over eight periods for $k^2{=}0.81$.}
\end{figure}
As can be checked, the initial conditions are
\begin{equation}
\Phi_1(0)=0\ ,\quad \dot\Phi_1(0)=1 \qquad\und\qquad
\Phi_2(0)=-1\ ,\quad \dot\Phi_2(0)=0\ ,
\end{equation}
which fixes the ambiguity of adding to $\Phi_2$ a piece proportional to~$\Phi_1$. Hence,
\begin{equation}
\widehat\Phi(0) \= \Bigl(\begin{smallmatrix} 0 & \!{-}1 \\[4pt] 1 & 0 \end{smallmatrix} \Bigr)
\qquad\Rightarrow\qquad
M\= \widehat\Phi(T)\, \Bigl(\begin{smallmatrix} 0 & 1 \\[4pt] {-}1 & 0 \end{smallmatrix} \Bigr)\ .
\end{equation}
We know that $\Phi_1\sim\dot\psi$ is $T$-periodic, and so is $\dot\Phi_1$, 
but not the second solution,
\begin{equation}
\Phi_2(\tau{+}T) \= \Phi_2(\tau) + \gamma\,T\,\Phi_1(\tau) \qquad\textrm{with}\quad
\gamma \= \frac{1}{T}\int_0^T\frac{\diff\sigma}{\Phi_1(\sigma)^2}\bigg|_{\textrm{reg}}
\ =:\ \sfrac{k^2}{\epsilon^6}\, \bigl\langle \dot\psi^{-2}\bigr\rangle_{\textrm{reg}}\ ,
\end{equation}
where the integral diverges at the turning points and must be regularized by subtracting
the Weierstra\ss\ $\wp$~function with the appropriate half-periods.
Since $\Phi_1$ has periodic zeros, $\Phi_2$ does return to~${-}1$ at integer multiples of~$T$.
It follows that the $\Phi_2$ oscillation linearly grows in amplitude with a rate (per period) of
\begin{equation}
\gamma \= \frac{1}{\epsilon^2}\,\Bigl[\,1 + \frac{2k^2{-}1}{1{-}k^2}\,\frac{E(k)}{K(k)} \,\Bigr]\ ,
\end{equation}
which is always larger than 7.629, attained at $k\approx0.882$.
\begin{figure}[h!]
\centering
\includegraphics[width = 0.5\paperwidth]{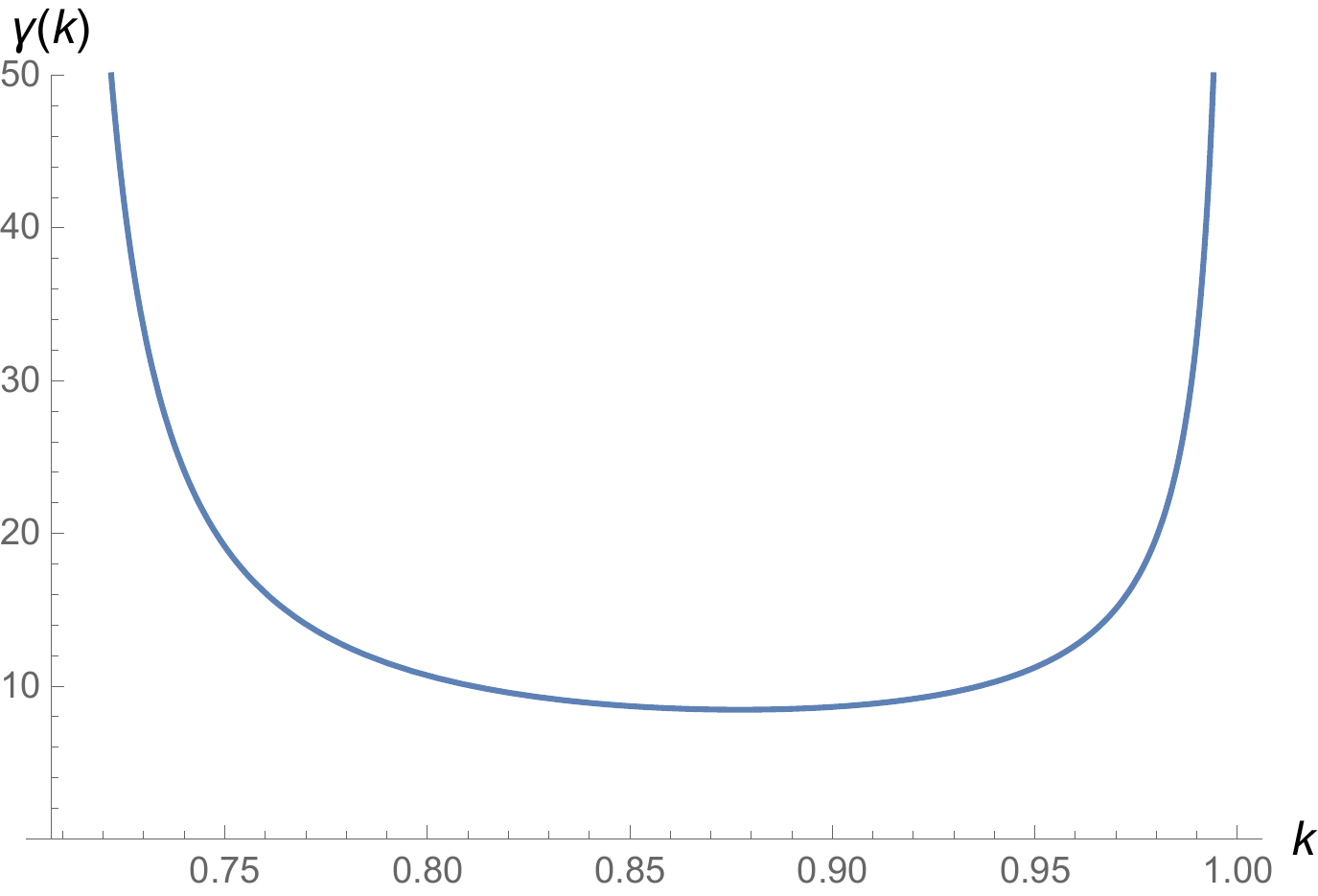} 
\caption{Plot of the linear growth rate~$\gamma$ as a function of~$k$.}
\end{figure}

In essence, we have managed to compute the monodromy
\begin{equation}
M \= \begin{pmatrix} {-}\Phi_2(T) & \Phi_1(T) \\[4pt] {-}\dot\Phi_2(T) & \dot\Phi_1(T) \ \end{pmatrix} 
\= \begin{pmatrix} \ 1 & 0 \\[4pt] \ \!\!{-}\gamma\,T & 1 \ \end{pmatrix} 
\= \exp \Bigl\{ {-}\gamma\,T\,\bigl(\begin{smallmatrix} 0 & 0 \\[4pt] 1 & 0 \end{smallmatrix}\bigr) \Bigr\}
\end{equation}
and thus easily obtain the Floquet representation,
\begin{equation}
R \= \begin{pmatrix} 0 & \gamma \\[4pt] 0 & 0 \end{pmatrix}  \qquad\Rightarrow\qquad
\ep^{\tau R} \= \begin{pmatrix} 1 & \gamma\,\tau \\[4pt] 0 & 1 \end{pmatrix}  \qquad\und\qquad
Q(\tau)\ = \begin{pmatrix} \Phi_1 & \Phi_2{-}\Phi_1\gamma\,\tau \\[4pt] 
\dot\Phi_1 & \dot\Phi_2{-}\dot\Phi_1\gamma\,\tau \end{pmatrix}\ .
\end{equation}
Obviously, we have encountered a marginally stable situation, since $M$ is of parabolic type. 
There is no exponential growth, and $\Phi_1$ is periodic thus bounded, but $\Phi_2$ 
grows without bound as long as one stays in the linear regime.
Note that we never made use of the form of our Newtonian potential.
In fact, this behavior is typical for a conservative mechanical system with oscillatory motion.

What to make of this linear growth? It can be (and actually is) easily overturned by nonlinear effects.
Going beyond the linear regime, though, requires expanding the Yang--Mills equation to higher orders
about our classical Yang--Mills solution~(\ref{Aansatz}).
While this is a formidable task in general, it can actually be done to all orders for the singlet perturbation!
The reason is that a singlet perturbation leaves us in the SO(4)-symmetric subsector,
thus connecting only to a neighboring ``cosmic background'', $\psi\to\tilde\psi$.
Since (\ref{backgrounds}) gives us analytic control over all solutions~$\psi(\tau)$,
the full effect of such a shift can be computed exactly. 
Splitting an exact solution~$\tilde{\psi}$ into a background part and its (full) deviation,
\begin{equation}
\tilde{\psi}(\tau) \= \psi(\tau)\ +\ \eta(\tau)\ ,
\end{equation}
inserting $\tilde{\psi}$ into the equation of motion~(\ref{Newton})
and remembering that $V$ is of fourth order, we obtain
\begin{equation} \label{nonlinear}
0 \= \ddot\eta + V''(\psi)\,\eta + \sfrac12 V'''(\psi)\,\eta^2 + \sfrac16 V''''(\psi)\,\eta^3
\= \ddot\eta + (6\psi^2{-}2)\,\eta + 6\psi\,\eta^2 + 2\,\eta^3\ ,
\end{equation}
extending the linear equation~(\ref{timeshift}) by two nonlinear contributions.
Perturbation theory introduces a small parameter~$\epsilon$ and formally expands
\begin{equation}
\eta \= \epsilon\eta_{(1)} + \epsilon^2\eta_{(2)} + \epsilon^3\eta_{(3)} + \ldots\ ,
\end{equation}
which yields the infinite coupled system
\begin{equation}
\begin{aligned}
& \bigl[\pa_\tau^2+(6\psi^2{-}2)\bigr]\,\eta_{(1)} \= 0 \ ,\\
& \bigl[\pa_\tau^2+(6\psi^2{-}2)\bigr]\,\eta_{(2)} \= -6\psi\,\eta_{(1)}^2 \ ,\\
& \bigl[\pa_\tau^2+(6\psi^2{-}2)\bigr]\,\eta_{(3)} \= -12\psi\,\eta_{(1)}\eta_{(2)} -2\,\eta_{(1)}^3\ ,\\
& \ldots \ ,
\end{aligned}
\end{equation}
which could be iterated with a seed solution~$\eta_{(1)}$ of the linear system.

However, we know that the exact solutions to the full nonlinear equation~(\ref{nonlinear}) 
is simply given by the difference
\begin{equation}
\eta(\tau) \= \tilde\psi(\tau) - \psi(\tau)
\end{equation}
of two analytically known backgrounds. The SO(4)-singlet background moduli space is parametrized by
two coordinates, e.g.~the energy~$E$ (or elliptic modulus~$k$) and the choice of an initial condition
which fixes the origin~$\tau{=}0$ of the time variable. In~(\ref{backgrounds}), we selected $\dot\psi(0)=0$,
but relaxing this we can reintroduce this collective coordinate by allowing shifts in~$\tau$. 
We may then parametrize the SO(4)-invariant Yang--Mills solutions as
\begin{equation}
\psi_{k,\ell}(\tau) \= \psi(\tau{-}\ell) 
\qquad\textrm{with}\qquad 2E=1/(2k^2{-}1)^2 \quad\und\quad \ell\in\R
\end{equation}
where $\psi$ is taken from~(\ref{backgrounds}). 
Note that $\dot\psi_{k,\ell}$ solves the background equation~(\ref{timeshift}) with a frequency-squared
$\omega_{k,\ell}^2=6\psi_{k,\ell}^2{-}2$.
Without loss of generality we assign $\psi=\psi_{k,0}$ and $\tilde\psi=\psi_{k{+}\delta k,\delta\ell}$,
hence
\begin{equation}
\begin{aligned}
\eta(\tau) &\= \delta k\,\pa_k\psi(\tau) - \delta\ell\,\dot\psi(\tau) 
+ \sfrac12(\delta k)^2\,\pa_k^2\psi(\tau) - \delta k \delta\ell\,\pa_k\dot\psi(\tau) 
+ \sfrac12(\delta\ell)^2\,\ddot\psi(\tau) + \ldots \\[4pt]
&\=  \delta k\,\pa_k\psi(\tau{-}\delta\ell) + \sfrac12(\delta k)^2\,\pa_k^2\psi(\tau{-}\delta\ell)
+ \sfrac16(\delta k)^3\,\pa_k^3\psi(\tau{-}\delta\ell) + \ldots\ ,
\end{aligned}
\end{equation}
because $\pa_\ell\psi=-\dot\psi$. 
Clearly, a shift in~$\ell$ only shifts the time dependence of the frequency and does not alter the energy~$E$,
which is not very interesting. Its linear part corresponds to the mode $\Phi_1\sim\dot\psi$ of the previous section.
A change in~$k$, in contract, will lead to a solution with an altered frequency and energy.
Its linear part is given by $\Phi_2$, which grows linearly in time. 
However, due to the boundedness of the full motion, the nonlinear corrections have to limit this growth and
ultimately must bring the fluctuation back close to zero. This is the familiar wave beat phenomenon:
the difference of two oscillating functions, $\tilde\psi$ and $\psi$, with slightly different frequencies,
will display an amplitude oscillation with a beat frequency given by the difference.
This is borne out in the following plots.
\begin{figure}[h!]
\centering
\includegraphics[width = 0.35\paperwidth]{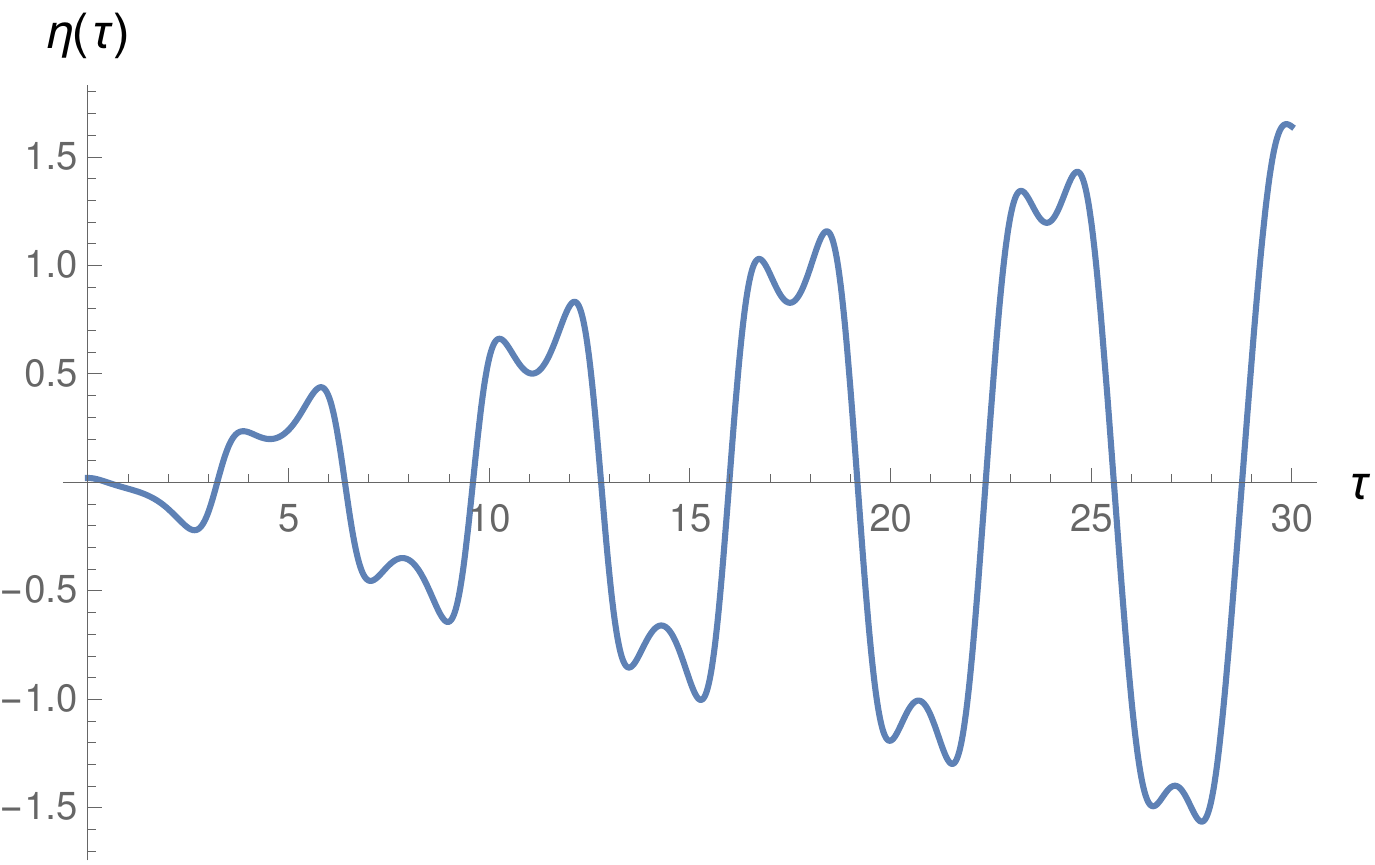} \qquad\quad
\includegraphics[width = 0.35\paperwidth]{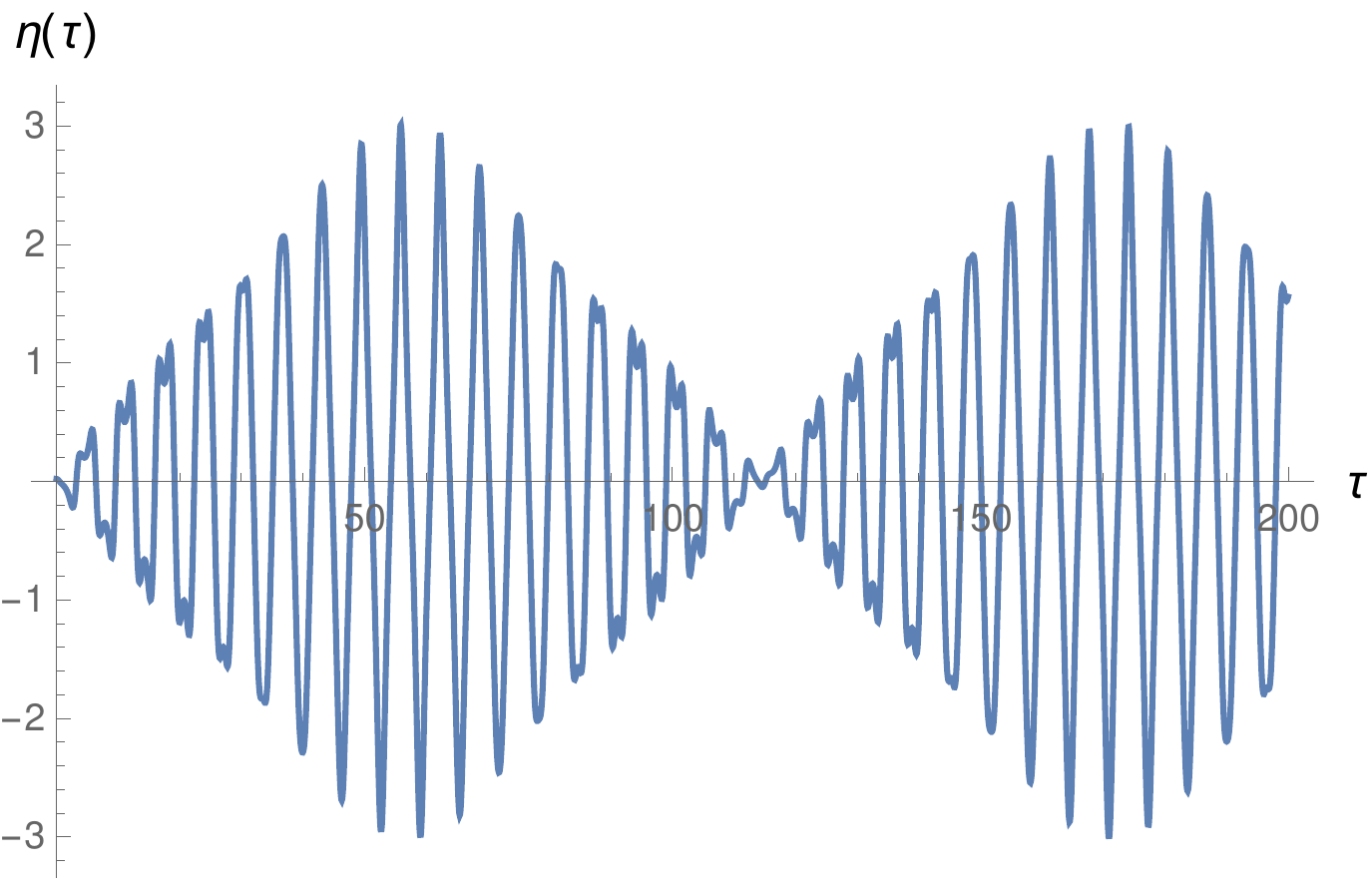}
\caption{Plots of the full perturbation $\eta$ at $k{=}0.95$ for $\eta(0){=}0.02$, $\dot{\eta}(0){=}0$, giving a beat ratio of ${\sim}19$.}
\end{figure}
As a result, we can assert a long-term stability of the cosmic Yang--Mills fields 
against the SO(4) singlet perturbation, even though on shorter time scales an excursion 
to a nearby solution is not met with a linear backreaction.

\section{Conclusions}

\noindent
We have revisited a cosmological scenario recently put forward by Friedan~\cite{Friedan} and based 
on the Standard Model plus gravity alone. An SO(4) symmetric sector is analytically solvable 
and reduces to three coupled anharmonic oscillators (for the metric, an SU(2) Yang--Mills field 
and the Higgs field, the latter being frozen to its vacuum state). We have presented a complete analysis 
of the linear gauge-field perturbations of the time-dependent Yang--Mills solution, by diagonalizing the 
fluctuation operator and studying the long-time behavior of the ensuing Hill's equations using
the stroboscopic map and Floquet theory. For parametrically large gauge-field energy 
(as is required in Friedan's setup) the natural frequencies and monodromies become universal, 
and some unstable perturbation modes survive even in this limit. This provides strong evidence that such 
oscillating cosmic Yang--Mills fields are unstable against small perturbations, although we have 
not yet included metric fluctuations here. Their influence will be analyzed in follow-up work. 


\subsection*{Acknowledgments}
\noindent
K.K.~is grateful to Deutscher Akademischer Austauschdienst (DAAD) for the doctoral research grant~57381412. 

\newpage 

\section*{Appendix} 

\begin{center}
\begin{tabular}{| p{10mm} | p{25mm} | p{45mm} | p{80mm} |} 
\hline
$j\ ,\ v $ & P($\lambda$) & Q($\lambda$)  & R($\lambda$) \\ [0.5ex] 
 \hline\hline
$0\ ,\ 0$ \phantom{\Big|} & N/A & $(-2 + 6 \psi^2) - \lambda$ & $-(2 - 6 \psi^2) + \lambda$ \\
 \hline
$0\ ,\ 1$ \phantom{\Big|} & 1 & $(2 + 2 \psi + 4 \psi^2) - \lambda$ & $(4 + 12\psi + 20 \psi^2 + 20 \psi^3 + 8 \psi^4 - 8 \dot{\psi}^2) - (4 + 6 \psi + 6 \psi^2) \lambda + \lambda^2$ \\
 \hline
$0\ ,\ 2$ \phantom{\Big|} & N/A & $(10 + 6 \psi) - \lambda$ & N/A \\
 \hline
$\sfrac12\ ,\ \sfrac12$ \phantom{\Big|} & $-(1 + 6 \psi^2) + \lambda$ & $-(1 + 2 \psi^2 + 24 \psi^4) + (2 + 10 \psi^2) \lambda - \lambda^2$ & $-(1 + 4 \psi^2 + 28 \psi^4 + 48 \psi^6 - 8 \dot{\psi}^2 - 48 \psi^2 \dot{\psi}^2) + (3 + 16 \psi^2 + 44 \psi^4 - 8 \dot{\psi}^2) \lambda - (3 + 12 \psi^2) \lambda^2 + \lambda^3$ \\ 
 \hline
$\sfrac12\ ,\ \sfrac32$ \phantom{\Big|} & $-(7 + 3 \psi) + \lambda$ & $-(49 + 42 \psi + 32 \psi^2 + 12 \psi^3)   + (14 + 6 \psi + 4 \psi^2) \lambda - \lambda^2$ & $-(343 + 588 \psi + 574 \psi^2 + 360 \psi^3 + 136 \psi^4 + 24 \psi^5 - 56 \dot{\psi}^2 - 24 \psi \dot{\psi}^2) + (147 + 168 \psi + 124 \psi^2 + 48 \psi^3 + 8 \psi^4 - 8 \dot{\psi}^2) \lambda - (21 + 12 \psi + 6 \psi^2) \lambda^2 + \lambda^3$ \\ 
 \hline
$\sfrac12\ ,\ \sfrac52$ \phantom{\Big|} & N/A & $(17 + 8 \psi) - \lambda$ & N/A \\
 \hline
$1\ ,\ 0$ \phantom{\Big|} & $1$ & $(2 - 2 \psi + 4 \psi^2) - \lambda$ & $( 4 - 12 \psi + 20 \psi^2 - 20 \psi^3 + 8 \psi^4 - 8 \dot{\psi}^2) - (4 - 6 \psi + 6 \psi^2) \lambda + \lambda^2$ \\
 \hline
$1\ ,\ 1$ \phantom{\Big|} & $(36 + 36 \psi^2) - (12 + 6\psi^2) \lambda + \lambda^2$ & $(216 + 192 \psi^2 + 104 \psi^4) - (108 + 92 \psi^2 + 24 \psi^4) \lambda + (18 + 10 \psi^2) \lambda^2 - \lambda^3$ & $(1296 + 1584 \psi^2 + 1008 \psi^4 + 208 \psi^6 - 288 \dot{\psi}^2 - 288 \psi^2 \dot{\psi}^2) - (864 + 960 \psi^2 + 432 \psi^4 + 48 \psi^6 - 96 \dot{\psi}^2 - 48 \psi^2 \dot{\psi}^2) \lambda + (216 + 188 \psi^2 + 44 \psi^4 - 8 \dot{\psi}^2) \lambda^2 - (24 + 12\psi^2) \lambda^3 + \lambda^4$ \\
 \hline
$1\ ,\ 2$ \phantom{\Big|} & $-(14 + 4 \psi) + \lambda$ & $-(196 + 112 \psi + 60 \psi^2 + 16 \psi^3) + (28 + 8 \psi + 4 \psi^2) \lambda - \lambda^2$ & $-(2744 + 3136 \psi + 2128 \psi^2 + 928 \psi^3 + 248 \psi^4 + 32 \psi^5 - 112 \dot{\psi}^2 - 32 \psi \dot{\psi}^2) + (588 + 448 \psi + 236 \psi^2 + 64 \psi^3 + 8 \psi^4 - 8 \dot{\psi}^2) \lambda - (42 + 16 \psi + 6 \psi^2) \lambda^2 + \lambda^3$ \\
 \hline
$1\ ,\ 3$ \phantom{\Big|} & N/A & $(26 + 10 \psi) - \lambda$ & N/A \\ 
 \hline
$\sfrac32\ ,\ \sfrac12$ \phantom{\Big|} & $-(7 - 3 \psi ) + \lambda$ & $-(49 - 42 \psi + 32 \psi^2 - 12 \psi^3) + (14 - 6 \psi+ 4 \psi^2) \lambda - \lambda^2$ & $-(343 - 588 \psi + 574 \psi^2 - 360 \psi^3 + 136 \psi^4 - 24 \psi^5 - 56 \dot{\psi}^2 + 24 \psi \dot{\psi}^2) + (147 - 168 \psi + 124 \psi^2 - 48 \psi^3 + 8 \psi^4 - 8 \dot{\psi}^2) \lambda - (21 - 12 \psi + 6 \psi^2) \lambda^2 + \lambda^3$ \\ 
 \hline
$\sfrac32\ ,\ \sfrac32$ \phantom{\Big|} & $(169 + 78 \psi^2) - (26 + 6 \psi^2) \lambda + \lambda^2$ & $(2197 + 962 \psi^2 + 216 \psi^4) - (507 + 204 \psi^2 + 24 \psi^4) \lambda + (39 + 10 \psi^2) \lambda^2 - \lambda^3$ & $(28561 + 16900 \psi^2 + 4732 \psi^4 + 432 \psi^6 - 1352 \dot{\psi}^2 - 624 \psi^2 \dot{\psi}^2) + (-8788 - 4628 \psi^2 - 936 \psi^4 - 48 \psi^6 + 208 \dot{\psi}^2 + 48 \psi^2 \dot{\psi}^2) \lambda + (1014 + 412 \psi^2 + 44 \psi^4 - 8 \dot{\psi}^2) \lambda^2 - (52 + 12 \psi^2) \lambda^3 + \lambda^4$ \\
 \hline
$\sfrac32\ ,\ \sfrac52$ \phantom{\Big|} & $-(23 + 5 \psi) + \lambda$ & $-(529 + 230 \psi + 96 \psi^2 + 20 \psi^3) + (46 + 10 \psi + 4 \psi^2) \lambda - \lambda^2$ & $-(12167 + 10580 \psi + 5566 \psi^2 + 1880 \psi^3 + 392 \psi^4 + 40 \psi^5 - 184 \dot{\psi}^2 + 40 \psi\dot{\psi}^2) + (1587 + 920 \psi + 380 \psi^2 + 80 \psi^3 + 8 \psi^4 - 8 \dot{\psi}^2) \lambda + (-69 - 20 \psi - 6 \psi^2) \lambda^2 + \lambda^3$ \\
 \hline
$\sfrac32\ ,\ \sfrac72$ \phantom{\Big|} & N/A & $(37 + 12 \psi) - \lambda$ & N/A \\ 
 \hline
$2\ ,\ 0$ \phantom{\Big|} & N/A & $(10- 6 \psi) - \lambda$ & N/A \\
 \hline
$2\ ,\ 1$ \phantom{\Big|} & $-(14 - 4 \psi) + \lambda$ & $-(196 - 112 \psi + 60 \psi^2 - 16 \psi^3) + (28 - 8 \psi + 4 \psi^2) \lambda - \lambda^2$ & $-(2744 - 3136 \psi + 2128 \psi^2 - 928 \psi^3 + 248 \psi^4 - 32 \psi^5 - 112 \dot{\psi}^2 + 32 \psi \dot{\psi}^2) + (588 - 448 \psi + 236 \psi^2 - 64 \psi^3 + 8 \psi^4 - 8 \dot{\psi}^2) \lambda - (42 - 16 \psi + 6 \psi^2) \lambda^2 + \lambda^3$ \\
 \hline
$2\ ,\ 2$ \phantom{\Big|} & $(484 + 132 \psi^2) - (44 + 6 \psi^2) \lambda + \lambda^2$ & $(10648 + 2816 \psi^2 + 360 \psi^4) - (1452 + 348 \psi^2 + 24 \psi^4) \lambda + (66 + 10 \psi^2) \lambda^2 - \lambda^3$ & $(234256 + 83248 \psi^2 + 13552 \psi^4 + 720 \psi^6 - 3872 \dot{\psi}^2 - 1056 \psi^2 \dot{\psi}^2) - (42592 + 13376 \psi^2 + 1584 \psi^4 + 48 \psi^6 - 352 \dot{\psi}^2 - 48 \psi^2 \dot{\psi}^2) \lambda + (2904 + 700 \psi^2 + 44 \psi^4 - 8 \dot{\psi}^2) \lambda^2 - 
 (88 + 12 \psi^2) \lambda^3 + \lambda^4$ \\
 \hline
$2\ ,\ 3$ \phantom{\Big|} & $-(34 + 6 \psi) + \lambda$ & $-(1156 + 408 \psi + 140 \psi^2 + 24 \psi^3) + (68 + 12 \psi + 4 \psi^2) \lambda - \lambda^2$ & $-(39304 + 27744 \psi + 11968 \psi^2 + 3312 \psi^3 + 568 \psi^4 + 48 \psi^5 - 272 \dot{\psi}^2 - 48 \psi \dot{\psi}^2) + (3468 + 1632 \psi + 556 \psi^2 + 96 \psi^3 + 8 \psi^4 - 8 \dot{\psi}^2) \lambda - (102 + 24 \psi + 6\psi^2) \lambda^2 + \lambda^3$ \\ 
 \hline
$2\ ,\ 4$ \phantom{\Big|} & N/A & $(50 + 14 \psi) - \lambda$ & N/A \\ 
 \hline
\end{tabular}
\end{center}

\pagebreak

\end{document}